%% file: paper.tex
\pdfoutput=1

\documentclass[a4paper,11pt]{article}

\usepackage{import}

\usepackage[utf8]{inputenc}
\usepackage[top=2cm, bottom=2cm, left=2.5cm, right=2.5cm]{geometry}
\usepackage{hyperref} 
\usepackage{enumitem} 
\usepackage{blindtext} 
\usepackage{fancyvrb} 
\usepackage{caption}
\usepackage{subcaption}
\usepackage{float} 
\renewcommand{\,}{\hspace{1pt}}
\renewcommand{\_}{\ul{\hspace{7pt}}}

\usepackage{tikz-cd} 
\usepackage{forest} 

\usepackage{amsmath} 
\usepackage{amssymb} 
\usepackage{stmaryrd} 
\usepackage{cmll}

\usepackage{amsthm}
\usepackage[capitalize]{cleveref}
\usepackage{thmtools}
\usepackage{thm-restate}
\declaretheorem[numberwithin=section]{theorem}
\declaretheorem[sibling=theorem]{corollary}
\declaretheorem[sibling=theorem]{proposition}

\declaretheorem[sibling=theorem]{example}
\declaretheorem[sibling=theorem, style=definition]{definition}
\declaretheorem[sibling=theorem, style=definition]{remark}

\usepackage[style=numeric, maxnames=10]{biblatex}
\bibliography{../Biblio/category_theory}
\bibliography{../Biblio/proof_assistants}
\bibliography{../Biblio/nested_datatypes}
\bibliography{../Biblio/modules_over_monads}
\bibliography{../Biblio/sigma-monoids}
\bibliography{../Biblio/hss}
\bibliography{../Biblio/unclassified}

\newcommand{\mc}[1]{\mathcal{#1}}
\newcommand{\mb}[1]{\mathbb{#1}}
\newcommand{\trm}[1]{\textrm{#1}}
\newcommand{\mrm}[1]{\mathrm{#1}}

\newcommand{\br}[1]{\ensuremath{\llbracket #1 \rrbracket}} 
\newcommand{\ol}[1]{\overline{#1}}
\newcommand{\ul}[1]{\underline{#1}}
\usepackage{scalerel}
\DeclareMathOperator*{\bigplus}{\scalerel*{+}{\sum}}

\newcommand{\N}{\mathbb{N}}

\newcommand{\Cmon}{(\mc{C}, \otimes, I, \alpha, \lambda, \rho)}
\newcommand{\Set}{\mrm{Set}}
\newcommand{\Mon}{\mrm{Mon}}
\newcommand{\Mod}{\mrm{Mod}}
\newcommand{\Model}{\mrm{Model}}
\newcommand{\Sig}{\mrm{Sig}}
\newcommand{\SigStrength}{\mrm{SigStrength}}
\newcommand{\TotMonMod}[1]{\int_{R : \Mon(#1)} \Mod(R)}
\newcommand{\TotSigModel}[1]{\int_{\Sigma : \Sig(#1)} \Model(\Sigma)}
\newcommand{\op}{\mrm{op}}
\newcommand{\Id}{\mrm{Id}}
\newcommand{\Lan}[2]{\mrm{Lan}_{#1}(#2)}

\newcommand{\List}{\mrm{List}}
\newcommand{\PCF}{\mrm{PCF}}
\newcommand{\Nat}{\mrm{Nat}}
\newcommand{\Bool}{\mrm{Bool}}
\newcommand{\Free}{\mrm{Free}}
\newcommand{\var}{\mrm{var}}
\newcommand{\app}{\mrm{app}}
\newcommand{\abs}{\mrm{abs}}
\newcommand{\inl}{\mrm{inl}}
\newcommand{\inr}{\mrm{inr}}
\newcommand{\swap}{\mrm{swap}}

\title{An Introduction to Different Approaches to Initial Semantics}
\author{Thomas Lamiaux \and Benedikt Ahrens}
\date{}

\begin{document}

\maketitle

\begin{abstract}
  In this work we consider initial semantics for programming languages --- in particular, untyped and simply-typed --- with variable binding.
  Initial semantics aims to characterize the syntax of a such a language, with its substitution structure, as an initial object in a category.
  Initiality then gives rise to a recursion principle that can be used to define substitution-safe translations between languages.

  Characterizing programming languages with variable binding as initial objects, was first achieved by Fiore, Plotkin, and Turi in their seminal paper published at LICS'99.
  To do so, in particular to prove initiality theorems, they developed a framework based on monoidal categories, functors with strengths, and $\Sigma$-monoids.
  An alternative approach using modules over monads was later introduced by Hirschowitz and Maggesi, for endofunctor categories, that is, for particular monoidal categories.
  This approach has the advantage of providing a more general and abstract definition of signatures and models;
  however, no general initiality result is known for this notion of signature.
  Furthermore, Matthes and Uustalu provided a categorical formalism for constructing (initial) monads via Mendler-style recursion, that can also be used for initial semantics.
  The different approaches have been developed further in several articles.
  However, in practice, the literature is difficult to access, and links
  between the different strands of work remain underexplored.

  In the present work, we give an introduction to initial semantics that encompasses the three different strands.
We develop a suitable ``pushout`` of Hirschowitz and Maggesi's framework with Fiore's, and rely on Matthes and Uustalu's formalism to provide modular proofs.
For this purpose, we generalize both Hirschowitz and Maggesi's framework, and Matthes and Uustalu's formalism to the general setting of monoidal categories studied by Fiore and collaborators.
Moreover, we provide fully worked out presentation of some basic instances of the literature, and an extensive discussion of related work explaining the links between the different approaches.

\end{abstract}
\newpage

\tableofcontents
\newpage

\import{Sections/}{all.tex}

\section{Acknowledgements}

We thank Ambroise Lafont for fruitful discussion on initiality theorems and their different proofs,
and Ralph Matthes for helpful discussions on heterogeneous substitution systems and their use to prove the initiality theorem.
We furthermore thank Marcelo Fiore for providing pointers to literature on the topic of initial semantics.
We are grateful to André Hirschowitz, Thea Li, Ambroise Lafont, and Ralph Matthes for valuable comments on early drafts.

\addcontentsline{toc}{section}{References}
\printbibliography

\end{document}

%% file: Sections/all.tex
\subimport{}{introduction.tex}
\subimport{}{preliminaries.tex}
\subimport{}{introduction_is.tex}

\subimport{}{models.tex}

\subimport{}{initiality_theorem.tex}

\subimport{}{building_initial_model.tex}

\subimport{}{example_Set_Set.tex}

\subimport{}{example_F_Set.tex}

\subimport{}{example_SetT_SetT.tex}
\subimport{}{related_work.tex}

\subimport{}{conclusion.tex}

%% file: Sections/introduction.tex
\section{Introduction}

\emph{Initial semantics} as a concept aims to characterize (higher-order)
programming languages as initial objects in well-chosen categories of
``models''.
Considering programming languages categorically, as appropriate initial
models, has many advantages.
By choosing models of a programming language to be appropriate algebras,
models provide us with a mathematical abstraction of the syntax of the
language, including variable-binding constructions.
Indeed, in a model, the terms and the constructors of the language are
represented as abstract categorical notions that capture their properties.
This enables us to manipulate the language by only considering these categorical notions, without
having to constantly deal with its syntax and implementation, in particular
with the implementation of binders.
Furthermore, the initiality of the model provides us with an abstract
recursion principle.
Indeed, informally, being initial for a model means that among the models
that abstract syntax, it is the ``smallest'' one, i.e. that it is
``embedded'' in any other model.
By unfolding the definitions, this provides us with a recursion principle in the
same categorical terms as the ones used for the abstract syntax,
thus enabling us to reason by recursion on the abstract syntax without
having to deal with any underlying implementation.
Initial models can therefore abstract the syntax with its recursion
principle.

Using more advanced notions of model, it is possible to account for syntax but
also for operational semantics.
In this work, we are concerned with accounting for the syntax of higher-order
languages with their substitution structures \cite{FPT99,HirschowitzMaggesi07,Hss04}.
This involves abstracting the syntax,
but also abstracting how simultaneous substitution works, its properties, and how it
interacts with constructors.
In particular, we refer to initial semantics as the theory of designing
models of languages, accounting for syntax and substitution, and proving initiality
theorems; such theorems assert, ideally under simple conditions, that a
language has an initial model.
Initiality theorems enable us to prove that a language has a well-defined
substitution structure, but by initiality, they also enable us to equip the
language with a substitution-safe recursion principle.
By this, we mean a recursion principle that respects the substitution structure by
construction, which is important if one wishes to preserve the operational semantics of
languages.
There are more advanced notions of initial semantics that enable us to
integrate meta-variables \cite{SecondOrderDep08}, equations between terms
\cite{FioreHur10,2Signatures19}, or reductions rules
\cite{UntypedRelativeMonads16,ReductionMonads20}, into the models, and to
obtain recursion principles that preserve the operational semantics by
construction, but we will not discuss them here.

\subsection{Different Approaches to Initial Semantics}

The starting point for this work is the observation that, while
prolific, the existing literature on initial semantics seems, to us,
difficult to access for newcomers to the field.
Indeed, there exist different, more or less known, approaches to initial
semantics, and the formal links between them are understudied, and
still largely unknown.
Moreover, many important notions are spread out over different papers,
often with small variations.
This is especially complicated as many papers have been published as
extended abstract or in conference proceedings, and as such, usually contain
limited related work, few fully worked-out examples, and often no proofs or
only sketches of proofs.

\subsubsection{Different Traditions}

The most well-known approach to initial semantics, but also the first one to
have been invented, was introduced by Fiore, Plotkin, and Turi in their
seminal paper \cite{FPT99}.
In \cite{FPT99}, focusing on untyped languages, the authors introduced
$\Sigma$-monoids to capture languages with their substitution, and suggested
a very general method based on monoidal categories and functors with strength
to prove initiality theorems.
This framework has been developed further to encompass richer type systems
\cite{Cbn02,MMCCS05,SecondOrderDep08,Polymorphism11,ListObjects17},
but also to add meta-variables and equations
\cite{HamanaMetavar04,SecondOrderDep08,HurPhd,PolymorphismEq13,FioreSzamozvancevPopl22}.
It has also recently been extended to skew-monoidal categories \cite{CellularHoweTheorem20}.

Another approach to initial semantics, alternative to $\Sigma$-monoids, was later developed by André
Hirschowitz and Maggesi \cite{HirschowitzMaggesi07}, using modules over
monads.
This approach has the advantage of providing a very general and abstract
definition of signatures and models
\cite{HirschowitzMaggesi12,ZsidoPhd10,PresentableSignatures21}, that can be
easier to manipulate.
For instance, it has been extended to account for equations
\cite{PresentableSignatures21,2Signatures19} and reduction rules
\cite{UntypedRelativeMonads16,TypedRelativeMonads19,ReductionMonads20,TransitionMonads22}.
Yet, modules over monads and associated notions have only been defined and
studied for endofunctor categories, that is, for particular monoidal categories, and only restricted and instance-specific initiality theorems have been proven \cite{HirschowitzMaggesi10,ZsidoPhd10},
for instance for the category $[\Set,\Set]$.

A third more marginal approach based on heterogeneous substitution systems,
abbreviated as hss, exists.
Hss were originally invented to prove that both wellfounded and
non-wellfounded syntax \cite{Hss04} have a monadic substitution structure.
It was applied in this way in the UniMath library, to construct untyped and
simply-typed higher-order languages \cite{HssUntypedUniMath19,HssTypedUnimath22}.
Yet, hss were also shown to provide a framework for initial semantics by
Ahrens and Matthes in \cite{HssRevisited15}.
This approach is based on functors with strength and endofunctor categories,
and has general initiality theorems, and seems to have a stronger notion
of models.

\subsubsection{Understanding and relation the different approaches}

Aside from the fragmentation of the literature, understanding and relating
the different approaches can be challenging as they have several structural
differences.

The different frameworks are not all defined for the same underlying structures.
The $\Sigma$-monoid tradition is defined for generic monoidal categories
\cite{SecondOrderDep08}, whereas the hss and modules over monads traditions
only deal with endofunctor categories \cite{Hss04,PresentableSignatures21},
very particular monoidal categories.
The different frameworks also have a different relationship towards initiality.
The $\Sigma$-monoid tradition provides a generic adjoint theorem
\cite{SecondOrderDep08}, from which an initiality theorem is deduced,
whereas the hss one only has an initiality theorem \cite{HssRevisited15},
and the modules over monads tradition only has initiality theorems for very
specific endofunctor categories like $[\Set,\Set]$ or $[\Set^T,\Set^T]$
\cite{HirschowitzMaggesi10},\cite[Section 6]{ZsidoPhd10}.
Different versions of the initiality theorems are also available in the
literature, whatever the tradition.
For instance, some rely on closedness \cite{SecondOrderDep08,HssRevisited15},
while others rely on $\omega$-cocontinuity \cite{ListObjects17,HssUntypedUniMath19}.
More generally, the different frameworks also use different notions of signatures
and models, and it is not always clear how these notions relate to each other.

Relating the different approaches has been investigated in the literature,
but the result are little known, do not cover the full extent of the variations
mentioned before, and are sometimes very specific.
For untyped algebraic signatures, the links between $\Sigma$-monoids on
$[\mb{F},\Set]$, and module based models on $[\Set,\Set]$, have been
investigated in Zsidó's dissertation \cite{ZsidoPhd10}.
Zsidó also investigated the links for simply-typed algebraic signatures, for
the categories $[\mb{F} \downarrow T,\Set]^T$ and $[\Set^T,\Set^T]$.
For the particular monoidal category $[\Set,\Set]$, signatures with strength
and $\Sigma$-monoids have been shown to be a particular case of signatures and
modules based models \cite{HirschowitzMaggesi12}.
Recently, and independently from us, heterogeneous substitution systems
have been shown to be an adequate abstraction to prove initiality result
for $\Sigma$-monoids, as briefly mentioned in \cite[Section 4.4]{HssNonWellfounded23}.

\subsection{Contributions}

\subsubsection{An Introduction to the Different Approaches}

As we have seen, the literature on initial semantics is not the easiest to
grasp, particularly as there exist different insufficiently related traditions,
with different small variations.
With this work, we aim at bridging this gap in the literature by providing an
introduction to the different approaches to initial semantics; that enables us
to understand each tradition on its own, but also in relation to each other.
Such that, from then on, one can explore more advanced notions of initial
semantics, for instance, initial semantics dealing with meta-variables or
reductions rules.

To do so, we have gathered and suitably combined and generalised the
different approaches mentioned before.
Our presentation is following the design choices below:
\begin{enumerate}
  \setlength\itemsep{-1pt}
  \item We base the framework on monoidal categories as these encompass the
        great majority of the existing instances.
        In particular, they encompass instances like $[\mb{F},\Set]$,
        that are not endofunctor categories.
  \item We present the framework, signatures and models, through modules
        over monoids, a variant of modules over monads, to provide an
        abstract and easy-to-manipulate framework.
  \item We exhibit signatures with strength and $\Sigma$-monoids as
        particular signatures and models, respectively, and use them to provide an initiality and an
        adjoint theorem.
  \item We use heterogeneous substitution systems to prove the
        initiality theorem and the adjoint theorem, efficiently and
        modularly.
\end{enumerate}
This enables us to recover the strength and results of each approach, while
still covering the different instances in the literature, which are detailed
in \cref{sec:example_set_set,sec:example_F_set,sec:example_setT_setT}.
To do so, we make several technical but important design choices, and
modifications to the existing frameworks. Most noticeably:
\begin{enumerate}
  \setlength\itemsep{-1pt}
  \item We generalise the existing frameworks on modules over monads from endofunctors categories
        to monoidal categories, and consequently switch from modules over monads to modules over monoids.
  \item We separate the initiality theorem from the adjoint theorem, and
        rely on $\omega$-cocontinuity rather than on closedness.
  \item We generalise hss from endofunctor categories to monoidal categories,
        and generalise and adapt a proof to prove our theorems.
\end{enumerate}
We fully justify our design choices in an extensive discussion of related work in
\cref{sec:related-work}, where we compare the different approaches to our presentation.
In particular, this enables us to shed light onto the literature, and to explain
the links between the different approaches.

\subsubsection{A Pedagogical Presentation}

As the literature is complicated to get into, with this work, we also aim at
giving an accessible and pedagogical presentation to initial semantics, and
its different approaches.

To that end, we give a complete introduction to all basic concepts, and
accompany them with an extensive discussion of related work in
\cref{sec:related-work}, discussing our choices of design, but also the
history of each tradition, and their variations.
We also fully work out and spell out different basic instances and
examples of the literature in
\cref{sec:example_set_set,sec:example_F_set,sec:example_setT_setT},

In this spirit, the paper is also as self-contained as possible, and only requires
an understanding of basic categorical notions such as functors, natural transformations,
and adjoints as found, e.g., in Riehl's book \cite{CategoryTheoryInContext14}.
All other required notions can be found in the preliminaries, in \cref{sec:prelims}.
We also provide in \cref{sec:intro_is}, a brief, gradual and pedagogical presentation
of the concept of initial semantics.

\subsection{Synopsis}
\label{sec:synopsis}

In \cref{sec:prelims}, we briefly review some definitions of category theory
that we use later in the paper.
The reader is encouraged to only consult this section when needed.
We also provide a brief introduction to the concept of initial semantics
in \cref{sec:intro_is}.

We present our framework for initial semantics, from
\cref{sec:models,sec:initiality_theorem,sec:building_initial_model}.
Specifically, we present signatures and models based on modules over monoids
in \cref{sec:models}.
As not all signatures have an initial model, we consider signatures with
strength and $\Sigma$-monoids to state an initiality theorem and an adjoint
theorem in \cref{sec:initiality_theorem}.
We then provide a detailed proof of the initiality theorem, based on
heterogeneous substitution systems, in \cref{sec:building_initial_model}.

We fully work out the basic instances of the literature, from
\cref{sec:example_set_set,sec:example_F_set,sec:example_setT_setT}.
We first explain how untyped higher-order languages are encompassed by
our framework, using infinite context i.e. the category $[\Set,\Set]$ in \cref{sec:example_set_set},
and using finite context i.e. the category $[\mb{F},\Set]$, in \cref{sec:example_F_set}.
We then consider how simply-typed higher-order languages are encompassed
using infinite contexts i.e. the category $[\Set^T,\Set^T]$, in
\cref{sec:example_setT_setT}.

We provide an extensive analysis of the literature, and of the links with our
presentation, in \cref{sec:related-work}.
We discuss limitations of the current framework, and open problems, in \cref{sec:conclusion}.

%% file: Sections/preliminaries.tex
\section{Preliminaries}
\label{sec:prelims}

The framework defined herein is entirely written in the language of category
theory.
We assume the reader is familiar with the notions of category, functor, and
natural transformation, as found, e.g., in Riehl's book \cite{CategoryTheoryInContext14}.
We recall here some of the more specific definitions and properties required
to understand the framework: monoidal categories, $\omega$-colimits and
their basic properties, and left Kan extensions and their characterization as
coends.

\subsection{Monoidal Categories}

To be able to define our notion of models, we need more structure on the
underlying category than the one provided by the mere definition of a
category.
Hence, throughout the paper, we work with monoidal categories.
A modern presentation of monoidal categories can be found in
\cite[Section 3.1]{2DimensionalCategories20}.

\begin{definition}[Monoidal Categories]
  \label{def:mon-cat}
  A \emph{monoidal category} is a tuple $\Cmon$, where $\mc{C}$ is a category, $\_
  \otimes \_ : \mc{C} \times \mc{C} \to C$ is a bifunctor called the
  monoidal product, and $I : \mc{C}$ an object called the unit.
  $\alpha,\lambda, \rho$ are natural isomorphisms called the associator, and
  left and right unitor:
  \begin{align*}
    \alpha_{X,Y,Z} : (X \otimes Y) \otimes Z \cong X \otimes (Y \otimes Z)
    &&
    \lambda_{X} : I \otimes X \cong X
    &&
    \rho_{X} : X \otimes I \cong X
  \end{align*}
  that respect the unit axiom and the pentagon axiom.
\end{definition}

\begin{remark}
  The above notations are fairly standard, however be careful that in some
  references, as in \cite{HssRevisited15,HssUntypedUniMath19}, $\rho_X$ and
  $\lambda_X$ are swapped.
\end{remark}

\begin{example}[Category of endofunctors]
  Given any category $\mc{C}$, the category of endofunctors $[\mc{C},\mc{C}]$
  is monoidal for the composition of functors as monoidal product, the
  identity functor $\Id$ as unit, and the identity natural transformation
  for $\alpha,\lambda,\rho$.
\end{example}

\begin{definition}[Monoidal functors]
  \label{def:monoidal-functor}
  A \emph{monoidal functor} $(\mc{C},I,\otimes) \to (\mc{D},J,\bullet)$ is a
  tuple $(F,F_0,F_2)$ such that $F : \mc{C} \to \mc{D}$ is a functor, $F_0 :
  J \to F(I)$ is a morphism of $\mc{D}$, and $F_2 : F(X) \bullet F(Y) \to
  F(X \otimes Y)$ is a natural transformation.
  Additionally, $(F,F_0,F_2)$ are required to respect associativity, and
  both unit laws.
\end{definition}

\subsection{$\omega$-colimits}

The notion of $\omega$-colimit is important for the construction of syntax,
since it abstractly formalizes the idea of building sets of abstract syntax trees by
recursion on the height of such trees.
We thus use $\omega$-colimits in the construction of initial models, and
we crucially use that signatures preserve $\omega$-colimits when we construct an initial model, as discussed
in \cref{sec:initiality_theorem}.

\begin{definition}[$\omega$-chains]
  An $\omega$-chains is a sequence $(C_i, c_i)_{i : \N}$ of objects and
  morphisms assembling as a left oriented infinite chain:
  \[
    \begin{tikzcd}
      C_0 \ar[r, "c_0"]
        & C_1 \ar[r, "c_1"]
        & C_2 \ar[r, "c_2"]
        & C_3 \ar[r, "c_3"]
        & \; ...
    \end{tikzcd}
  \]
\end{definition}

\begin{example}
  Given a category $\mc{C}$ with initial object $0 : \mc{C}$, for any
  endofunctor $F : \mc{C} \to \mc{C}$  on $\mc{C}$, there is a canonical
  $\omega$-chain associated to $F$ denoted by $\mrm{chn}_F$:
  \[
    \begin{tikzcd}
      0 \ar[r, "\star"]
        & F(0) \ar[r,   "F(\star)"]
        & F^2(0) \ar[r, "F^2(\star)"]
        & F^3(0) \ar[r, "F^3(\star)"]
        & \; ...
    \end{tikzcd}
  \]
\end{example}

\begin{definition}[$\omega$-cocontinuous functors]
  \label{def:omega-cocontinuous}
  The colimit of an $\omega$-chain is called an $\omega$-colimit.
  A functor $F : \mc{C} \to \mc{D}$ is \emph{$\omega$-cocontinuous} if it
  preserves all $\omega$-colimits.
  $\omega$-cocontinuous functors of $[\mc{C}, \mc{D}]$ form a full subcategory
  of $[\mc{C}, \mc{D}]$.
\end{definition}

\begin{example}
  The identity functor $\Id : [\mc{C}, \mc{C}]$ is $\omega$-cocontinuous.
\end{example}

\subsubsection*{Closure properties}

We aim to build our signatures modularly, i.e. out of
smaller signatures.
As being $\omega$-continuous is a key feature of our signatures, we need
$\omega$-continuous functors to be closed under some operations as
colimits, limits and composition.
Such results can be proven using different results in
\cite[Section 3.8]{CategoryTheoryInContext14}.

\begin{proposition}[Closure under colimits]
  \label{prop:omega-colimits}
  If $\mc{C}$ is cocomplete, then the category of $\omega$-cocontinuous
  functors of $[\mc{C}, \mc{C}]$ is closed under colimits.
\end{proposition}

\noindent While $\omega$-continuous functors are closed under all
small colimits for a wide variety of base categories, they are usually only closed under finite limits.
For our purpose, this will suffice.

\begin{proposition}[Closure under limits]
  \label{prop:omega-limits}
  If $\mc{C}$ admits a class of limits that commutes with $\omega$-colimits
  in $\mc{C}$, then the category of $\omega$-cocontinuous functors of
  $[\mc{C}, \mc{C}]$ is closed under this particular class of limits.
\end{proposition}

\begin{example}
  \label{ex:presheaves-limits}
  In the categories $\Set$ and $[\mc{C}, \Set]$, finite limits commute with
  $\omega$-colimits.
\end{example}

\begin{proposition}[Closure under composition]
  \label{prop:omega_comp}
  The category of $\omega$-cocontinuous endofunctors $[\mc{C},\mc{C}]$ is
  closed under composition of endofunctors.
\end{proposition}

\subsection{Left Kan extensions and Coends}

In \cref{sec:example_F_set}, we apply our framework to the category $[\mb{F},\Set]$,
which requires us to build a monoidal structure on $[\mb{F},\Set]$.
For this purpose, we rely on left Kan extensions and their expression as
coends that we recall here.

\begin{definition}[Left Kan extensions]
  Given a functor $G : \mc{C} \to \mc{D}$ and a functor $F : \mc{C} \to \mc{E}$,
  a \emph{left Kan extensions} of $F$ along $G$ is a functor
  $\Lan{G}{F} : \mc{D} \to \mc{E}$ and a natural transformation
  $\eta : F \to \Lan{G}{F} \circ G$ such that for any other tuple
  $(H : \mc{C} \to \mc{D}, \gamma : F \to H \circ G)$,
  $\gamma$ factors uniquely through $\eta$ as shown below:
  \[
    \begin{tikzcd}[column sep=large, row sep=large]
      \mc{C} \ar[rr, "F" {name=A}] \ar[dr, swap, "G"]
        &
        & \mc{E} \\
        & \mc{D} \ar[from=A, Rightarrow, swap, "\eta" {inner sep=4pt}, shorten=15pt, shift right=10pt]
                 \ar[ur, bend left, "\Lan{G}{F}" {description, name=B} ]
                 \ar[ur, swap, bend right, "H" {name=C} ]
                 \ar[from=B, to=C, Rightarrow, swap, "\exists!" {inner sep=3pt}, shorten=6pt ]
        &
    \end{tikzcd}
  \]
\end{definition}

\noindent The universal property of a left Kan extension can be difficult to use.
Thankfully, under certain conditions, left Kan extensions can be computed as
coends {{\cite[Sections 9.5,9.6,10.4]{CatWorkingMath}}}; this provides a
universal property that is easier to work with:

\begin{definition}[Coends]
  Let $F : \mc{C}^{\textrm{op}} \times \mc{C} \to \mc{D}$ a bifunctor. The
  \emph{coend} of $F$ is an object of $\mc{D}$ denoted by $\int^{C : \mc{C}}
  F(C,C)$ together with a family of morphism $i_C : F(C,C) \to \int^{C :
  \mc{C}} F(C,C)$ making the diagram below commute, and such that for any
  other pair $(X, (x_C)_{C : \mc{C}})$, there is a unique map $h : \int^{C :
  \mc{C}} F(C,C) \to X$ as below:
  \[
    \begin{tikzcd}
              & F(C,C) \ar[dr, start anchor=south east, end anchor=north west, "i_C"]
                       \ar[drr, start anchor=5, end anchor=north west, bend left=20, "x_C"]
              &
              & \\
      F(C',C)   \ar[ur, start anchor=north east, end anchor=south west, "F(f{,} \Id)"]
                \ar[dr, swap, start anchor=south east, end anchor=north west, "F(\Id{,} f)"]
              &
              & \int^{C : \mc{C}} F(C,C) \ar[r, dashed, "\exists ! h"]
              & X \\
              & F(C',C') \ar[ur, start anchor=north east, end anchor=south west, swap, "i_{C'}"]
                         \ar[urr, start anchor=-5, end anchor=south west, swap, bend right=20, "x_{C'}"]
              &
              &
    \end{tikzcd}
    \]
\end{definition}

\begin{proposition}[Left Kan extensions and coends]
  \label{prop:LKE_and_coends}
  Let $G : \mc{C} \to \mc{D}$ be a functor such that $\mc{C}$ is small and
  $\mc{D}$ is locally small.
  Given a category $\mc{E}$, if $\mc{E}$ is cocomplete, then the global left Kan
  extension $\Lan{G}{\_} : [\mc{C},\mc{E}] \to [\mc{D},\mc{E}]$ exists, and it
  is the left adjoint to precomposition by $G$.
  Furthermore, it computes pointwise as the following coend:
  \begin{align*}
    \begin{tikzcd}[ampersand replacement = \&, column sep=large]
        [\mc{C}{,}\mc{E}]
            \ar[r, bend left, "\Lan{G}{\_}"]
            \ar[r, phantom, "\perp"]
          \& {[}\mc{D}{,}\mc{E}{]} \ar[l, bend left, "G^*"]
    \end{tikzcd}
    &&
    \Lan{G}{F}(d) \;:= \int^{c : \mc{C}} \bigsqcup_{\mc{D}(G(c),d)} F(c)
  \end{align*}
\end{proposition}

\begin{example}[${[}\mb{F}{,}\Set{]}$]
  \label{prop:LKE-coends-F-Set}
  The embedding $J : \mb{F} \to \Set$ admits a global left Kan extension
  $\Lan{J}{\_} : [\mb{F},\Set] \to [\Set,\Set]$, which is left adjoint to
  precomposition by $J$, and computes pointwise as the following coend:
  \begin{align*}
    \begin{tikzcd}[ampersand replacement=\&, column sep=large]
        [\mb{F}{,}\Set]
            \ar[r, bend left, "\Lan{J}{\_}"]
            \ar[r, phantom, "\perp"]
          \& {[}\Set{,}\Set{]} \ar[l, bend left, "J^*"]
    \end{tikzcd}
    &&
    \Lan{J}{F}(X) \;:= \int^{n : \mb{F}} F(n) \times X^n
  \end{align*}
\end{example}

%% file: Sections/introduction_is.tex
\section{A Brief Introduction to Initial Semantics}
\label{sec:intro_is}

In this section, we give a brief and gradual introduction to initial semantics.
We introduce initial semantics through the example of natural numbers and
lists, two particularly simple programming languages, in \cref{subsec:is-inductive-types}.
We explain how variable-binding constructors and higher-order programming
languages can be integrated into initial semantics in
\cref{subsec:is-binders}.
We then discuss the interests and the challenges of integrating substitution into
initial semantics in \cref{subsec:is-subst}.

\subsection{Initial Algebra Semantics}
\label{subsec:is-inductive-types}

To illustrate how initial semantics works and its usefulness, we introduce
initial semantics for natural numbers and lists, basic inductive types
that can also be seen as particularly simple programming languages as they
do not have binding constructors.
See, \cite{AlgebraOfProgramming97} or Vene's dissertation \cite{VeneThesis},
for an introduction to initiality and categorical programming.

\subsubsection{Initial Semantics for Natural Numbers}

Let's first consider the type of natural numbers.
Naturals numbers are usually implemented as the following inductive type of
``unary'' natural numbers, which expresses exactly how we think of natural
numbers: each number is either zero or the successor of a natural number.
\begin{verbatim}
  data Nat = zero : Nat | succ : Nat -> Nat
\end{verbatim}

This implementation is naturally modelled as a type/set \verb|X|, and
operation \verb|s : X -> X| and a constant \verb|z : X| which can be merged
into one operation \verb|r := [z,s] : 1 + X -> X| by identifying \verb|Z|
and \verb|1 -> Z|.
In categorical language, the unary natural numbers can be modelled by
$F$-algebras for the functor $F : \Set \to \Set$ defined as $F(X) := 1 + X$,
that is by sets $X : \Set$ equipped with an operation $r : F(X) \to X$ i.e. $r
: 1 + X \to X$.
We call any such $F$-algebra a ``model of the natural numbers''.
Indeed, such an $F$-algebra provides an abstraction for any equivalent
implementation of the natural numbers, e.g. the binary numbers:
\begin{verbatim}
  data BinNat = one : BinNat -> BinNat | zero : BinNat -> BinNat
                | end: BinNat
\end{verbatim}
Having such abstractions for datatypes is important as there are often many
slightly different implementations of a same datatype, e.g. the integers.

Moreover, the unary natural numbers form the initial model, i.e. the initial
$F$-algebra for the functor $F(X) := 1 + X$.
This means that for any other $F$-algebra $(X,z,s)$, there is a unique
morphism of $F$-algebras from $\mrm{Nat}$ to $X$, that is a function $f :
\mrm{Nat} \to X$ that commutes with the constructors, i.e. such that the two
following diagrams commute:
\begin{align*}
  \begin{tikzcd}[ampersand replacement=\&]
    1 \ar[r, "\mrm{zero}"] \ar[d] \& \mrm{Nat} \ar[d, "f"] \\
    1 \ar[r, swap, "z"]\& \mrm{X}
  \end{tikzcd}
  &&
  \begin{tikzcd}[ampersand replacement=\&]
    \mrm{Nat} \ar[r, "\mrm{succ}"] \ar[d, swap, "f"] \& \mrm{Nat} \ar[d, "f"] \\
    X         \ar[r, swap, "s"]           \& \mrm{X}
  \end{tikzcd}
\end{align*}
In other words, there exists a unique function $f : \mrm{Nat} \to X$ that
recursively computes on the constructors, that is, such that $f(\mrm{zero}) =
z$ and $f(\mrm{succ}(n)) = s(f(n))$.
Therefore, being initial for a model of the natural numbers exactly provides
it with a recursion principle.
Indeed, initiality asserts that to build a function $\mrm{Nat} \to X$,
it suffices to provide functions $z : 1 \to X$ and $s : X \to X$,
which yields a recursively computed function as one would expect.

\subsubsection{Initial Semantics for Lists}

We have seen that initial $F$-algebras, for a suitable functor $F$, enable us to give an abstract account of the natural
numbers with its recursion principle.
This is, more generally, possible for inductive types by appropriately choosing
the functor $F$.

As another example, consider the datatype of lists over a fixed base type $A$.
Lists are usually implemented as the following inductive type:
\begin{verbatim}
  data List A = nil : List A | cons : A * (List A) -> (List A)
\end{verbatim}
This leads us to model \verb|List A| by $F$-algebras for the functor $F : \Set
\to \Set$ defined as $F(X) := 1 + A \times X$;
that is, by sets $X : \Set$ equipped with a function $r : F(X) \to A$ i.e. $r
: (1 + A \times X) \to X$, hence with two functions $n : 1 \to X$ and $c : A
\times X \to X$.

Moreover, as for the natural numbers, \verb|List A|, with the two constructors, forms the initial such
$F$-algebra providing it with a recursion principle.
Indeed, by initiality, it suffices to build functions $n : 1 \to X$ and $c :
A \times X \to X$ to obtain a unique function $f : \mrm{List}\; A \to X$ that
computes recursively on the constructors such that $f(\mrm{nil}) = n$ and
$f(\mrm{cons}(a,l)) = c(a,f(l))$.
Alternatively, in categorical terms, we ask that the following diagrams commute:
\begin{align*}
  \begin{tikzcd}[ampersand replacement=\&]
    1 \ar[r, "\mrm{nil}"] \ar[d] \& \mrm{List}\; A \ar[d, "f"] \\
    1 \ar[r, swap, "n"]\& \mrm{X}
  \end{tikzcd}
  &&
  \begin{tikzcd}[ampersand replacement=\&]
    A \times \mrm{List}\;A \ar[r, "\mrm{cons}"] \ar[d, swap, "f"]
      \& \mrm{Nat} \ar[d, "f"] \\
    A \times X             \ar[r, swap, "c"]
      \& \mrm{X}
  \end{tikzcd}
\end{align*}

Therefore, initial models in the form of appropriate $F$-algebras also
enable us to provide an abstract account of lists and and an abstraction to program with them.
This is especially interesting in computer science where specifying functions
as \texttt{fold}s allows various \emph{optimizations}, such as fusing
together functions.
See for instance, Hutton's tutorial \cite{DBLP:journals/jfp/Hutton99} for an
excellent introduction to fusion for functions on lists.

\subsection{Nested Datatypes for Variable Binding}
\label{subsec:is-binders}

Higher-order programming languages are notoriously tedious to work with in
proof assistant, as attested by the existence of the POPLmark challenge \cite{PoplMarkChallenge}.
This mostly stems from variable binding that forces us to work up to
$\alpha$-equivalence, which complicates substitution and consequently semantics.
Therefore, we would like to apply the principle of initial semantics to
higher-order programming languages to deal with free and bound variables,
as well as with substitution.

\subsubsection{Extrinsic and Intrinsic Definitions}

Higher-order languages like the untyped lambda calculus are often
implemented using a simple inductive type, using integers as de Bruijn
indices or levels to represent and deal with variables.
\begin{figure}[H] \centering
\begin{BVerbatim}
data Lam :=
| var : Nat -> Lam
| app : Lam -> Lam -> Lam
| abs : Nat -> Lam -> Lam
\end{BVerbatim}
\end{figure}
\noindent Such implementations are called ``extrinsic'' as contexts, bindings and
well-scopedness are not hardcoded into \verb|Lam| itself, but instead
defined on top of \verb|Lam|.
For instance, terms like \verb|abs 0 (var 4)| are always well-defined terms of
type \verb|Lam|, but are only well-scoped, or not, in a given context.
Even though this is the usual presentation of the lambda calculus, and it does
have an initial model as an inductive types, it is not very well suited for our
purpose.
Indeed, as binding and well-scopedness are external to \verb|Lam|, the
initial model abstracting \verb|Lam| would not tell us much about free and bound variables.

Therefore, we are interested in definitions of higher-order languages where
contexts, binding, well-scopedness and, if applicable, typing are \emph{internal to the language definition}.
Such definitions are called ``intrinsic'', and can be provided through nested
inductive types \cite{NestedDataTypes98}, using the host language type system to enforce the
constraints statically.
For instance, the untyped lambda calculus can be defined as the nested inductive type:
\begin{figure}[H] \centering
\begin{BVerbatim}
data Lam A :=
| var : A -> Lam A
| app : Lam A -> Lam A -> Lam A
| abs : Lam (A + 1) -> Lam A
\end{BVerbatim}
\end{figure}
\noindent where \verb|A + 1| is the type \verb|A| plus one element.
Compared to the previous definition, we do not define one type of all lambda
terms, but a family of types \verb|Lam A|.
Seeing the type \verb|A| as a context of variables, \verb|Lam A| then
corresponds to the type of well-scoped lambda terms in context \verb|A|.
That is, if \verb|t : Lam A|, then \verb|t| is well-scoped in context \verb|A|.
This is possible as contexts \verb|A| are explicit and part of the definition,
enabling us to define binders simply as \verb|abs : Lam (A + 1) -> Lam A|,
that is, a constructor that takes a term in a context with a fresh variable
\verb|A+1| and returns a term in a context without that fresh variable, i.e. \verb|A|.

\subsubsection{Initial Semantics}

To capture binding, we want to characterize the family \verb|Lam A| as an
initial model.
For parametrised inductive types, like \verb|List A|, using that
the \verb|List a| can be defined independently, we associated an initial
model --- an $F$-algebra on $\Set$ --- for each parameter \verb|A|.
This is no longer possible as the family \verb|Lam A| is interdependent,
due to the inductive type constructor \verb|_+1| that is nested in the
constructor \verb|abs : Lam(A + 1) -> Lam A|.
This issue arises more generally for all nested data types, that is, for inductive
types as above where an inductive type constructor appears nested in the argument of
a constructor.

Consequently, to capture higher-order languages, defined intrinsically using
nested data types like \verb|Lam A|, we need to change our notion of models.
To account for the interdependency, we change from a family of $F$-algebras
on $\Set$ as the list one $(\verb|List A|)_{\verb|A|}$, to one $F$-algebra
defined on the functor category $[\Set,\Set]$ \cite{DeBruijnasNestedDatatype99}.
Indeed, such an $F$-algebra consists of a functor $R : \Set \to \Set$ that
models the family \verb|Lam|, with a transformation $r : F(R) \to R$ that
provides a function $r_A : F(R)(A) \to R(A)$ for each $A$.
Hence, working in a functor category enables us to capture the interdependency by choosing $F$ appropriately.
For instance, the untyped lambda calculus can be captured by the following functor.
\[
\begin{array}{clcll}
  F_\Lambda  &: & [\Set,\Set] & \longrightarrow & [\Set, \Set]                              \\
  F_\Lambda  &:=& R           & \longmapsto     & (A \longmapsto A + R(A) \times R(A) + R(A+1))
\end{array}\]

\noindent Indeed, such a model is a functor $R : \Set \to \Set$, with for
all $A : \Set$, constructors $\mrm{var}_A' : A \to R(A)$, $\mrm{app}_A' :
R(A) \to R(A) \to R(A)$ and $\mrm{abs}_A' : R(A+1) \to R(A)$.
Moreover, the functorial action captures renaming of variables without
variable capture, proof that it correctly models binding.
We refer to \cref{subsec:model_set_set} for details on the subject.

As before, \verb|Lam a| forms the initial such $F$-algebra which provides it
with a recursion principle.
Given another $F$-algebra $R,\mrm{var}',\mrm{app}',\mrm{abs}'$, there is a
unique natural transformation $f : \mrm{Lam} \to R$ commuting with the
constructors; that is, for all $A : \Set$ a function $f_A : \mrm{Lam}\;A \to
R\;A$ such that:
\begin{align*}
  \begin{tikzcd}[ampersand replacement=\&]
    A \ar[r, "\mrm{var}_A"] \ar[d]
      \& \mrm{Lam}\; A \ar[d, "f_A"] \\
    A \ar[r, swap, "\mrm{var}_A'"]
      \& R\;a
  \end{tikzcd}
  &&
  \begin{tikzcd}[ampersand replacement=\&]
    \mrm{Lam}\; A \times \mrm{Lam}\; A \ar[r, "\mrm{app}_A"] \ar[d, swap, "f_A \times f_A"]
      \& \mrm{Lam}\; A \ar[d, "f_A"] \\
    R\; A \times R\; A \ar[r, swap, "\mrm{app}_A'"]
      \& R\;A
  \end{tikzcd}
  &&
  \begin{tikzcd}[ampersand replacement=\&]
    \mrm{Lam}\; (A+1) \ar[r, "\mrm{abs}_A"] \ar[d, swap, "f_{A+1}"]
      \& \mrm{Lam}\; A \ar[d, "f_A"] \\
    R\; (A+1 )  \ar[r, swap, "\mrm{abs}_A'"]
      \& R\;A
  \end{tikzcd}
\end{align*}
But furthermore, this recursion principle is providing us with added safety
conditions.
Indeed, it ensures by construction that we translate \verb|abs : Lam (A+1) -> Lam A|,
that is, a binder, by a constructor \verb|R (A+1) -> R a|, i.e., by another
binder.
In other words, the recursion principle respects variable binding by construction.

\subsection{Integrating Substitution into Initiality}

\label{subsec:is-subst}

Substitution is an important operation on syntax as it is an essential part of
operational semantics through $\beta$-reduction.
In particular, to be able to reason about semantics, e.g. to prove
confluence or normalisation, we need to be able to reason about substitution.
Yet, while ubiquitous, substitution is not trivial to implement; typical
aspects to consider are $\alpha$-equivalence and capture of variables.
Moreover, the different properties of substitution, which enable us to check
the correctness of the implementation but also to reason about it,
are tedious to establish.
For instance, if $x \notin \mathrm{free}(u)$ and $y \notin \mathrm{free}(v)$,
then we expect the equation $t[x := v][y := u] = t[y := u][x := v]$ to hold.

As initial semantics aims to provide an abstract interface to work with
higher-order languages, we are interested in integrating a finitely
axiomatized substitution into initial semantics, in order to be able to
reason about semantics directly.
Doing so enables us to reason about substitution in terms of
properties without having to deal with a particular implementation.
In particular, using a common axiomatization of substitution
enables us to build reusable libraries about substitution.
Moreover, by initiality, integrating substitution into models provides us
with a recursion principle compatible with substitution; this
means that any map defined using the recursion principle is automatically
compatible with substitution.
This is important when studying semantics-preserving translations between
languages.

In practice, we are interested in axiomatizing \emph{unary substitution} as
it is the one appearing in $\beta$-reduction.
Denoted $t[x := u]$, unary substitution replaces each occurrence of a
variable $x$ in a term $t$ by a term $u$ without capture of variables.
Yet, unary substitution is not easy to axiomatize; any axiomatization,
such as substitution algebras \cite[Section 3]{FPT99}, needs to deal
explicitly with variables.
Thankfully, unary substitution is a particular case of \emph{simultaneous
substitution}.
Given an assignment $f : \Gamma \to \Lambda(\Delta)$, simultaneous
substitution $\sigma(f) : \Lambda(\Gamma) \to \Lambda(\Delta)$ replaces all
the variables $x : \Gamma$ in a term $t : \Lambda(\Gamma)$ by the associated
terms $f(x)$.
Compared to unary substitution, simultaneous substitution has the advantage
to have a well-established axiomatization as particular monoids
\cite{BellegardeHook94}, that can be easily formulated and manipulated in category
theory.
This makes it easier to integrate it into initial semantics.
Consequently, we will axiomatize simultaneous substitution, and deduce the
unary one as a particular case, rather than axiomatizing unary substitution
directly.

The properties of simultaneous substitution can be axiomatized by particular
monoids, but it is not enough to integrate it into initial semantics.
Indeed, it remains to capture how substitution commutes with substitution,
and to prove initiality theorems.
This is not trivial.
For instance, as explained in \cite{DeBruijnasNestedDatatype99}, the
``flattening operation'' of the monoid can not be built directly out of the
induction principle; it first requires one to strengthen the induction principle \cite{GeneralisedFold99}.
Therefore, several different approaches have been considered to integrate
simultaneous substitution into initial semantics.
The different approaches can be roughly divided into three traditions,
depending on the mathematical structures they are using: $\Sigma$-monoids
\cite{FPT99,SecondOrderDep08,ListObjects17}, modules over monads
\cite{HirschowitzMaggesi07,ZsidoPhd10,PresentableSignatures21}, and
heterogeneous substitution systems (hss) \cite{Hss04,HssRevisited15,HssNonWellfounded23}.

As discussed in the introduction, the literature on initial semantics is not
easiest to access; particularly as the different traditions have different
variations, and the formal links between them are insufficiently know.
In the remainder of the paper, we provide an introduction to the different
approaches mentioned above, and discuss the links between them.

%% file: Sections/models.tex
\section{Modelling the Syntax}
\label{sec:models}

We aim to define a general notion of model encompassing the substitution
structure of untyped and typed higher-order languages, that is, languages with variable binding.
Yet different kinds of type systems and contexts have different structures
and as such require different kinds of categories to model them.
Moreover, there can be more than one suitable category to model a class of
higher-order languages, as illustrated in
\cref{sec:example_set_set,sec:example_F_set} in the untyped case.
Thus, to account for different options, we define a category of models for a
generic monoidal category, principle already introduced in \cite{FPT99} for
$\Sigma$-monoids, such that it can then just be appropriately instantiated
depending on the context.

As we aim to characterise our languages as initial models, our models
inherently depend on term constructors and as such on the signatures
specifying them.
It is common to use syntactic signatures as binding signatures, also known as algebraic signatures.
However, syntactic signatures are defined for specific type systems, and
as such would not be parametric enough for our purpose.
Hence, we rely on a generic mathematical notion of signatures, that can be
instantiated differently depending on the context.
Similarly, how constructors commute with substitution can not directly be
hardcoded in models as variable binding is type dependent and models
need to deal with substitution generically.
Therefore, signatures need to specify how term constructors commute with
substitution.

To do so, we adopt the view of modules over monads and associated notions,
that we generalise to monoidal categories using modules over monoids.
Indeed, once generalised to monoidal categories, modules over monads
provide us with a fully abstract framework, which, as discussed in
\cref{subsec:sigstrength-to-sig}, encompasses $\Sigma$-monoids as a particular
important particular case.
Modules over monads and initial semantics have been introduced, and mostly
studied with small variations on the particular monoidal category
$[\Set,\Set]$, e.g., in
\cite{HirschowitzMaggesi07,HirschowitzMaggesi10,ZsidoPhd10,HirschowitzMaggesi12,PresentableSignatures21}.
We refer to \cref{subsec:rw-modules-over-monoids} for a detailed
discussions on modules over monads, and the differences with our
approach.

In the following, we first review monoids and modules over monoids in
\cref{subsec:monoids,subsec:modules} used to model simultaneous substitution.
This enables us to define our category of signatures in \cref{subsec:signatures}.
Afterwards, given a signature, we define and describe its associated category of
models in \cref{subsec:models}.
Models and how they work is exemplified in detail in \cref{sec:example_set_set,sec:example_F_set}
for untyped languages, and in \cref{sec:example_setT_setT} for simply-typed ones.

\subsection{Monoids}
\label{subsec:monoids}

Monads --- monoids in a category of endofunctors --- axiomatize the behaviour
of simultaneous substitution on languages as identified in
\cite{BellegardeHook94,DeBruijnasNestedDatatype99,AltenkirchReus99}
for the untyped lambda calculus and detailed in \cref{subsec:model_set_set}.
Informally, a monad accounts for the existence of variables and simultaneous
substitution, and axiomatizes their properties.
Yet, there are also interesting monoids that are not trivially monads,
axiomatizing simultaneous substitution as seen in \cite{FPT99} and detailed
in \cref{subsec:model_F_set}.
To be able to account directly for both possibilities, we rely on monoids in
a monoidal category \cite[Section 1.2]{2DimensionalCategories20} to model
the substitution structure of languages.

\begin{definition}[Monoids]
  \label{def:monoids}
  Given a monoidal category $\Cmon$, a \emph{monoid} on $\mc{C}$ is a tuple
  $(R,\mu,\eta)$ where $R$ is an object of $\mc{C}$ and $\mu : R \otimes R
  \to R$ and $\eta : I \to R$ are morphisms of $\mc{C}$ such that the
  following diagrams commute.
  \begin{align*}
    \begin{tikzcd}[ampersand replacement=\&]
      (R \otimes R) \otimes R \ar[r, "\alpha"] \ar[d, swap, "\mu \otimes R"]
        \& R \otimes (R \otimes R) \ar[r, "R \otimes \mu"]
        \& R \otimes R \ar[d, "\mu"] \\
      R \otimes R \ar[rr, swap, "\mu"]
        \&
        \& R
    \end{tikzcd}
  &&
    \begin{tikzcd}[ampersand replacement=\&]
      I \otimes R \ar[r, "\eta \otimes R"] \ar[dr, swap, "\lambda_R"]
        \& R \otimes R \ar[d, "\mu"]
        \& R \otimes I \ar[l, swap, "R \otimes \eta"] \ar[dl, "\rho_R"] \\
        \& R
        \&
    \end{tikzcd}
  \end{align*}
  We call $\mu$ the \emph{multiplication} and $\eta$ the \emph{unit} of the
  monoid.
\end{definition}

\begin{example}
  In the category $[\Set,\Set]$ and $[\mb{F},\Set]$, the untyped lambda
  calculus can be equipped with a monoid structure, where $\eta$ is the
  variable constructor and $\mu$ is a flattening operation.
\end{example}

\begin{definition}[Morphisms of Monoids]
  \label{def:morphisms-monoids}
  A \emph{morphism of monoids} $(R,\mu,\eta) \to (R',\mu',\eta')$ is a morphism $f
  : R \to R'$ in $\mc{C}$ preserving multiplication and the unit i.e such that
  the following diagrams commute:
  \begin{align*}
    \begin{tikzcd}[ampersand replacement=\&]
      R \otimes R \ar[r, "f  \otimes f"] \ar[d, swap, "\mu"]
        \& R' \otimes R' \ar[d, "\mu'"] \\
      R \ar[r, swap, "f"]
        \& R'
    \end{tikzcd}
    &&
    \begin{tikzcd}[ampersand replacement=\&]
      \& I \ar[dl, swap, "\eta"] \ar[dr, "\eta'"]
      \& \\
      R \ar[rr, swap, "f"]
      \&
      \& R'
    \end{tikzcd}
  \end{align*}
\end{definition}

\begin{proposition}[Category of Monoids]
  \label{prop:cat-monoids}
  Given a monoidal category $\mc{C}$, monoids in $\mc{C}$ and their morphisms
  form a category denoted $\Mon(\mc{C})$.
\end{proposition}

\subsection{Modules}
\label{subsec:modules}

Monoids enable us to model variables and simultaneous substitution, but they
fail to account, on their own, for the behaviour of substitution on term
constructors.
Indeed, term constructors such as application ($\app$) can not be characterized as
morphism of monoids as they \emph{do not} preserve variables.
Therefore, we rely on modules over a monoid \cite[Chapter 4]{ZsidoPhd10}, a
more liberal variant of monoids and monoid morphisms, and characterise
constructors as morphisms of suitable modules.
This idea was introduced in \cite{HirschowitzMaggesi07,HirschowitzMaggesi10},
using the slightly different notion of modules over monads, on the category
$[\Set,\Set]$.
The link between the two notions is discussed in detail
in \cref{subsubsec:module_monads_vs_monoids}.

\subsubsection{Modules over Monoids}

\begin{definition}[Modules over a Monoid]
  \label{def:modules}
  Given a monoid $R : \Mon(\mc{C})$, a (left) $R$-module is a tuple $(M,
  p^M)$ where $M$ is an object of $\mc{C}$ and $p^M : M \otimes R \to M$ is
  a morphism of $\mc{C}$ (called \emph{module substitution}), compatible
  with the multiplication and the unit of the monoid:
  \begin{align*}
    \begin{tikzcd}[ampersand replacement=\&]
      (M \otimes R) \otimes R \ar[r, "\alpha_{M,R,R}"] \ar[d, swap, "p^M \otimes R"]
        \& M \otimes (R \otimes R) \ar[r, "M \otimes \mu"]
        \& M \otimes R \ar[d, "p^M"] \\
      M \otimes R \ar[rr, swap, "p^M"]
        \&
        \& M
    \end{tikzcd}
    &&
    \begin{tikzcd}[ampersand replacement=\&]
      M \otimes I \ar[r, "M \otimes \eta"] \ar[dr, swap, "\rho_M"]
        \& M \otimes R \ar[d, "p^M"] \\
        \& M
        \&
    \end{tikzcd}
  \end{align*}
\end{definition}

\noindent Intuitively, modules give us more freedom than monoids, as they
do not have a unit constructor $\eta$, and enable us to specify --- via the module
substitution --- how the substitution should be done on them.
And indeed, all monoids are trivially modules over themselves:

\begin{example}
  $(R,\mu)$ is a module over itself denoted $\Theta$.
\end{example}

\begin{example}
  Given an object $D : \mc{C}$, for any $R$-module $(M,p^M)$, there is an
  associated $R$-module with object $D \otimes M : \mc{C}$ and module
  substitution:
  \[
    \begin{tikzcd}
      (D \otimes M) \otimes R \ar[r, "\alpha"]
        & D \otimes (M \otimes R) \ar[r, "D \otimes p^M"]
        & D \otimes M.
    \end{tikzcd}
  \]
\end{example}

\noindent For our purpose, modules enable us to specify how substitution
behaves at the input and output of term constructors, with the trivial
module $\Theta$ corresponding to the term language itself.
Module morphisms are then just morphisms respecting the module
substitutions.
As such, constructors can be modelled as particular module morphisms.
In our case, our constructors will always return one term, without introducing
fresh variables, hence they will always be of the form $M \to \Theta$.

\begin{definition}[Morphisms of Modules]
  \label{def:morphisms-modules}
  Given a monoid $R : \Mon(\mc{C})$, a \emph{morphism of $R$-modules} $(M, p^M) \to
  (M', p^{M'})$ is a morphism $r : M \to M'$ of $\mc{C}$ commuting with
  the respective module substitutions:
  \[
    \begin{tikzcd}
      M \otimes R \ar[r, "r \otimes R"] \ar[d, swap, "p^M"]
        & M' \otimes R \ar[d, "p^{M'}"] \\
      M \ar[r, swap, "r"] & M'
    \end{tikzcd}
  \]
\end{definition}

\begin{example}
  For appropriate categories and monoids, the constructor $\app$ of the
  untyped lambda calculus can be characterised as a module morphism $\Theta
  \times \Theta \to \Theta$; as it takes two terms as arguments and returns a
  new term without binding any variable.
\end{example}

\begin{proposition}[Category of Modules]
  \label{prop:cat-modules}
  Given a monoid $R : \Mon(\mc{C})$, the modules over $R$ and their
  morphisms form a category denoted $\Mod(R)$.
\end{proposition}

\noindent To specify term constructors as module morphisms $M \to \Theta$,
we must be able to build modules specifying the inputs of term constructors
e.g. $\Theta \times \Theta$.
Some modules are instance dependent and can not be built generically, such
as variable binding.
However, it is generically possible to build modules \emph{modularly} as
the category of modules is complete and cocomplete under reasonable conditions:

\begin{proposition}[Closure under (co)limits]
  \label{prop:modules-(co)complete}
  Given a monoid $R : \Mon(\mc{C})$, if $\mc{C}$ is (co)complete, and
  $\_ \otimes R$ preserves (co)limits, then the category $\Mod(R)$ is
  (co)complete.
\end{proposition}

\begin{example}
  \label{prop:module-language-cst}
  Under the above assumptions, $\Mod(R)$ has a terminal module $\Theta^0$,
  and is closed under products and coproducts.
\end{example}

\noindent Those language constructors are of particular importance.
The terminal module enables us to represent constants, i.e. constructors
without input, as module morphisms $\Theta^0 \to \Theta$.
The closure under products enables us to represent term constructors with
several independent inputs as $\Theta \times \Theta \to \Theta$;
whereas coproducts allow us to specify languages with several independent
term constructors e.g. $(\Theta^0 + \Theta \times \Theta) \to \Theta$ as it
amounts, by universal property, to a morphism $\Theta^0 \to \Theta$ and a
morphism $\Theta \times \Theta \to \Theta$

\subsubsection{The Total Category of Modules}

A model of a given signature $\Sigma$ should be a pair consisting of a
monoid $(R,\eta,\mu)$ axiomatizing variables and simultaneous substitution,
and an $R$-module morphism $M \to \Theta$  axiomatizing term constructors.
Hence a signature $\Sigma$ should associate, to any monoid $(R,\eta,\mu)$, a
module $\Sigma(R) : \Mod(R)$ over it representing the input module.
This monoid dependency cannot be expressed directly since modules form a
category over a fixed monoid $\Mon(R)$.
Thankfully, modules over different monoids can be assembled into a total
category as discussed in \cite[Chapter 4]{ZsidoPhd10}, and \cite[Section 2.2]{PresentableSignatures21}.

\begin{proposition}[The pullback functor]
  Given a morphism of monoids $f : R \to R'$, there is a functor
  $f^* : \Mod(R') \to \Mod(R)$ that associates to each module
  $(M',p^{M'}) : \Mod(R')$ a $R$-module structure defined as
  \[
    \begin{tikzcd}
      M' \otimes R \ar[r, "M' \otimes f"]
        & M' \otimes R' \ar[r,"p^{M'}"]
        & M'
    \end{tikzcd}
  \]
  The module $f^*M'$ is called the \emph{the pullback module}.
\end{proposition}

\begin{proposition}[The module functor]
  There is a contravariant functor associating to each monoid $R$ the category
  $\Mod(R)$, and to each morphism of monoid $f : R \to R'$, the functor
  $f^* : \Mod(R)' \to \Mod(R)$:
  \[ \Mod : \Mon(\mc{C})^\op \longrightarrow \mrm{Cat} \]
\end{proposition}

\begin{proposition}[The total category of modules]
  \label{prop:cat-totalcat-modules}
  There is a \emph{total category of modules} $\TotMonMod{\mc{C}}$.
  Its objects are tuples $(R,M)$ where $R : \Mon(\mc{C})$ is a monoid, and
  $M : \Mod(R)$ a module over it.
  Its morphisms $(R,M) \to (R',M')$ are tuples $(f,r)$ where $f : R \to R'$
  is a morphism of monoids and $r : M \to f^*M'$ a morphism of $R$-monoids.
\end{proposition}

\begin{remark}
  The forgetful functor from the total category of modules to the base
  category of monoid is a Grothendieck fibration.
  \[ U : \TotMonMod{\mc{C}} \longrightarrow \Mon(\mc{C}) \]
\end{remark}

\subsection{Signatures}
\label{subsec:signatures}

Following \cite{HirschowitzMaggesi12,PresentableSignatures21}, we define
signatures and prove basic properties using the total category of modules.
We refer to \cref{subsubsec:rw-modules-sig} for a discussion on signatures
and modules.
We define signatures as functors $\Mon(\mc{C}) \longrightarrow
\TotMonMod{\mc{C}}$ that return a module over the same monoid as the input
one.

\begin{definition}[Signatures]
  A \emph{signature} is a functor $\Sigma : \Mon(\mc{C}) \longrightarrow \TotMonMod{\mc{C}}$
  making the following diagram commute:
  \[
    \begin{tikzcd}
      \Mon(\mc{C}) \ar[rr, "\Sigma"] \ar[dr, swap, equal]
        &
        & \TotMonMod{\mc{C}} \ar[dl, "U"] \\
      & \Mon(\mc{C}) &
    \end{tikzcd}
  \]
\end{definition}

\noindent In other words, signatures are sections of the forgetful functor
$U : \TotMonMod{\mc{C}} \longrightarrow \Mon(\mc{C})$.
As the first component is always the identity, we will never denote it.
\begin{remark}
  Our definition of signatures is sometimes refereed as parametric modules,
  see, e.g., \cite{TransitionMonads22}.
\end{remark}

Many constructions on modules extend to signatures.
For instance, the trivial module can be extended into a signature:

\begin{example}
  There is a trivial signature, also denoted $\Theta$, associating to every
  monoid $(R,\eta,\mu)$ the trivial module $\Theta := (R,\mu)$ over it.
\end{example}

\begin{example}
  Given an object $D : \mc{C}$, for any signature $\Sigma$, there is an
  associated signature $D \otimes \Sigma$ that associates to any monoid $R :
  \Mon(\mc{C})$ the $R$-module $D \otimes \Sigma(R)$.
\end{example}

\begin{example}
  In appropriate categories, the untyped lambda calculus, first-order logic,
  simply-typed lambda calculus or PCF are representable by signatures.
\end{example}

\begin{definition}[Morphism of Signatures]
  A morphism of signature $\Sigma \to  \Sigma'$ is a natural transformation
  $h : \Sigma \to  \Sigma'$ that is the identity when composed with the
  forgetful functor $U : \int_{R : \Mon} \Mod(R) \longrightarrow \Mon$
\end{definition}

\begin{proposition}[Category of Signatures]
  \label{prop:cat-signatures}
  Signatures and their morphisms form a category $\mrm{Sig}(\mc{C})$.
\end{proposition}

\noindent We must be able to build signatures to be able to represent languages.
As for modules, some signatures are not definable generically as they are
type dependent.
However, it is possible to build signatures \emph{modularly}, since signatures
inherit their (co)limits from modules, being functors preserving the
input monoid.

\begin{proposition}[Closure under (co)limits]
  \label{prop:sig-(co)complete}
  If $\mc{C}$ is (co)complete, and for all $R : \Mon(\mc{C})$, $\_ \otimes
  R$ preserves (co)limits, then the category of signatures $\Sig(\mc{C})$ is
  (co)complete.
\end{proposition}

\begin{example}
  Under the above assumptions, $\Sig(\mc{C})$ has a terminal signature, also
  denoted $\Theta^0$, and is closed under products and coproducts.
\end{example}

\noindent As before, the terminal signature enables us to represent constants,
products to represent constructors with different independent inputs, and
coproducts to represent languages with different independent constructors.
For instance, on an appropriate category, a language with one constant and a
binary constructor can be represented by the signature $\Theta^0 + \Theta \times \Theta$.

\subsection{Models}
\label{subsec:models}

We have identified that to abstract the substitution structure of
higher-order languages, a model of a signature should consist of a monoid
and a module morphism.
However, it is not sufficient to fully abstract higher-order languages as it
does not provide them with a recursion principle.
Hence, following the classical method characterising inductive objects as as
initial algebra \cite{Goguen76,GoguenEtAl75}, we characterise our languages
as initial models to provide them with a recursion principle.
This concept was introduced for higher-order languages in \cite{FPT99}.
Modules based models were introduced in \cite{HirschowitzMaggesi07}, and its
current form in \cite{HirschowitzMaggesi12,PresentableSignatures21}, in both
cases for $[\Set,\Set]$.

\subsubsection{Models of a signature}

\begin{definition}[Models]
  \label{def:models}
  Given a signature $\Sigma : \Sig(\mc{C})$, a \emph{model} of $\Sigma$ is a
  tuple $(R,r)$ where $R : \Mon(\mc{C})$ is a monoid and $r : \Sigma(R) \to
  R$ is a morphism of $R$-modules.
\end{definition}

\begin{definition}[Morphism of Models]
  \label{def:morphims-models}
  A morphism of $\Sigma$-models $(R,r) \to (R',r')$ is a morphism of monoids
  $f : R \to R'$ compatible with the module morphism $r$ and $r'$, i.e. such
  that the left-hand side digram of $R$-modules commutes:
  \begin{align*}
    \begin{tikzcd}[ampersand replacement=\&]
      \Sigma(R) \ar[r, "r"] \ar[d, swap, "\Sigma(f)"]
        \& R \ar[d, "f"] \\
      f^* \Sigma(R') \ar[r, swap, "f^*r'"]
        \& f^*R'
    \end{tikzcd}
    &&
    \begin{tikzcd}[ampersand replacement=\&]
      (R,\Sigma(R)) \ar[r, "(id{,}r)"] \ar[d, swap, "\Sigma(f)"]
        \& (R,R) \ar[d, "\Theta(f)"] \\
      (R', \Sigma(R')) \ar[r, "(id{,}r')"]
        \& (R',R')
    \end{tikzcd}
  \end{align*}
  Here, for simplicity, we simply denote the module $\Theta(R)$ by $R$.
\end{definition}

\begin{remark}
  The pullback along $f^*$ is here for homogeneity as $\Sigma(R')$ is an
  $R'$-module on its own.
  It can be understood by viewing the left-hand side diagram as a diagram in
  the total category of modules, as on the right-hand side.
\end{remark}

\begin{proposition}
  \label{prop:morph-models-alg}
  As the forgetful functor $U : \Mod(R) \to \mc{C}$ is faithful, morphisms
  of models are exactly morphisms of algebras that have been strengthened to
  also be morphisms of monoids.
\end{proposition}

\begin{proposition}[Category of Models]
  \label{prop:cat-mod-sig}
  Given a signature $\Sigma : \Sig(\mc{C})$, its models and their morphisms
  form a category denoted $\Model(\Sigma)$.
\end{proposition}

\subsubsection{Representable Signatures}

In general, there is no reason why all signatures would admit an initial model.
Hence, we focus on the ones that do, which we call \emph{representable}.
For representable signatures, we provide a fixpoint property and use it to
give an example of a non-representable signature, as done in
\cite[Section 5.1]{PresentableSignatures21} for $[\Set,\Set]$.

\begin{definition}[Representable Signatures]
  A signature $\Sigma$ is \emph{representable} if its category of models
  $\Model(\Sigma)$ has an initial model.
  In this case, its initial model is denoted $\ol{\Sigma}$, and called
  the \emph{syntax} associated to $\Sigma$.
\end{definition}

\begin{example}
  In appropriate categories, the associated signatures to the untyped lambda
  calculus, first-order logic, simply-typed lambda calculus or PCF are
  representable, i.e. have an initial model.
\end{example}

\begin{proposition}
  \label{prop:fixpoint-models}
  Let $\mc{C}$ be a monoidal category with binary coproducts, and such
  that for all $Z : \mc{C}$, $\_ \otimes Z$ preserves coproducts.
  If a signature $\Sigma$ has a model $M$, then $\Sigma(M) + I$ can be
  turned into a model of $\Sigma$.
  Moreover, if $\Sigma$ is representable, there is an isomorphism of models
  $\Sigma(\ol{\Sigma}) + I \cong \ol{\Sigma}$.
\end{proposition}

\begin{example}
  \label{ex:not-representable}
  In the monoidal category $[\Set,\Set]$, the signature $\mc{P} \circ
  \Theta$ is not representable.
  Indeed, otherwise there would be an isomorphism of models $(\mc{P} \circ \Theta)
  (\ol{\mc{P} \circ \Theta}) + \Id \cong \ol{\mc{P} \circ \Theta}$.
  Yet, as there is a forgetful functor $\Model([\Set,\Set]) \to
  [\Set,\Set]$, for all $X : \Set$, there would be an isomorphism of sets
  $\mc{P} ((\ol{\mc{P} \circ \Theta})(X)) + X \cong (\ol{\mc{P} \circ
  \Theta}) (X)$.
\end{example}

\subsubsection{The Total Category of Models}

By definition, the models over different signatures $\Sigma,\Sigma'$ belong to
different categories $\Model(\Sigma)$ and $\Model(\Sigma')$, and as such can
not be directly related.
Fortunately, as for modules, models assemble in a total category, that can
be used to express some modularity, as introduced in
\cite{HirschowitzMaggesi12,PresentableSignatures21} on $[\Set,\Set]$.

\begin{proposition}[The pullback functor]
  Given a morphism of signatures $h : \Sigma \to \Sigma'$, there is a functor
  $h^* : \Model(\Sigma') \to \Model(\Sigma)$ that associates to to each model
  $(R',r') : \Model(\Sigma')$ a $\Sigma$-model structure defined as
  \[
    \begin{tikzcd}
      \Sigma(R') \ar[r, "h_R"]
        & \Sigma'(R') \ar[r, "r'"]
        & R'
    \end{tikzcd}
  \]
  The model $h^*(R',r')$ is called the \emph{the pullback model} of
  $(R',r')$ along $h$.
\end{proposition}

\begin{proposition}[The module functor]
  There is a contravariant functor associating to each signature $\Sigma :
  \Sig(\mc{C})$ the category $\Model(\Sigma)$, and to each morphism of
  signatures $h : \Sigma \to \Sigma'$, the functor $h^* : \Model(\Sigma') \to
  \Model(\Sigma)$:
  \[ \Model : \Sig(\mc{C})^\op \longrightarrow \mrm{Cat} \]
\end{proposition}

\begin{proposition}[The total category of models]
  \label{prop:cat-totalcat-models}
  There is a \emph{total category of models} $\TotSigModel{\mc{C}}$. Its
  objects are tuples $(\Sigma,(R,r))$ where $\Sigma : \Sig(\mc{C})$ is a
  signature, and $(R,r) : \Model(\Sigma)$ a model over it.
  Its morphisms $(\Sigma,(R,r)) \to (\Sigma',(R',r'))$ are tuples $(h,f)$
  where $h : \Sigma \to \Sigma'$ is a morphism of signatures and
  $f : (R,r) \to f^*(R',r')$ a morphism of $\Sigma$-models.
\end{proposition}

\begin{remark}
  The forgetful functor from the total category of models to the base
  category of signatures is a Grothendieck fibration.
  \[ \TotSigModel{\mc{C}} \longrightarrow \Sig(\mc{C}) \]
\end{remark}

\begin{proposition}[Modularity]
  \label{prop:modularity-models}
  Given a pushout of representable signature $\Sigma$,$\Sigma_1$,$\Sigma_2$,
  $\Sigma_{12}$, their initial models form a pushout above it, in the total
  category of models:
  \begin{align*}
    \begin{tikzcd}[ampersand replacement=\&]
      \Sigma \ar[r] \ar[d] \arrow[dr, phantom, "\ulcorner", very near end]
        \& \Sigma_2 \ar[d] \\
      \Sigma_1 \ar[r]
        \& \Sigma_{12}
    \end{tikzcd}
    &&
    \begin{tikzcd}[ampersand replacement=\&]
      (\Sigma, \ol{\Sigma}) \ar[r] \ar[d] \arrow[dr, phantom, "\ulcorner",
                                                very near end]
        \& (\Sigma_2, \ol{\Sigma_2}) \ar[d] \\
      (\Sigma_1, \ol{\Sigma_1}) \ar[r]
        \& (\Sigma_{12}, \ol{\Sigma_{12}})
    \end{tikzcd}
  \end{align*}
\end{proposition}

%% file: Sections/initiality_theorem.tex
\section{The Initiality Theorem}
\label{sec:initiality_theorem}

The general notion of signature is built to be abstract and practical to
use.
As such, it enables us to define models and abstractly reason about them,
e.g., to prove a \hyperref[prop:modularity-models]{modularity result}, or
add equations \cite{2Signatures19} or reduction rules \cite{ReductionMonads20}.
However, it lacks many desirable properties.
As seen in \cref{ex:not-representable}, not all signatures are
representable, and more generally there is no known criterion for
a signature to have, or not to have, an initial model.
Similarly, there is no known theorem asserting that the product or coproduct
of two representable signatures is representable.

Hence, to provide an initiality theorem we must turn to different signatures.
As seen in \cref{prop:fixpoint-models}, the initial model of a representable
signature is a fixpoint of models for the functor $\ul{\Id} + \Sigma$.
As we are looking for an initial model with an underlying algebra, by
Lambek's theorem and \hyperref[thm:adamek]{Adámek's theorem}, it strongly
suggests to consider $\omega$-cocontinuous functors $\Sigma : \mc{C} \to \mc{C}$.
Yet, turning $\Sigma$ into a signature requires more data.
Indeed, given a monoid $R : \Mon(\mc{C})$, equipping $\Sigma(R)$ with a module
structure means giving an appropriate morphism $\Sigma(R) \otimes R \to \Sigma(R)$.
Using the available monoid multiplication $\mu : R \otimes R \to R$ that yields a
morphism $H(R \otimes R) \to H(R)$ by functoriality, it suffices to provide an appropriate morphism
$\Sigma(R) \otimes R \to \Sigma(R \otimes R)$.
Unfolding the requirement of such a morphism leads us to consider left strengths,
and the associated signatures with strength.
And indeed, signatures with strength will enable us to prove initiality theorems.

Intuitively, signatures with strength are better behaved than signatures as
the module substitution associated to them is of a particularly constrained
form; this enables us to prove initiality theorem for them.
Historically, functors with strength were first applied to prove initiality theorems in the seminal work of Fiore, Plotkin, and Turi \cite{FPT99}.
Note, however, that while modules over monoids naturally lead one to consider strength, they were actually employed a posteriori.
They have since been studied extensively in particular in \cite{SecondOrderDep08,ListObjects17}.
In the following, we provide an initiality theorem and adjoint theorem that are direct
counterpart to those in \cite{FPT99,SecondOrderDep08,ListObjects17}, up to technical details.
Compared to \cite{FPT99,SecondOrderDep08}, akin to \cite{ListObjects17}
we use $\omega$-cocontinuity rather than closedness.
Compared to \cite{FPT99,SecondOrderDep08,ListObjects17}, we deduce the
adjoint theorem from the initiality theorem rather than the opposite.
While those differences may appear as minor, they are important to
relate the different approaches.
We discuss Fiore et al.'s work on strengths and initiality theorems in
\cref{subsec:rw-sigma-mon}, and those differences in \cref{subsubsec:rw-co-vs-adj}.

In the remainder of this section, we start by reviewing signatures with
strength in \cref{subsec:sig_with_strength}, before discussing the links
between signatures with strength and $\Sigma$-monoids with signatures and models
in \cref{subsec:sigstrength-to-sig}.
We then state an initiality theorem and an adjoint theorem for signatures
with strength in \cref{subsec:initiality_theorem}, that we prove in \cref{sec:building_initial_model}.

\subsection{Signatures with Strength}
\label{subsec:sig_with_strength}

Using strengths to prove initiality theorems was introduced in \cite{FPT99}
for $\Sigma$-monoids, and signatures as a formal notion in \cite{Hss04}. We
follow a similar presentation to \cite{Hss04}, but a more general one
defined in terms of actegories can also be found, e.g., in \cite{SecondOrderDep08}
and \cite{HssNonWellfounded23}.

\begin{definition}[Pointed Object]
  \label{def:pointed-object}
  In a monoidal category $\mc{C}$, a \emph{pointed object} is a tuple
  $(Z,e)$ where $Z : \mc{C}$ and $e : I \to Z$.
  A morphism of pointed objects $(Z,e) \to (Z',e')$ is a morphism $f : Z \to
  Z'$ such that $f \circ e = e'$.
  Pointed objects form a category $\mathrm{Ptd}(\mc{C})$.
\end{definition}

\begin{definition}[Signatures with Strength]
  \label{def:sig-strength}
  A \emph{signature with (pointed) strength} on a monoidal category $\Cmon$ is a pair
  $(H,\theta)$ where $H : \mc{C} \to \mc{C}$ is an endofunctor with a
  \emph{strength} $\theta$, a natural transformation such that for all
  $A : \mc{C}$ and pointed object $b : I \to B$:
  \[ \theta_{A,b} : H(A) \otimes B \longrightarrow H(A \otimes B) \]
  and such that $\theta$ is compatible with the monoidal associativity and
  unit, i.e. such that for all $A, b : I \to B, c : I \to C$ the following
  diagrams commute:
  \begin{align*}
    \begin{tikzcd}[ampersand replacement=\&, column sep=large]
      (H(A) \otimes B) \otimes C \ar[d, swap, "\theta_{A,b} \otimes C"]
        \& H(A) \otimes (B \otimes C) \ar[l, swap, "\alpha^{-1}_{(H(A),B,C)}"]
                                     \ar[dd, "\theta_{A, b \otimes c}"] \\
      H(A \otimes B) \otimes C \ar[d, swap, "\theta_{A \otimes B, c}"]
        \& \\
      H((A \otimes B) \otimes C) \ar[r, swap, "H(\alpha_{A,B,C})"]
        \& H(A \otimes (B \otimes C))
    \end{tikzcd}
    &&
    \begin{tikzcd}[ampersand replacement=\&]
      H(A) \otimes I \ar[r, "\theta_{A,id}"] \ar[d, swap, "\rho_{H(A)}"]
        \& H(A \otimes I) \\
      H(A) \ar[ur, swap, "H(\rho^{-1}_A)"]
    \end{tikzcd}
  \end{align*}
  Here, $b \otimes c$ is used as an abbreviation for
  $I \xrightarrow{\lambda^{-1}_I} I \otimes I
     \xrightarrow{b \otimes c} B \otimes C$.
\end{definition}

\begin{example}
  There is a trivial signature with strength $(\Id,\Id)$ denoted $\Theta$.
\end{example}

\begin{example}
  \label{ex:left-comp}
  Given an object $D : \mc{C}$, for any signature with strength
  $(H,\theta)$, there is an associated signature with strength, for the
  endofunctor $D \otimes H(\_) : \mc{C} \to \mc{C}$ and the strength:
  \[
    \begin{tikzcd}
      (D \otimes H(A)) \otimes B \ar[r, "\alpha"]
        & D \otimes (H(A) \otimes B) \ar[r, "D \otimes \theta"]
        & D \otimes H(A \otimes B)
    \end{tikzcd}
  \]
\end{example}

\begin{definition}[Morphism of signatures with strength]
  A morphism of signatures with strength $(H, \theta) \to (H', \theta')$ is a natural
  transformation $h : H \to H'$ compatible with the strengths, i.e., such that
  for all $A$ and $b : I \to B$:
  \[
    \begin{tikzcd}
      H(A) \otimes B \ar[r, "h_A \otimes B"] \ar[d, swap, "\theta_{A,b}"]
        & H'(A) \otimes B \ar[d, "\theta'_{A,b}"] \\
      H(A \otimes B) \ar[r, swap, "h'_{A \otimes B}"]
        & H'(A \otimes B)
    \end{tikzcd}
  \]
\end{definition}

\begin{proposition}[Category of signatures with strength]
  Signatures with strength and their morphisms form a category denoted
  $\SigStrength(\mc{C})$.
  Composition and identity in this category are inherited from the category
  of functors and natural transformations.
\end{proposition}

\noindent As for signatures, to be modular, we require signatures with
strength to be closed under some limits and colimits.
This follows under the same requirement as for signatures:

\begin{proposition}[Closure under (co)limits]
  \label{prop:sigstrength-(co)complete}
  If $\mc{C}$ is (co)complete, and for all $B : \mc{C}$, $\_ \otimes B$
  preserves (co)limits, then the category of signatures with strength
  $\SigStrength(\mc{C})$ is (co)complete.
\end{proposition}

\begin{example}
  Under the above assumptions, $\SigStrength(\mc{C})$ has a terminal
  signature, and is closed under products and coproducts.
\end{example}

\noindent Another way of combining signatures with strength is to
\emph{compose} them.
Composition of signatures with strength can be used to iterate constructions, for instance, to iterate the
binding of one variable to get a binding of $n$ variables.
However, note that this construction is not available for signatures, and as
such signatures with strength provide more modularity than mere signatures.

\begin{proposition}[Closure under composition]
  \label{prop:sigstrength_comp}
  The category of signatures with strength $\SigStrength(\mc{C})$ is closed
  under composition.
\end{proposition}

\subsection{Signatures with strength as signatures}
\label{subsec:sigstrength-to-sig}

Signatures with strength \cite{FPT99,Hss04} have been introduced before
modules over monoids and signatures \cite{HirschowitzMaggesi07,HirschowitzMaggesi12},
and some authors solely rely on them as they seem enough for most applications.
For instance, the remainder of this article could mostly be read in terms of
signatures with strength.
However, some notions like presentable signatures \cite{PresentableSignatures21}
or transition monads \cite{TransitionMonads22} do not seem to
have clear counterparts in terms of strengths yet.
Moreover, modules over monoids are important conceptually, and seem to allow
for more modularity.
To relate the existing frameworks, it is therefore particularly interesting
to see signatures with strength as particular signatures.

The category of signatures with strength is not by definition a subcategory
of signatures.
Yet, as identified in \cite{HirschowitzMaggesi12} for $[\Set,\Set]$, every
signature with strength yields a signature.
In this regard, as signatures are functors associating to a monoid a module
over it, signatures with strength can be understood as signatures where the
association has been restricted to be of a particular form:

\begin{proposition}
  \label{prop:sigstrength_to_sig}
  There is a functor $\SigStrength(\mc{C}) \to \Sig(\mc{C})$ from signatures
  with strength to signatures.
  Given a signature with strength $(H,\theta)$,
  it associates the signature (i.e., a functor) that associates to a monoid
  $(R,\eta,\mu)$ the $R$-module $H(R)$ with the following module substitution:
  \[
    \begin{tikzcd}
      H(R) \otimes R \ar[r, "\theta_{R,\eta}"]
        & H(R \otimes R) \ar[r, "H(\mu)"]
        & H(R)
    \end{tikzcd}
  \]
  This signature associates to a monoid morphism $f$, the morphism of modules
  $H(f)$.
  Given a morphism of signatures with strength $h$, the functor associates
  the morphism of signatures $(\Id,h)$.
\end{proposition}

\begin{remark}
  \label{remark:model-sigma-monoids}
  Unfolding the definition of model, one can see that a model for a
  signature with strength, as defined in \cref{subsec:models}, is exactly
  the same as the definition of model used in \cite{FPT99,SecondOrderDep08,ListObjects17}.
  Indeed, an $H$-monoid is an $H$-algebra structure $(R,r)$ equipped with
  a monoid structure $(R,\eta,\mu)$ such that
  \[
    \begin{tikzcd}
      H(R) \otimes R \ar[r, "\theta"] \ar[d, swap, "\tau \otimes R"]
        & H(R \otimes R) \ar[r, "H(\mu)"]
        & H(R) \ar[d, "\tau"]\\
      R \otimes R \ar[rr, swap, "\mu"] & & R.
    \end{tikzcd}
  \]
  i.e a monoid $(R,\eta,\mu)$ with a $R$-module morphism $r : H(R) \to
  \Theta$, that is a model.
\end{remark}

\noindent The above functor yields signatures from signatures with strength.
However, prima facie we could get different signatures whether we take a
construction at the level of signatures or at the level of signatures with strength.
For instance, $\Theta : \SigStrength(\mc{C})$ could not map to $\Theta :
\Sig(\mc{C})$, or taking the (co)product of signatures with strength and
translating to signatures, or translating and then taking the (co)product as
signatures could yield different results.
Thankfully, this is not the case, and we can indeed see signatures with
strength as particular signatures.

\begin{proposition}
  The above defined functor $\SigStrength(\mc{C}) \to \Sig(\mc{C})$ maps the
  signature with strength $\Theta : \SigStrength(\mc{C})$ to the signature
  $\Theta : \Sig(\mc{C})$.
  It also maps the construction $D \otimes \_$ of signatures with strength
  to the corresponding construction $D \otimes \_$ on signatures.
\end{proposition}

\begin{proposition}
  \label{prop:sigstrength_sig_(co)limits-preserved}
  If $\mc{C}$ is (co)complete, and for all $B : \mc{C}$, $\_ \otimes B$
  preserves (co)limits, then the above defined functor $\SigStrength(\mc{C})
  \to \Sig(\mc{C})$ preserves (co)limits.
\end{proposition}

\subsection{The Initiality Theorem}
\label{subsec:initiality_theorem}

As for signatures, not all signatures with strength admit an initial model.
For instance, the signature $\mc{P} \circ \Theta$ can be equipped with a
strength by \cref{ex:left-comp}, but is not representable on $[\Set,\Set]$
by \cref{ex:not-representable}.
Thankfully, as signatures with strength $(H,\theta)$ decompose into an
endofunctor $H : \mc{C} \to \mc{C}$ and a strength $\theta$, it is easier to
provide an initiality theorem and an adjoint theorem by requiring conditions
on $\mc{C}$ and $H$.
We do so similarly to \cite{FPT99,SecondOrderDep08,ListObjects17}, except
that compared to \cite{FPT99,SecondOrderDep08} but similarly to
\cite{ListObjects17}, we rely on $\omega$-cocontinuity rather than
closedness; and prove the adjoint theorem from the initiality theorem rather
than the opposite.
Those technical changes are discussed in detail in \cref{subsubsec:rw-co-vs-adj}.

\begin{restatable}[The Initiality Theorem]{theorem}{initialitytheorem}
  \label{thm:initiality-theorem}
  Let $\mc{C}$ be a monoidal category, with initial object, binary coproducts,
  $\omega$-colimits and such that for all $Z : \mc{C}$, $\_ \otimes Z$
  preserves initiality, binary coproducts and $\omega$-colimits.
  Then, given a signature with strength $(H,\theta)$, if $H$ is
  $\omega$-cocontinuous, then the associated signature has an initial model
  $\ol{H}$, with underlying object $\mu A.(I + H(A))$.
\end{restatable}

\noindent This theorem is very powerful as it provides a single initiality
theorem applicable to different input monoidal categories, hence handling
different kinds of contexts and different type systems.
Moreover, the conditions are relatively mild, as in practice, we will always
work with cocomplete categories, and with nice monoidal products on the left.
Under one more hypothesis, the initiality theorem can be extended to an adjoint:

\begin{restatable}[The Adjoint Theorem]{theorem}{adjointtheorem}
  \label{thm:adjoint-theorem}
  Let $\mc{C}$ be a monoidal category, with initial object, binary coproducts,
  $\omega$-colimits and such that for all $Z : \mc{C}$, $\_ \otimes Z$
  preserves initiality, binary coproducts and $\omega$-colimits.
  If additionally, for all $X : \mc{C}$, $X \otimes \_$ preserves $\omega$-colimits,
  then, for any $\omega$-cocontinuous signature with strength $(H,\theta)$,
  the forgetful functor $U : \Model(H) \to \mc{C}$ has a left adjoint
  $\mrm{Free} : \mc{C} \to \Model(H)$, where $\Free(X)$, for $X : \mc{C}$,
  has $\mu A.(I + H(A) + X \otimes A)$ as underlying object.
\end{restatable}

\noindent In practice, one works with a fixed monoidal category $\mc{C}$
satisfying the hypotheses of the initiality theorem.
Hence, it is particularly interesting to consider the subcategory of
$\omega$-cocontinuous signatures with strength:

\begin{definition}
  We define the category $\SigStrength_\omega(\mc{C})$ as the subcategory of
  signatures with strength that preserve $\omega$-colimits.
\end{definition}

\begin{corollary}
  \label{coro:all-pres}
  Under the hypotheses of the initiality theorem, all signatures of
  $\SigStrength_\omega(\mc{C})$ are representable, and there is a functor
  associating its initial model to each signature:
  \[\ol{(\_)} : \SigStrength_\omega(\mc{C}) \longrightarrow \TotSigModel{\mc{C}}\]
\end{corollary}

\noindent Previously considered closure properties enable us to build
signatures (with strength) representing languages.
By \cref{coro:all-pres}, by applying them to $\SigStrength_\omega(\mc{C})$,
we can modularly build an initial model for our signatures.

\begin{proposition}[Closure under colimits]
  \label{prop:sigstrength_omega_cocomplete}
  By \cref{prop:omega-colimits,prop:sigstrength-(co)complete}, if $\mc{C}$
  is cocomplete and for all $B : \mc{C}$, $\_ \otimes B$ preserves colimits,
  then the category $\SigStrength_\omega(\mc{C})$ is cocomplete.
\end{proposition}

\begin{proposition}[Closure under limits]
  \label{prop:sigstrength_omega_complete}
  By \cref{prop:omega-limits,prop:sigstrength-(co)complete}, if $\mc{C}$
  admits a class of limits that commute with $\omega$-colimits, and that is
  preserved by $\_ \otimes B$ for all $B : \mc{C}$, then the category
  $\SigStrength_\omega(\mc{C})$ is closed under this class of limits.
\end{proposition}

\begin{example}
  Admitting a suitable monoidal structure on $\Set$ and $[\mc{C},\Set]$, the
  categories $\SigStrength_\omega(\Set)$ and $\SigStrength_\omega([\mc{C},\Set])$
  have a terminal object, are closed under \emph{finite} products, and are
  closed under coproducts.
\end{example}

\begin{proposition}[Closure under composition]
  \label{prop:sigstrength_omega_comp}
  By \cref{prop:omega_comp,prop:sigstrength_comp}, the category
  $\SigStrength_\omega(\mc{C})$ is closed under composition.
\end{proposition}

%% file: Sections/building_initial_model.tex
\section{Building an Initial Model}
\label{sec:building_initial_model}

Modules over monoids provide a flexible framework, and signatures
with strength provide an \hyperref[thm:initiality-theorem]{initiality
theorem}.
To be as complete as possible as an introduction, we provide a detailed
proof of the initiality theorem.
If one only wishes to apply the framework, the proof can be skipped and
considered as a black box, in which case, we refer to
\cref{sec:example_set_set,sec:example_F_set,sec:example_setT_setT} for
examples.

To prove the initiality theorem, we rely on heterogeneous substitution systems (hss).
Hss were originally designed to prove that both wellfounded and non-wellfounded
syntax \cite{Hss04} have a monadic substitution structure.
In the wellfounded case, they were also shown to provide a form of initial
semantics \cite{HssRevisited15}.
In the following, we present a proof that is a generalisation from
endofunctor categories to monoidal categories, and an adaptation to our
purpose, of a proof scattered throughout
\cite{Hss04,DeBruijnasNestedDatatype99,HssRevisited15,HssUntypedUniMath19}.
We stress that hss and a variant of the proof below have been recently and
independently generalised to monoidal categories by Matthes, Wullaert, and
Ahrens, as mentioned in \cite[Section 4.4]{HssNonWellfounded23}.
We provide a full discussion on hss in \cref{subsec:rw-hss}.

The proof can be presented in three steps.
We start by building an initial algebra using Adámek's Theorem.
Then, using generalized iteration in Mendler's style we enrich the initial
algebra into a heterogeneous substitution system; structure that, in turn, is
used to derive the structure of a model.
Lastly, one shows that the obtained model is initial using the initiality of
the algebra.
To convey more intuition and to better motivate heterogeneous substitution
systems, we choose a slightly different presentation.
We first review hss and explain why they allow us to construct models in \cref{subsec:hss_models}.
Then, we explain how to build a heterogeneous substitution system from our
assumptions in \cref{subsec:building_hss}, and why the model constructed from
it is initial as a model in \cref{subsec:building_initial_model}.

In this section, we suppose that $\mc{C}$ has coproducts, and that they are
preserved by $\_ \otimes Z$ for all $Z : \mc{C}$, as it is needed to state
our definitions.
As before, these hypotheses are by no means a restriction as they are also hypotheses of the
initiality theorem.
For readability, we write the functor $I + H\_$, i.e. $A \mapsto I + H(A)$,
as $I + H$ when appearing as an index.

\subsection{Heterogeneous Substitution Systems and Models}
\label{subsec:hss_models}

The recursion principle arising from initial algebras is not sufficient to
prove that a higher-order language, such as the lambda calculus, forms a monad.
As explained, for $[\Set,\Set]$, in \cite{DeBruijnasNestedDatatype99,GeneralisedFold99},
a stronger recursion principle is required; in their case, one named
``generalised fold''.
Heterogeneous substitution systems, introduced in \cite{Hss04,HssRevisited15}
on endofunctor categories, are inspired by generalised folds, and are
designed to prove that higher-order languages form monads when considered
with their substitutions.
Following them, hss strengthen the concept of algebra by offering the
possibility to precompose $R$ by a pointed object $Z$, and to build a unique
map $\{ f \} : R \otimes Z \to R$ for any morphism $f : Z \to R$.

\begin{definition}[Heterogeneous Substitution System]
  \label{def:hss}
  For a signature with strength $(H, \theta)$, a heterogeneous substitution
  system (hss) is a tuple $(R, \eta, r)$ where $R : \mc{C}$ is an object
  of $\mc{C}$ and $\eta : I \to R$ and $r : H(R) \to R$ are morphisms
  of $\mc{C}$ such that, for all $(Z , e) : \trm{Ptd}(\mc{C})$ and $f : Z
  \to R$, there is a unique morphism $\{ f \} : R \otimes Z \to
  R$ making the following diagram commute:
  \[
    \begin{tikzcd}[column sep=large]
      I \otimes Z \ar[r, "\eta \otimes Z"] \ar[dd, swap, "\lambda_Z"]
        & R \otimes Z \ar[dd, "\{f\}"]
        & H(R) \otimes Z \ar[l, swap, "r \otimes Z"]
                         \ar[d, "\theta_{R,e}"] \\
        &
        & H(R \otimes Z) \ar[d, "H(\,\{f\}\,)"] \\
      Z \ar[r , swap, "f"]
        & R
        & H(R) \ar[l, "r"]
    \end{tikzcd}
  \]
\end{definition}

\begin{remark}
  In previous works on hss \cite{Hss04,HssRevisited15,HssUntypedUniMath19,HssTypedUnimath22},
  it was required for $f$ to be a morphism of pointed objects $f : (Z,e) \to (R,\eta)$.
  It was later realised independently by Matthes, Wullaert, and Ahrens in \cite{HssNonWellfounded23},
  and by Lafont that this hypothesis is superfluous for our purpose.
\end{remark}

\noindent Indeed, in a strict monoidal category, if one fixes $(Z,e) := (I,\Id)$
and $f := \Id$, then $\theta_{I,\Id} = \Id$ and we get an algebra for $H$.
As done in \cite{Hss04} for endofunctor categories, the added freedom of choosing
$Z$ as we wish will enable us to derive a monoid, and later a model, from a
hss:

\begin{proposition}[Monoids from Hss]
  \label{prop:hss_to_monoid}
  If a signature $(H, \theta$) has a heterogeneous substitution system $(R,
  \eta, r)$, then $(R, \eta, \{ \Id_{(R,\eta)} \})$ is a monoid.
\end{proposition}
\begin{proof}
  Let's denote $\{\Id_{(R,\eta)}\} : R \otimes R \to R$ by $\mu$.
  By definition, $(R, \eta, \mu)$ is well-typed.
  Moreover, the left-unit law of the monoid $\lambda_R = \mu \circ (\eta \otimes R)$
  holds by definition of $\mu$.
  Hence, there are only the right-unit law and the associativity of $\mu$
  to check.

  To prove the second unit law, $\rho_R = \mu \circ (R \otimes \eta)$, let's
  apply the hss for $f := \eta : I \to R$.
  Both morphisms make the diagram commute, thus are equal by uniqueness.
  The morphism $\rho_R$ can be shown to satisfy the diagram using that
  $\lambda_I = \rho_I$, the definition of $\theta_{R,\Id}$, and the
  naturality of $\rho$.
  The morphism $\mu \circ (R \otimes \eta)$ can be shown to satisfy the
  diagram using the left-unit law, the definition of $\mu$, and the
  naturality of $\lambda$, and of $\theta$ in the second argument.
  Hence by uniqueness $\{ \eta \} = \mu \circ (R \otimes \eta)$.
  Thus $\rho_R = \mu \circ (R \otimes \eta)$.

  For homogeneity reasons, to prove the associativity of $\mu$, we prove $\mu
  \circ (R \otimes \mu) = \mu \circ (\mu \otimes R) \circ \alpha^{-1}$.
  To do so, let's apply the hss in $\mu : R \otimes R \to R$, where $R \otimes R$ is pointed by
  $I \xrightarrow{\lambda_I} I \otimes I \xrightarrow{\eta \otimes \eta} R \otimes R$.
  Both morphisms make the diagram commute, thus are equal by uniqueness.
  The morphism $\mu \circ (\mu \otimes R)$ can be shown to satisfy it using
  the unit law, the naturality of $\theta$, and the definition of $\mu$.
  The morphism $\mu \circ (\mu \otimes R) \circ \alpha^{-1}$ can be shown to
  satisfy the diagram using the naturality of $\theta$ and $\alpha$, the
  definition of $\mu$, and the definition of $\theta$ in $e \otimes e'$.
\end{proof}

\begin{proposition}[Models from Hss]
  \label{prop:hss_to_model}
  If a signature $(H, \theta$) has a heterogeneous substitution system
  $(R,\eta,r)$, then $((R,\eta,\mu), r)$ is a model of $H$.
\end{proposition}
\begin{proof}
  By \cref{prop:hss_to_monoid}, $(R,\eta,\mu)$ is a monoid.
  It remains to prove that $r$ is a morphism of modules $H(R) \to R$ i.e.
  to prove that $\mu \circ (r \otimes R) = r \circ H(\mu) \circ \theta_{R,\Id}$.
  This is true by the definition of $\mu$.
\end{proof}

\subsection{Building Heterogeneous Substitution Systems}
\label{subsec:building_hss}

The previous result implies that it suffices to build a heterogeneous
substitution system for a signature with strength to get a model for it.
As described in \cite{HssUntypedUniMath19} for endofunctor categories, we
can do so using generalized iteration in Mendler's style. As the proof of the
theorem relies on a particular way of building initial algebras, we will
first recall, and give a proof of, Adámek's Theorem.

\begin{theorem}[Adámek's Theorem \cite{Adamek74}]
  \label{thm:adamek}
  Let $\mc{C}$ be a category with an initial object $0$, and
  $\omega$-colimits. Let $F : \mc{C} \to \mc{C}$ be an $\omega$-cocontinuous
  functor.
  Then there is an initial $F$-algebra $(R,r)$ with $R$ built as the
  $\omega$-colimit of:
  \[
    \begin{tikzcd}[column sep=large]
      0 \ar[r, "i"] \ar[dr, swap, "t_0"]
        & F(0) \ar[r, "F(i)"] \ar[d, "t_1"]
        & F^2(0) \ar[r, "F^2(i)"] \ar[dl, "t_2"]
        & ... \\
      & R
        &
        &
        &
    \end{tikzcd}
  \]

\end{theorem}
\begin{proof}
  As $F$ preserves $\omega$-colimits, $F(R)$ is the colimit of the image by
  $F$ of the chain $F(\trm{chn}_F)$.
  Using the initiality of $0$, it can be completed into a cocone for the
  full chain $\trm{chn}_F$.
   Given any other cocone $A$ of $\trm{chn}_F$, $A$ is, in particular, a
   cocone for $F(\trm{chn}_F)$, which yields a map $h : F(R) \to A$
   satisfying the universal property of $F(\trm{chn}_F)$.
   Using initiality, $h$ also verifies the one of $F(\trm{chn}_F)$.
  Given any other such map $h'$, in particular both $h$ and $h'$ satisfy the
  universal property $F(\trm{chn}_F)$, hence as $F(R)$ is a colimit of this
  chain, $h = h'$.
  Thus $F(R)$ is a colimit of $\trm{chn}_F$, which yields a unique map
  $r : F(R) \to R$.

  Given an $F$-algebra $(A,a)$, we build a cocone on $A$.
  We define $a_n$ recursively: the morphism $a_0$ is the unique morphism
  from $0 \to A$; for a natural number $n$, $a_{n+1}$ is defined as the
  composition $F^{n+1}(0) \xrightarrow{a_n} F(A) \xrightarrow{a} A$.
  The commutativity holds by the uniqueness in the $0$ case and recursively
  for $n+1$.
  Then by the universal property we obtain a unique map $h : C \to A$.
  Both $a\circ F(h)$ and $h \circ c$ satisfy the universal property of the
  colimit, as such they are equal and so $h$ is a morphism of algebras.
  Given another morphism of $F$-algebras $h'$, it can be shown by induction
  that it verifies the universal property, hence by uniqueness $h = h'$, and
  so $h$ is unique.
  In consequence, $(R,r)$ is an initial $F$-algebra.
\end{proof}

\begin{restatable}[Generalized Mendler's style Iteration {{\cite[Theorem 1]{GeneralisedFold99}}}]{theorem}{Mendler}

  \label{thm:gen-mendler}
  Let $\mc{C}$ and $\mc{D}$ be two categories with initial object and
  $\omega$-colimits, $F : \mc{C} \to \mc{C}$ and $L : \mc{C} \to \mc{D}$ be
  two functors.
  If $F$ and $L$ preserve $\omega$-colimits, and $L$ preserves initiality,
  then for all $X : \mc{D}$ and natural transformation
  \[ \Psi : \mc{D}(L\_,X) \to \mc{D}(L(F\_),X) \]
  there is a unique morphism $\mrm{It}^L_F(\Psi) : L(R) \to X$ in $\mc{D}$
  --- where $R$ is the initial $F$-algebra obtained by Adámek's Theorem ---
  making the following diagram commute:
  \[
    \begin{tikzcd}[ampersand replacement=\&]
      L(F(R)) \ar[r, "L(r)"] \ar[dr, swap, "\Psi_R(\mrm{It}^L_F(\Psi))"]
        \& L(R) \ar[d, "\mrm{It}^L_F(\Psi)"] \\
      \& X
    \end{tikzcd}
  \]
\end{restatable}
\begin{proof}
  As a shorthand, we will denote $\mrm{It}^L_F(\Psi)$ by $h$ in the proof of existence and uniqueness of $\mrm{It}^L_F(\Psi)$.
  First, let's construct $h$.
  As $L$ preserves initiality, $L(0)$ is initial and there is a unique map
  $x : L(0) \to X$.
  Iterating $\Psi$ yields a cocone $(X, (\Psi^n(x))_{n : \mb{N}})$ for the
  diagram $(LF^n(0), LF^n(i))_{n : \mb{N}}$.
  Yet as $L$ preserves $\omega$-colimits, $L(R)$ is the colimit of this
  diagram, and by the universal property of colimits, there is a unique $h :
  R(L) \to X$ such that $\forall n.\; h \circ L(t_n) = \Psi^n(x)$, i.e.,
  such that the following diagram commutes:
  \[
    \begin{tikzcd}[column sep=large]
      L(0) \ar[r, "L(i)"] \ar[dr, swap, "L(t_0)"] \ar[ddr, bend right, swap, "\Psi^0(x)"]
        & LF(0) \ar[r, "LF(i)"] \ar[d, "L(t_1)"]
        & LF^2(0) \ar[r, "LF^2(i)"] \ar[dl, "L(t_2)"] \ar[ddl, bend left, "\Psi^2(x)"]
        & ... \\
      & L(R) \ar[d, "h"]
        &
        &
        & \\
      & X
        &
        &
        &
    \end{tikzcd}
  \]
  We have built a map $h$, it remains to prove that it satisfies the
  desired property:  $h \circ L(r) = \Psi(h)$.
  To do so we are going to prove that $\forall n.\; \Psi(h) \circ
  L(r^{-1}) \circ L(t_n) = \Psi^n(x)$.
  Indeed by the uniqueness of $h$, we can conclude that $h = \Psi(h) \circ
  L(r^{-1})$, i.e. $L(r) \circ h = \Psi(h)$.
  We prove $\forall n.\; \Psi(h) \circ L(\alpha^{-1}) \circ L(t_n) =
  \Psi^n(x)$ by induction on $n$.
  The $n = 0$ case holds by initiality of $L(0)$.
  The $n+1$ case holds by the following chain of equalities:
  \[
  \begin{array}{cclc}
    \Psi(h) \circ L(r^{-1}) \circ L(t_{n+1})
      &=& \Psi(h) \circ L(r^{-1}) \circ L(r) \circ LF(t_{n})
          & (\textrm{definition of }r) \\
      &=& \Psi(h) \circ LF(t_{n}) & \\
      &=& \Psi(h \circ L(t_{n})) & (\textrm{naturality of }\Psi) \\
      &=& \Psi(\Psi^n(x)) & (\textrm{induction hypothesis})
  \end{array}\]
  Now that we have proven existence, we need to prove uniqueness.
  To do so, suppose we have a $h : L(R) \to X$ such that $h \circ L(r) =
  \Psi(h)$; we are going to show that $\forall n.\; h \circ L(t_n) =
  \Psi^n(x)$.
  As a consequence, such an $h$ verifies the universal property of $L(R)$ and
  hence $h$ is unique. We prove the equations by induction on $n$.
  The $n = 0$ case holds by initiality of $L(0)$.
  The $n+1$ case holds by the following chain of equation:
  \[
  \begin{array}{cclc}
    h \circ L(t_{n+1})
      &=& h \circ L(r \circ t_n)
          & (\textrm{definition of }r) \\
      &=& h \circ L(\alpha) \circ L(t_n) & \\
      &=& \Psi(h) \circ L(t_n) & (\textrm{definition of }h) \\
      &=& \Psi(h \circ LF(t_n)) & (\textrm{naturality of }\Psi) \\
      &=& \Psi(\Psi^n(x)) & (\textrm{induction hypothesis})
  \end{array}\]
\end{proof}

\begin{proposition}[Building an Hss]
  Let $\mc{C}$ be a monoidal category, with initial object, coproducts,
  $\omega$-colimits and such that for all $Z : \mc{C}$, $\_ \otimes Z$
  preserves initial objects, coproducts, and $\omega$-colimits.
  Let $(H,\theta)$ be a signature with strength such that $H$ is
  $\omega$-cocontinuous.
  Then $(R, \eta + r)$ --- the initial algebra of $I + H\_$ obtained by
  Adámek's Theorem --- is a heterogeneous substitution system for
  $(H,\theta)$.
\end{proposition}

\begin{proof}
  Let $(Z,e)$ be a pointed object, and $f : Z \to R$ a morphism.
  We are going to construct $\{ f \}$ and show that it is uniquely defined,
  by a suitable application of generalized Mendler's style iteration.
  Let's apply the theorem for $F := \Id + H$, $L := \_ \otimes Z$, $X := R$
  and $\Psi \;h \mapsto (f + r) \circ (\Id + H(h)) \circ (\lambda_Z +
  \theta_{R,e})$.
  This will give us a unique map $\mrm{It}^{\_ \otimes Z}_{I + H}(\Psi) : R
  \otimes Z \to R$, such the following diagram:
  \[
  \begin{tikzcd}[column sep=4cm]
    I \otimes Z + H(R) \otimes Z
                        \ar[r, "(\eta + r) \otimes Z"]
                        \ar[d, swap, "\lambda_Z + \theta_{R,e}"]
                        \ar[ddr, start anchor=-25, shorten <=8pt, end anchor=north west,
                             "\Psi(\mrm{It}^{\_ \otimes Z}_{I + H}(\Psi))"]
      & R \otimes Z     \ar[dd, "\mrm{It}^{\_ \otimes Z}_{I + H}(\Psi)"]\\
    Z + H(R \otimes Z)  \ar[d, swap, "\Id + H(\mrm{It}^{\_ \otimes Z}_{I + H}(\Psi))"]
      & \\
    Z + H(R)            \ar[r, swap, "{[}f {,} r{]}"]
      & R \\
  \end{tikzcd}
  \]
  This diagram is exactly the definition of an hss, if one denotes
  $\mrm{It}^{\_ \otimes Z}_{I + H}(\Psi)$ by $\{ f \}$.
  Hence it suffices to verify the hypothesis for the given input to get the
  result.
  The only non-obvious hypothesis is that the above defined $\Psi$ is natural
  in $R$, which follows by naturality of $\theta$ in its first argument.
\end{proof}

\subsection{Building an Initial Model}
\label{subsec:building_initial_model}

The previous work has enabled us to construct a model of a given signature.
It remains to prove that this model is initial.
To do so, we adapt a proof for hss on endofunctor categories introduced in
\cite{HssRevisited15} to models on monoidal categories.
This crucially relies on a fusion law for generalized Mendler's style iteration
which enables us to factorise a generalized Mendler's style iteration followed
by a natural transformation into a Mendler's style iteration.
This is not surprising as we are relying on a generalized recursion principle
to build our model, and fusion laws are ubiquitous in computer science to
simplify recursion principles.

\begin{theorem}[Fusion Law for Generalized Mendler's style Iteration {{\cite[Lemma 9]{HssRevisited15}}}]
  \label{thm:fusion-law}
  Let $\mc{C,D},F,L, X, \Psi$ be objects satisfying the hypotheses of the
  \hyperref[thm:gen-mendler]{generalized Mendler's style iteration} theorem.
  For the same $\mc{C,D},F$, suppose given other $L',X',\Psi'$ satisfying
  the hypotheses.
  If there is a natural transformation $\Phi : \mc{D}(L\_,X) \to
  \mc{D}(L'\_,X')$ such that $\Phi_{F(\mu F)} \circ \Psi_{\mu F} =
  \Psi'_{\mu F} \circ \Phi_{\mu F}$ --- where $\mu F$ denotes the
  $F$-algebra built by Adámek's theorem --- then
  \[ \Phi_{\mu F}(\mrm{It}^L_F(\Psi)) = \trm{It}^{L'}_F(\Psi') \]
\end{theorem}
\begin{proof}
  By uniqueness, it suffices to prove that $\Phi_{\mu F}(\trm{It}^L_F(\Psi))$
  satisfies the defining diagram of $\trm{It}^{L'}_F(\Psi')$.
  This can be done using the definition of the assumption, the definition of
  $\trm{It}^L_F(\Psi)$ and the naturality of $\Phi$.
\end{proof}

We are now ready to prove the initiality theorem:

\initialitytheorem*
\begin{proof}
  By \cref{prop:hss_to_model}, $((R,\eta, \mu), r)$ is a model of $H$.
  We need to prove that this model is actually initial.
  Let $((R',\eta', \mu'), r')$ be another model.
  We need to prove there is a unique morphism of models $((R,\eta, \mu),
  r) \to ((R',\eta',\mu'),r')$.
  In other words, we need to prove there is a unique morphism of monoids $f
  : (R,\eta,\mu) \to (R',\eta',\mu')$ --- i.e., respecting $\eta$ and $\mu$
  --- making the following diagram commute:
  \[
    \begin{tikzcd}
      H(R) \ar[r, "r"] \ar[d, swap, dashed, "H(f)"]
        & R \ar[d, dashed, "\exists ! f"] \\
      f^*H(R') \ar[r, swap, "f^*r'"] & f^*R'
    \end{tikzcd}
  \]
  To prove both uniqueness and existence, we are going to use that $(R',
  \eta' + r')$ is a $(I + H\_)$-algebra and that $(R, \eta + r)$ is the
  initial one. \medskip

  To prove uniqueness, suppose we have a morphism of models $f$, in
  particular it is a morphism of $(I + H\_)$-algebras $f : (R, \eta + r) \to
  (R', \eta' + r')$ and as such is unique by the initiality of $(R, \eta
  + r)$. \medskip

  To prove the existence of such a model morphism, we are going to show that
  the morphism of algebras $f : (R, \eta + r) \to (R', \eta' + r')$
  existing by initiality of $(R, \eta + r)$ is a morphism of models.
  The morphism $f$ respects $\eta$ as it is a morphism of algebras.
  The commutativity of the diagram of module morphisms holds if it holds for
  the underlying morphism of $\mc{C}$, which is also verified as $f$ is a
  morphism of algebras.

  It remains to prove that it respects $\mu$, i.e., that $f \circ \mu = \mu'
  \circ f \otimes f$.
  To do so, we are going to use that $\mu := \{ \Id \} := \mrm{It}^{\_
  \otimes Z}_{I + H}(\Psi)$  to factorise $f \circ \mu$ into another
  iteration $\mrm{It}^{\_ \otimes Z}_{I + H}(\Psi')$ as shown below.
  The point is that iterations are the unique morphisms making their
  associated diagrams commute.
  In consequence, it will suffice to prove that $\mu' \circ f \otimes f$
  satisfies it to prove the equalities.
  \begin{align*}
    \begin{tikzcd}[ampersand replacement=\&]
      R \otimes R \ar[r, "f \otimes f"]
                  \ar[d, swap, "\mrm{It}^{\_ \otimes Z}_{I + H}(\Psi)"]
        \& R' \otimes R' \ar[d, "\mu'"] \\
      R \ar[r, swap, "f"]
        \& R'
    \end{tikzcd}
    &&
    \begin{tikzcd}[ampersand replacement=\&]
      R \otimes R \ar[r, "f \otimes f"] \ar[dr, swap, bend right, "\mrm{It}^{\_ \otimes Z}_{I + H}(\Psi')"]
        \& R' \otimes R' \ar[d, "\mu'"] \\
        \& R'
    \end{tikzcd}
  \end{align*}
  Here $\Psi \;h := (\Id + r) \circ (\Id + H(h)) \circ (\lambda_Z +
  \theta_{R,e})$ and $\Psi' \;h := (f + r') \circ (\Id + H(h)) \circ
  (\lambda_Z + \theta_{R,e})$.
  To do the factorisation, we apply the fusion law for $\Phi := f^*$ i.e.
  precomposition by $f$, for which the assumption is satisfied as $f$ is a
  morphism of algebras.
  Finally, proving that $\mu' \circ f \otimes f$ satisfies the diagram of
  $\mrm{It}^{\_ \otimes Z}_{I + H}(\Psi')$ unfolds to proving:

  \begin{align*}
    \begin{tikzcd}[ampersand replacement=\&]
      I \otimes R \ar[ddd, swap, "\lambda_R"] \ar[rr, "\eta \otimes R"]
                  \ar[dr, swap, "I \otimes f"]
        \&
        \& R \otimes R \ar[d, "R \otimes f"] \\
        \& I \otimes R' \ar[r, "\eta \otimes R'"]
                       \ar[dr, swap, near end, "\eta' \otimes R'"]
                       \ar[ddr, swap, bend right, near start, "\lambda_{R'}"]
        \& R \otimes R' \ar[d, "f \otimes R'"] \\
        \&
        \& R' \otimes R' \ar[d, "\mu'"] \\
      R \ar[rr, "f"]
        \&
        \& R'
    \end{tikzcd}
    &&
    \begin{tikzcd}[ampersand replacement=\&]
      H(R) \otimes R \ar[rr, "r \otimes R"] \ar[dr, "H(f) \otimes f"]
                     \ar[d, swap, "\theta_{R,\Id}"]
        \&
        \& R \otimes R \ar[dd, "f \otimes f"] \\
      H(R \otimes R) \ar[d, swap, "H(f \otimes f)"]
        \& H(R') \otimes R' \ar[dl, "\theta_{R',\Id}"] \ar[dr, "r' \otimes R'"]
        \& \\
      H(R' \otimes R') \ar[d, swap, "H(\mu')"]
        \&
        \& R' \otimes R' \ar[d, "\mu'"]\\
      H(R') \ar[rr, swap, "r'"]
        \&
        \& R'
    \end{tikzcd}
  \end{align*}
  The left diagram commutes using the naturality of $\lambda$, the fact that
  $f$ is a morphism of algebras, and by the monoid laws.
  The right diagram commutes by naturality of $\theta$ in both arguments,
  the fact that $f$ is a morphism of algebras and because $r$ is a
  morphism of modules.
\end{proof}

\subsection{Building an Adjoint}
\label{subsec:building-adjoint}

While it is possible to directly prove the adjoint theorem, and deduce the
initiality theorem from it, it is possible and actually better to do the
opposite.
Indeed, by doing so, one can save in the initiality theorem, the hypothesis
that $X \otimes \_$ is $\omega$-cocontinuous:

\adjointtheorem*
\begin{proof}
  First, let's prove the existence of $\Free : \mc{C} \to \Model(H)$.
  Given $X : \mc{C}$, $X \otimes \_$ can be equipped with the strength
  $\alpha : (X \otimes A) \otimes B \to X \otimes (A \otimes B)$.
  Hence, by \cref{prop:sigstrength_omega_cocomplete}, $H + X \otimes \_$
  is an $\omega$-cocontinuous signature with strength, and so has an initial
  model by the initiality theorem denoted $\Free(X)$, with carrier $\mu A.
  (I + H(A) + X \otimes A)$.
  We then get a model of $H$ by forgetting the extra $X \otimes \_$ structure.
  Given a morphism $X \to Y$, $\Free(Y)$ can be equipped with a $H + X
  \otimes \_$ structure.
  Hence, by initiality, there is a unique $H + X \otimes \_$ morphism of
  model $\Free(X) \to \Free(Y)$, which is in particular a $H$ morphism of
  models.

  Given an object $X : \mc{C}$, and a model of $H$, $M : \Model(H)$, we need
  to build an natural isomorphism $\Model(H)(\Free(X),\, M) \cong C(X,\,
  M)$, where we identify $U(M)$ with $M$.
  To build $K : C(X,\, M) \to \Model(H)(\Free(X),\, M)$, given $f : X \to
  M$, we turn $M$ from an $H$ model to an $H + X\otimes \_$ model.
  Indeed, by initiality of $\Free(X)$, this will provide a unique morphism
  of $H + X \otimes \_$ models, which is in particular a morphism of
  $H$-models.
  To do so, it suffices to provide a module morphism $X \otimes \Theta \to
  \Theta$, which can be defined as $X \otimes M \xrightarrow{f \otimes M} M
  \otimes M \xrightarrow{\mu_M} M$.
  We can build an inverse $L : \Model(H)(\Free(X),\, M) \to C(X,\, M)$ using
  that $\Free(X)$ is a $H + X \otimes \_$ model, and as such is equipped
  with morphism $\sigma : X \otimes \Free(X) \to \Free(X)$.
  Given $f^\# : \Free(X) \to M$, we define $L(f^\#)$ as
  $X \xrightarrow{\rho_X}         X \otimes I
     \xrightarrow{X \otimes \eta} X \otimes \Free(X)
     \xrightarrow{\sigma}         \Free(X)
     \xrightarrow{f^\#}           M$.
  Proving that $L \circ K (f) = f$, follows the universal property of $G(f)$,
  the definition of $L$ and the monoid laws.
  By definition, $K(L(f^\#))$ is the unique morphism of $H + X \otimes \_$
  models from $X$ to $M$ when equipped with $L(f)$.
  Hence, to prove that $K(L(f^\#)) = f^\#$, it suffices to prove that $f^\#$
  is such a morphism.
  As $h$ is already a morphism of $H$ models, it suffices to verify that it
  is compatible with the $X \otimes \_$ constructor.
  It follows by the definition of the strength $X \otimes \_$ and the
  different monoid and monoidal laws.
  Lastly, the naturality in both arguments follows from the definition.
\end{proof}

%% file: Sections/example_Set_Set.tex
\section{Untyped languages: The instance $[\Set,\Set]$}
\label{sec:example_set_set}

We have defined a category of models for signatures on a monoidal category
$\mc{C}$, and provided an initiality theorem for $\omega$-cocontinuous
signatures with strength.
However, the theorem would be of little use if $\omega$-cocontinuous signatures
with strength were too restrictive to represent higher-order languages, or
if our models did not appropriately modelled the syntax.
In this section, we justify in detail that the initiality theorem does
enable us to abstractly reason about \emph{untyped} higher-order languages.
To do so, we focus on the monoidal category $[\Set,\Set]$, a popular
instance in the literature for its simplicity \cite{HirschowitzMaggesi07,ZsidoPhd10}.
Very roughly, the idea is that a language is represented by a functor that,
given a set of variables $\Gamma : \Set$ (i.e., a context), returns the set of
well-scoped terms in context $\Gamma$.

We first explain, in \cref{subsec:assumptions_set_set}, why the initiality
theorem can be applied; and explain, in \cref{subsec:sig_set_set}, how
untyped higher-order languages such as the lambda calculus are encompassed
by $\omega$-cocontinuous signatures with strength.
We then justify, in \cref{subsec:model_set_set}, that our notion of model
does indeed axiomatise the substitution structure of untyped higher-order
languages; and detail, in \cref{subsec:rec_set_set}, how initiality
provide us with substitution-safe recursion principle.

\subsection{The instance $[\Set,\Set]$}
\label{subsec:assumptions_set_set}

To instantiate the \hyperref[thm:initiality-theorem]{initiality theorem} (\cref{thm:initiality-theorem})
with the base category $[\Set,\Set]$, the category must must have a monoidal
structure $\Cmon$, have an initial object, binary coproducts, and
$\omega$-colimits and precomposition by $\_ \otimes Z$ for $Z : [\Set,\Set]$
must preserve them.
As a category of endofunctors, $[\Set,\Set]$ is a strict\footnote{
  A strict monoidal category is a monoidal category such that
  $\alpha,\lambda,\rho$ are the identity natural transformation $\Id$.
}
monoidal category for the composition of endofunctors $\_\circ\_$ as the
monoidal product, and the identity functor $\Id$ as the unit $I$.
As a functor category, limits and colimits in $[\Set,\Set]$ are inherited
from $\Set$ and computed pointwise.
As $\Set$ is complete and cocomplete, so is $[\Set,\Set]$.
Thus, as particular colimits, $[\Set,\Set]$ admits an initial object,
coproducts and $\omega$-colimits.
Moreover, as (co)limits are computed pointwise, they are preserved by
precomposition $\_ \circ Z$ for $Z : [\Set,\Set]$.
Thus, the hypotheses of the initiality theorem are verified and there is
always an initial model for the large class of $\omega$-cocontinuous
signatures with strength.

\subsection{Signatures on $[\Set,\Set]$}
\label{subsec:sig_set_set}

As a direct consequence, it suffices to give an $\omega$-cocontinuous
signature with strength $(H,\theta)$ representing a language to obtain an
initial model for it.
Hence, we need to justify that $\omega$-cocontinuous signatures with
strength are expressive enough to specify untyped higher-order languages.
Thanks to \cref{prop:sigstrength_omega_cocomplete}, using coproducts, we can
build such signatures \emph{modularly}, by glueing together ``smaller''
independent signatures; each independent constructor gives rise to a summand
in the signature of the language.
Hence our expressive power depends on the constructors we can express.
We illustrate this, by considering algebraic signatures, presentable
signatures, and non-linear signatures.

\subsubsection{Algebraic Signatures}

The first class of signatures we consider are \emph{algebraic} signatures,
also known as \emph{binding} signatures; they are the simplest signatures
enabling us to specify untyped languages with variable binding such as the
untyped lambda calculus:

\begin{definition}[Algebraic Signatures]
  \label{def:untyped-alg-sig}
  An untyped algebraic signature is given by a set $I : \Set$, with an arity
  function $\mrm{ar} : I \to \List(\N)$.
\end{definition}

\noindent For all $i : I$, $\mrm{ar}(i) : \List(\N)$ specify the arity of an
independent constructor.
The length of $\mrm{ar}(i)$ specifies the number of independent input of the
constructor, and the content of the list specifies the number of variables
bound in the corresponding input.

By closure under coproducts, it suffices to be able to represent
single constructors to represent algebraic signatures.
Moreover, by \cref{ex:presheaves-limits,prop:omega-limits,prop:sigstrength_omega_complete},
$\omega$-cocontinuous signatures with strength are closed under finite
products.
Hence, as the inputs are independent and in finite numbers, it suffices to
be able to represent zero-ary constructors, i.e constants, and  unary
constructors, i.e. the binding of $n$ variables, to be able to represent
algebraic signatures.
Constants are represented by the terminal signature $\Theta^0$.
A unary constructor that binds no variable is represented by the trivial
signature $\Theta$.
Intuitively, a constructor binding a variable should take as input a term
defined in a context with one fresh variable, i.e. defined in context
$\Gamma + 1$, and return a term where it is has been bound, i.e. a term in
context $\Gamma$.
For instance, the abstraction of the lambda calculus should be of the
type $\abs : \Lambda(\Gamma + 1) \to \Lambda(\Gamma)$.
This intuition directly translates to a signature with strength representing
variable binding:

\begin{proposition}[Variable binding]
  In the category $[\Set,\Set]$, there is an $\omega$-cocontinuous signature
  with strength $(\delta,\theta)$ denoted $\Theta^{(1)}$ representing
  variable binding:
  \begin{align*}
    \begin{array}{lcl}
      \delta &:& [\Set,\Set] \to [\Set,\Set] \\
      \delta\; X\; n &=& X(n + 1)
    \end{array}
    &&
    \begin{array}{lcl}
      \theta_{X,Y}\; n &:& X(Y(n)+1) \to X(Y(n+1)) \\
      \theta_{X,Y}\; n &=& X(Y(\inl_{n+1}) + y_{n+1} \circ \inr_{n+1} )
    \end{array}
  \end{align*}
  where by the definition of a strength, $X,Y : [\Set,\Set]$ are two
  functors, and $y : \Id \to Y$ is a natural transformation.
\end{proposition}

\noindent Then, as by \cref{prop:sigstrength_omega_comp}, $\omega$-cocontinuous
signatures with strength are closed under composition, the binding of $n$
variables is represented by $\Theta^{(n)}$ defined as $\Theta^{(1)} \circ
... \circ \Theta^{(1)} \circ \Theta$.
It follows that we can represent algebraic signatures as:

\begin{example}[Algebraic Signatures]
  An algebraic signature $(I,\mrm{ar})$ is represented by the following
  $\omega$-cocontinuous signature with strength, where $\mrm{ar}(i) :=
  [n_1,...,n_{m_i}]$:
  \[ \bigplus_{\substack{i:I }} \; \Theta^{(n_1)} \times ... \times \Theta^{(n_{m_i})} \]
\end{example}

As algebraic signatures can be represented by $\omega$-cocontinuous
signatures with strength, by the initiality theorem, algebraic signatures
are representable.
As instances of algebraic signatures, we can conclude that many untyped
higher-order languages are representable:

\begin{example}[Untyped Lambda calculus]
  \label{ex:ULC}
  The untyped lambda calculus can be represented by the $\omega$-cocontinuous
  signature with strength:
  \[ \Sigma_\Lambda := \Theta \times \Theta + \Theta^{(1)} \]
\end{example}

\begin{example}[First order logic]
  \label{ex:first-order-logic}
  First order logic composed of $\top, \bot, \neg, \lor, \wedge,
  \Rightarrow, \forall, \exists$ can be represented by an
  $\omega$-cocontinuous signature with strength:
  \[ \Sigma_{\mrm{FOL}} := 2\Theta^0 + 1\Theta^1 + 3\Theta^2 + 2\Theta^{(1)} \]
\end{example}

\begin{example}[Linear Logic]
  \label{ex:linear-logic}
  The syntax of linear logic is composed of four constants $\top,\bot,0,1$,
  two unary constructors $!,?$, five binary constructors
  $\&,\invamp,\otimes,\oplus,\multimap$, and two unary constructors
  $\exists,\forall$ binding one variable.
  Hence, it can be represented by:
  \[ \Sigma_{\mrm{FOL}} := 4\Theta^0 + 2\Theta^1 + 5\Theta^2 + 2\Theta^{(1)} \]
\end{example}

\subsubsection{Presentable Signatures}

In algebraic signatures, the different inputs of a constructor are independent.
As such, they do not enable us to define constructors with semantic properties
between the inputs, such as a commutative constructor $\mrm{cons}$ that satisfies
$\mrm{cons}(a,b) = \mrm{cons}(b,a)$.
To represent such constructors, on $[\Set,\Set]$, in \cite{PresentableSignatures21}, the
authors introduced presentable signatures :
\begin{definition}[Presentable signatures]
  \label{def:pres-sig}
  A signature $\Sigma$ is presentable iff there exists an algebraic signature
  $\Upsilon$ with an epimorphism $\Upsilon \to \Sigma$.
\end{definition}
\noindent They then prove that all presentable signatures are representable,
and detail different examples of presentable signatures, such as lists
invariant under repetitions or an $n$-ary coherent explicit substitution.
While this result might be generalisable to monoidal categories, in practice,
most examples are already covered by $\omega$-cocontinuous signatures with
strength; particularly by colimits of algebraic signatures, which are still
representable by \cref{prop:sigstrength_omega_cocomplete}.
For instance:

\begin{example}[Commutative Binary Operator]
  There is an $\omega$-cocontinuous signature with strength representing
  commutative binary operators, defined as the coequalizer of:
  \[
    \begin{tikzcd}[column sep=large]
      \Theta \times \Theta \ar[r, shift left=.75ex, "\mrm{swap}"]
                           \ar[r, shift right=.75ex, swap,"\Id"]
        & \Theta \times \Theta
    \end{tikzcd}
  \]
\end{example}

\noindent However, note that even though presentable signatures are powerful,
they can only enforce equations between the inputs of a constructor.
For instance, they can not be used to define associative constructors satisfying
$\mrm{cons}(a,\mrm{cons}(b,c)) = \mrm{cons}(\mrm{cons}(a,b),c)$.
Indeed, $\mrm{cons}$ needs to exist to appear in the input, which is impossible
since we are currently defining it.
To add semantics between constructors, one needs to depart from pure
signatures, for instance by considering 2-signatures as done in \cite{2Signatures19}.

\subsubsection{Non Linear Signatures}

The previously considered signatures are already very powerful but
still do not fully exploit the the power of computing a higher-order fixed
point.
Indeed, our signatures are functors $\Sigma : [\Set,\Set] \to [\Set,\Set]$,
but given $X : \Set \to \Set$, we have not exploited that $X$ can be
composed with itself.
We call signatures involving a self-application ``\emph{non-linear}''.
Non linear signatures are not particularly common.
However, one example of a non-linear signature is \emph{explicit
flattening} (see, e.g., \cite{Hss04}):
\begin{proposition}[Explicit flattening]
  There is a signature with strength on $[\Set,\Set]$, $(\trm{ExF},\theta)$
  representing explicit flattening defined as:
  \begin{align*}
    \begin{array}{lcl}
      \mrm{ExF} &:& [\Set,\Set] \to [\Set,\Set] \\
      \mrm{ExF}\; X &=& X \circ X
    \end{array}
    &&
    \begin{array}{lcl}
      \theta_{X,Y} &:& X \circ X \circ Y \to X \circ Y \circ X \circ Y \\
      \theta_{X,Y} &=& X \circ y \circ X \circ Y
    \end{array}
  \end{align*}
\end{proposition}
\noindent Note, however, that this signature is not $\omega$-cocontinuous on
$[\Set,\Set]$.
Indeed, it additionally requires for the input functors of the colimits to
be $\omega$-cocontinuous. Hence, to add explicit substitution, one needs to
restrict to the subcategory of $\omega$-cocontinuous functors of
$[\Set,\Set]$.

\subsection{Models on $[\Set,\Set]$}
\label{subsec:model_set_set}

We have shown that $\omega$-cocontinuous signatures with strength and the
initiality theorem encompass a wide class of untyped higher-order languages,
e.g., the untyped lambda calculus.
It remains to justify that our notion of model appropriately axiomatises
the syntactic structure of higher-order languages.
To justify this claim properly, we unfold the structure of a model for a
general signature $\Sigma : \Sig(\mc{C})$, using the untyped lambda calculus
as a running example.

\subsubsection{Language and Constructors}
A model of a signature $\Sigma : \Sig(\mc{C})$ is a monoid $(R,\eta,\mu)$
and a morphism of modules $r : \Sigma(R) \to R$.
Let's forget temporarily the substitution structure --- the monoid
multiplication $\mu$ and the module structure ---, and let's consider the
underlying structure on $[\Set,\Set]$.
A signature $\Sigma$ is represented by a functor $R : \Set \to \Set$
assigning well-scoped terms to contexts, and by two natural transformations,
$\eta : \Id \to R$ representing variables, and $r : \Sigma(R) \to R$
representing the language-specific terms constructors.

Consider the untyped lambda calculus, defined by $\Sigma_\Lambda := \Theta^2
+ \Theta^{(1)}$, as an example.
It is modelled by a functor $\Lambda : \Set \to \Set$ associating, to any
context $\Gamma : \Set$, the set of well-scoped lambda terms in context
$\Gamma$.
The natural transformation $\eta$ of the monoid corresponds to the variable
constructor $\var_\Gamma : \Gamma \to \Lambda(\Gamma)$, associating to any
label a term.
Meanwhile, the natural transformation $r : \Lambda(\Gamma + 1) + \Lambda(\Gamma)
\times \Lambda(\Gamma) \to \Lambda(\Gamma)$ splits into the two natural
transformations $\app_\Gamma : \Lambda(\Gamma) \times \Lambda(\Gamma) \to
\Lambda(\Gamma)$ and $\abs_\Gamma : \Lambda(\Gamma + 1) \to
\Lambda(\Gamma)$, representing constructors.

\subsubsection{Renaming}

As $\Lambda$ is a functor and $\eta, \app ,\abs$ are natural transformations,
they also act on morphisms.
Given a function between contexts $c : \Gamma \to \Delta$ renaming
labels, $\Lambda(c) : \Lambda(\Gamma) \to \Lambda(\Delta)$ is the
associated renaming of lambda terms.
The functorial laws are then ensuring that renaming is well behaved:
renaming by the identity is the identity $\Lambda(\Id) = \Id$, and renaming
twice is renaming for the composition $\Lambda(c') \circ \Lambda(c) =
\Lambda(c' \circ c)$.
Then, the naturality of constructors asserts exactly how renaming computes
on --- commutes with --- constructors:
\begin{align*}
  \begin{tikzcd}[ampersand replacement=\&]
    \Gamma \ar[r, "\var_\Gamma"] \ar[d, swap, "c"]
      \& \Lambda (\Gamma) \ar[d, "\Lambda(c)"]\\
    \Delta \ar[r, swap, "\var_\Delta"]
      \& \Lambda (\Delta)
  \end{tikzcd}
  &&
  \begin{tikzcd}[ampersand replacement=\&]
    \Lambda(\Gamma) \times \Lambda(\Gamma) \ar[r, "\app_\Gamma"]
      \ar[d, swap, "\Lambda(c) \times \Lambda(c)"]
      \& \Lambda(\Gamma) \ar[d, "\Lambda(c)"] \\
    \Lambda(\Delta) \times \Lambda(\Delta) \ar[r, swap, "\app_\Delta"]
      \& \Lambda(\Delta)
  \end{tikzcd}
  &&
  \begin{tikzcd}[ampersand replacement=\&]
    \Lambda(\Gamma + 1) \ar[r, "\abs_\Gamma"] \ar[d, swap, "\Lambda(c + 1)"]
      \& \Lambda(\Gamma) \ar[d, "\Lambda(c)"] \\
    \Lambda(\Delta + 1) \ar[r, swap, "\abs_\Delta"]
      \& \Lambda(\Delta)
  \end{tikzcd}
\end{align*}
Renaming of $\var$ is renaming of the label, renaming of $\app$ is renaming
of the inputs, and renaming for $\abs$ is renaming of the input but without
modifying the locally nameless variable, the one that is bound by $\abs$.
The same principles apply to any model of any signature.

\subsubsection{Flattening}

Coming back to the substitution structure, the untyped lambda calculus ---
as any language --- comes equipped with a natural transformation $\mu_\Gamma
: \Lambda (\Lambda(\Gamma)) \to \Lambda(\Gamma)$ that corresponds to
flattening.
Roughly, given a term that has lambda terms as label, i.e. an object of
$\Lambda (\Lambda (\Gamma))$, $\mu$ removes a layer of variables to yield a lambda
term, i.e. an object of $\Lambda(\Gamma)$.
As before, the naturality asserts that $\mu$ commutes with renaming.
The rest of the monoid structure, i.e., the monoid laws, specify the
behaviour of $\mu$ regarding associativity and variables:
\begin{align*}
  \begin{tikzcd}[ampersand replacement=\&]
    \Lambda (\Lambda (\Lambda (\Gamma))) \ar[r, "\Lambda (\mu_\Gamma)"]
                                    \ar[d, swap, "\mu_{\Lambda(\Gamma)}"]
      \& \Lambda (\Lambda (\Gamma)) \ar[d, "\mu_\Gamma"] \\
    \Lambda (\Lambda (\Gamma)) \ar[r, swap, "\mu_\Gamma"]
      \& \Lambda (\Gamma)
  \end{tikzcd}
  &&
  \begin{tikzcd}[ampersand replacement=\&]
    \Lambda (\Gamma) \ar[r, "\var_{\Lambda (\Gamma)}"] \ar[dr, swap, "\Id"]
      \& \Lambda (\Lambda (\Gamma)) \ar[d, "\mu_{\Lambda(\Gamma)}"] \\
    \& \Lambda (\Gamma)
  \end{tikzcd}
  &&
  \begin{tikzcd}[ampersand replacement=\&]
    \Lambda (\Lambda (\Gamma)) \ar[d, swap, "\mu_{\Lambda(\Gamma)}"]
      \& \Lambda (\Gamma) \ar[l, swap, "\Lambda (\var_\Gamma)"] \ar[dl, "\Id"] \\
    \Lambda (\Gamma)
      \&
  \end{tikzcd}
\end{align*}
The first law asserts that the order of flattening in the structure does not
matter.
The second law asserts that flattening computes on variables
by returning the input, $\mu (\var(t)) = t$.
The third law asserts that replacing all labels $x : \Gamma$ in a term $t :
\Lambda(\Gamma)$ by $\var(x)$, then flattening it yields the original term $t$.

The monoid structure in itself does not fully specify how $\mu$ computes, as
it does not specify how $\mu$ commutes with $\app$ and $\abs$, the language-specific
constructions.
This is what is specified by the module structure, as module morphisms
commute with their associated module substitutions.
In the case of signatures with strength, this provides us with a recursive
algorithm, since module substitutions are built out of $\mu$.
For instance, for $\app$ and $\abs$, $\mu$ computes as:
\begin{align*}
  \begin{tikzcd}[ampersand replacement=\&]
    \Lambda(\Lambda(\Gamma)) \times \Lambda(\Lambda(\Gamma))
          \ar[r, "\app_{\Lambda(\Gamma)}"]
          \ar[d, swap, "\theta^\app := \Id"]
      \& \Lambda(\Lambda(\Gamma)) \ar[dd, "\mu_\Gamma"] \\
    \Lambda(\Lambda(\Gamma)) \times \Lambda(\Lambda(\Gamma))
        \ar[d, swap, "\mu_\Gamma \times \mu_\Gamma"]
      \& \\
    \Lambda(\Gamma) \times \Lambda(\Gamma) \ar[r, swap, "\app_\Gamma"]
      \& \Lambda(\Gamma)
  \end{tikzcd}
  &&
  \begin{tikzcd}[ampersand replacement=\&]
    \Lambda(\Lambda(\Gamma) + 1) \ar[r, "\abs_{\Lambda(\Gamma)}"]
                            \ar[d, swap, "\theta^\abs"]
      \& \Lambda(\Lambda(\Gamma)) \ar[dd, "\mu_\Gamma"] \\
    \Lambda(\Lambda(\Gamma + 1)) \ar[d, swap, "\mu_{\Gamma+1}"]
      \& \\
    \Lambda(\Gamma + 1) \ar[r, swap, "\abs_\Gamma"]
      \& \Lambda(\Gamma)
  \end{tikzcd}
\end{align*}
Flattening for $\app$ is flattening of the inputs.
By definition of $\theta^{\mrm{abs}}$, flattening for $\abs(t)$ with $t :
\Lambda(\Lambda(\Gamma) +1)$, consists in replacing all labels $u :
\Lambda(\Gamma)$ by their weakening in $\Lambda(\Gamma + 1)$, and the fresh
variable $*$ by $\var{(*)}$, and then by flattening.
Hence, it flattens the lambda term, but such that the locally nameless
variable that is bound is preserved.
This, combined with the previous laws fully specify how flattening computes
on lambda terms.

\subsubsection{Simultaneous Substitution}

We have entirely unfolded the definition of model, yet we have not justified
that models actually model simultaneous substitution.
This is because so far, we have been relying on the general definition of
monoids, while substitution is encompassed by an equivalent definition of
monoid in a category of endofunctors, namely Kleisli Triples.
\begin{definition}[Kleisli Triples]
  \label{def:kleisli_triples}
  A \emph{Kleisli triple} on $\mc{C} \to \mc{C}$ is given by
  \begin{itemize}[label=$-$]
    \setlength\itemsep{-1pt}
    \item A function $R : \mrm{ob}(\mc{C}) \to \mrm{ob}(\mc{C})$ on objects
          of $\mc{C}$.
    \item For all $\Gamma : \mc{C}$, a morphism $\eta_{\,\Gamma} : \Gamma \to R(\Gamma)$
    \item For all $\Gamma,\Delta : \mc{C}$, a morphism
    $\sigma_{\, \Gamma,\Delta} : \mc{C}(\Gamma,\, R(\Delta)) \to \mc{C}(R(\Gamma),\, R(\Delta))$
  \end{itemize}
  satisfying the equations:
  \begin{align*}
    \begin{tikzcd}[ampersand replacement=\&]
      \Gamma \ar[r, "\eta_{\,\Gamma}"] \ar[dr, swap, "c"]
        \& R(\Gamma) \ar[d, "\sigma_{\,\Gamma,\Delta}(c)"] \\
        \& R(\Delta)
    \end{tikzcd}
    &&
    \begin{tikzcd}[ampersand replacement=\&]
      R(\Gamma) \ar[dr, bend left, "\sigma_{\,\Gamma,\Gamma}(\eta_{\,\Gamma})"]
                \ar[dr, swap, bend right, "\Id"]
        \& \\
        \& R(\Gamma)
    \end{tikzcd}
    &&
    \begin{tikzcd}[ampersand replacement=\&]
      R(\Gamma) \ar[r, "\sigma_{\,\Gamma,\Delta}(c)"]
                \ar[dr, swap, "\sigma_{\,\Gamma,\Upsilon}(\sigma_{\Delta,\Upsilon}(c') \circ c)"]
        \& R(\Delta) \ar[d, "\sigma_{\Delta,\Upsilon}(c')"] \\
        \& R(\Upsilon)
    \end{tikzcd}
  \end{align*}
\end{definition}

\noindent As in the previous section, $R$ represents the language and $\eta$
the variable constructors.
However, here $\sigma$ corresponds to simultaneous substitution.
For instance, for the lambda calculus, $\sigma$ is of type $(\Gamma \to
\Lambda(\Delta)) \to (\Lambda(\Gamma) \to \Lambda(\Delta))$, and associates
to any renaming of labels by terms the associated substitution of terms.
The rest of the structure specifies the behaviour of simultaneous substitution.
The first equation specifies how $\sigma$ computes on variables, the second
one that substituting variables by variables is the identity, and the third
one asserts how sequential simultaneous substitutions compose.

Monads --- monoids in a category of endofunctors --- and Kleisli triples are
equivalent, so our model based on flattening enables us to define
simultaneous substitution.
This is done by first renaming the labels by terms, and then by removing
the extra layer of variables introduced by renaming using flattening.
\[
  \begin{tikzcd}
    \Lambda(\Gamma) \ar[r, "\Lambda(c)"]
      & \Lambda(\Lambda(\Delta)) \ar[r, "\mu_\Delta"]
      & \Lambda(\Delta)
  \end{tikzcd}
\]
This can be done without capture of variables as both renaming and
flattening are well defined and avoid capture of variables.
The different equations of the Kleisli triple's definition can then be
derived using the ones satisfied by $\mu$ as a monoid multiplication.

To correctly axiomatise simultaneous substitution, we also need to specify
its behaviour on constructors such as $\app$ and $\abs$.
As simultaneous substitution is defined in terms of flattening and renaming,
it computes as renaming, followed by the module substitution.
For instance, for a signature with strength $(H,\theta)$, we get:
\[
  \begin{tikzcd}[column sep=huge]
    H(T)(\Gamma)         \ar[r, "r_{\vspace{1pt}\Gamma}"] \ar[d, swap, "H(T)(c)"]
                         \ar[ddd, swap, shift right=27pt, shorten=-5pt,
                             bend right=42, "\sigma(?)", red]
      & T(\Gamma)        \ar[d, "T(c)"]
                         \ar[ddd, shift left=15pt, shorten=-5pt, bend left=40, "\sigma_{\,\Gamma,\Delta}(c)"] \\
    H(T)(T(\Delta))      \ar[r, dashed, "r_{T(\Delta)}"]
                         \ar[d, swap, "\theta"]
      & T(T(\Delta))     \ar[dd, "\mu_\Delta"]\\
    H(T \circ T)(\Delta) \ar[d, swap, "H(\mu_\Delta)"]
      & \\
    H(T)(\Delta)         \ar[r, swap, "r_\Delta"]
      & T(\Delta)
  \end{tikzcd}
\]
However, unlike for renaming and flattening, this does not provide us with a
recursive algorithm to compute simultaneous substitution, as there is
reason why the left-hand side would factor as a simultaneous substitution.
This is an additional reason why our notions are defined in terms of
flattening.
Yet for some signatures, e.g. algebraic signatures, the left-hand side can
be factorised, and $\sigma$ can be defined recursively.
For instance, for the lambda calculus, simultaneous substitution computes
recursively on $\app, \abs$ as:
\begin{align*}
  \begin{tikzcd}[ampersand replacement=\&]
    \Lambda(\Gamma) \times \Lambda(\Gamma) \ar[r, "\app_\Gamma"]
                                           \ar[d, swap, "\sigma_{\,\Gamma,\Delta}(c) \times \sigma_{\,\Gamma,\Delta}(c)"]
      \& \Lambda(\Gamma)                    \ar[d, "\sigma_{\,\Gamma,\Delta}(c)"] \\
    \Lambda(\Delta) \times \Lambda(\Delta) \ar[r, swap, "\app_\Delta"]
      \& \Lambda(\Delta)
  \end{tikzcd}
  &&
  \begin{tikzcd}[ampersand replacement=\&]
    \Lambda(\Gamma + 1) \ar[r, "\abs_\Gamma"]
                        \ar[d, swap, "\sigma(\ol{c})"]
      \& \Lambda(\Gamma) \ar[d, "\sigma(c)"] \\
    \Lambda(\Delta+1)   \ar[r, swap, "\abs_\Delta"]
      \& \Lambda(\Delta)
  \end{tikzcd}
\end{align*}
where
$\ol{c} = [\Lambda(\mrm{inl}_{\Delta,1}) \circ c, \eta_{\Delta+1} \circ
  \mrm{inr}_{\Delta,1}] : X+1 \to  \Lambda (\Delta + 1)$.
We recognise here the usual rules of untyped lambda calculus.
Substitution for $\app$ is substitution in the input, and substitution for
$\abs$ is substitution for the input while preserving the fresh variable.

\subsection{The Recursion Principle on $[\Set,\Set]$}
\label{subsec:rec_set_set}

We have justified that models abstract untyped higher-order languages with
their substitution structure, but such an abstractions would be of little
use without a recursion principle.
Such a recursion principle is provided by the initiality of the model, as
provided by the initiality theorem.
We unfold and exemplify this concept, and explain how in our case, the
recursion principle obtained is strengthened / restricted to respect the
substitution structure.

\subsubsection{Initiality and Recursion}

Given a signature $\Sigma : \Sig([\Set,Set])$, a
\hyperref[def:morphims-models]{morphism of models} $f : M_1 \to M_2$ is, by
\cref{prop:morph-models-alg}, a morphism of $[\Set,Set]$, i.e. a natural
transformation, that is also a morphism of monoids and a morphism of
algebras.
As a morphism of algebras, $f$ respects the language-specific constructors
represented by $r$, and as a morphism of monoids, it also respects the
variable constructor represented by $\eta$.
Hence, a morphism of models $f$ respects all of the language constructors.
For instance, for the lambda calculus, one gets:
\begin{align*}
  \begin{tikzcd}[ampersand replacement=\&]
    \Gamma \ar[r, "\var_1"] \ar[dr, swap, "\var_2"]
      \& \Lambda_1(\Gamma) \ar[d, "f"] \\
      \& \Lambda_2(\Gamma)
  \end{tikzcd}
  &&
  \begin{tikzcd}[ampersand replacement=\&]
    \Lambda_1(\Gamma) \times \Lambda_1(\Gamma) \ar[r, "\app_1"]
              \ar[d, swap, "f \times f"]
      \& \Lambda_1(\Gamma) \ar[d, "f"] \\
    \Lambda_2(\Gamma) \times \Lambda_2(\Gamma) \ar[r, swap, "\app_2"]
      \& \Lambda_2(\Gamma)
  \end{tikzcd}
  &&
  \begin{tikzcd}[ampersand replacement=\&]
    \Lambda_1(\Gamma + 1) \ar[r, "\abs_1"]
              \ar[d, swap, "f"]
      \& \Lambda_1(\Gamma) \ar[d, "f"] \\
    \Lambda_2(\Gamma+1) \ar[r, swap, "\abs_2"]
      \& \Lambda_2(\Gamma)
  \end{tikzcd}
\end{align*}
Moreover, $f$ respects renaming as it is natural transformation, and respects
flattening as it is a monoid morphism.
Hence, as $\sigma_{\,\Gamma,\Delta} := \mu_\Delta \circ \Lambda(c)$, $f$
respects simultaneous substitution:
\[
  \begin{tikzcd}
    \Lambda_1(\Gamma) \ar[r, "\Lambda_1(c)"] \ar[d, swap, "f"]
                    \ar[rr, shift left=8pt, shorten=-5pt, bend left=25, "\sigma_1(c)"]
      & \Lambda_1(\Lambda_1(\Delta)) \ar[r, "\mu_1"] \ar[d, dashed, "f \circ f"]
      & \Lambda_2(\Delta ) \ar[d, "f"]\\
    \Lambda_2(\Gamma) \ar[r, swap, "\Lambda_2(f \circ c)"]
              \ar[rr, swap, shift right=8pt, shorten=-5pt, bend right=25, "\sigma_2(f \circ c)"]
      & \Lambda_2(\Lambda_2(\Delta)) \ar[r, swap, "\mu_2"]
      & \Lambda_2(\Delta)
  \end{tikzcd}
\]
Thus, a model morphism has the same structure as a map built by the
recursion principle but strengthened to respect simultaneous substitution.

Initiality provides us with an automatic procedure to build such morphisms.
Indeed, by definition, if a model $\ol{\Sigma} : \Model(\Sigma)$ is initial,
than for any other model $M : \Model(\Sigma)$, there is a unique morphism of
models $\ol{\Sigma} \to M$.
So it suffices to build models of $\Sigma$ to build model morphisms.
By unfolding the definition of a model, one can then see that we get exactly
the usual recursion principle but strengthened to respect substitution by
design, as one needs to provide a functor $M$ with constructors, but also a
substitution structure $\mu$ and prove that the constructors respect it.
As such, one also gets a more restricted recursion principle as it only
applies to functors with a substitution structure and constructors
respecting it.

\subsubsection{Example: Translating languages}

One of the most interesting application of the substitution-safe recursion
principle is using it to translate languages.
Indeed, such a translation would respect substitution, which is of particular
importance when substitution plays a part in semantics.
As an example of translation between languages, we reproduce here a
substitution-safe translation, specified in \cite[Section 9.1]{PresentableSignatures21},
from \hyperref[ex:first-order-logic]{first-order intuitionistic logic} into
\hyperref[ex:linear-logic]{linear logic}.
By the initiality theorem, first-order logic and linear logic have an
initial model, hence, by initiality, it suffices to provide a model of
first-order logic on top of the monoid of linear logic to get a translation
respecting substitution.
Consequently, we have to build a module morphism $\Sigma_{\mrm{FOL}}(\Theta)
\to \Theta$ over the monoid of linear logic.
In other words, we need to build the morphism of modules corresponding to
first-order logic, denoted by $r'$, using the morphism of modules
corresponding to linear logic, denoted by $r$.
\begin{align*}
  r'_\top,r'_\bot & : \Theta^0 \to \Theta
    & r_\top,r_\bot & : \Theta^0 \to \Theta \\
  r'_\neg &: \Theta \to \Theta
    & r_!,r_? &: \Theta \to \Theta \\
  r'_\land, r'_\lor, r'_\Rightarrow &: \Theta \times \Theta \to \Theta
    & r_\&,r_\invamp,r_\otimes,r_\oplus,r_\multimap &: \Theta \times \Theta \to \Theta \\
  r'_\exists, r'_\forall &: \Theta' \to \Theta
    & r_\exists,r_\forall &: \Theta' \to \Theta
\end{align*}

\begin{example}[First-order logic to linear logic]
  \label{ex:first-order-to-linear}
  There is a substitution-safe translation $L_\mrm{FOL} \to L_\mrm{LL}$ from
  first-order logic to linear logic given by:
  \begin{align*}
    r'_\top &= r_\top
      & \top^0 &:= \top \\
    r'_\bot &= r_\bot
      & \bot^0 &:= \bot \\
    r'_\neg  & = r_\multimap \circ (r_! \times r_0)
      & (\neg A)^o &:= (!A^o) \multimap 0 \\
    r'_\land & = r_\&
      & (A \land B)^o &:= A^o \,\&\, B^o \\
    r'_\lor  & = r_\oplus \circ( r_! \times r_!)
      & (A \lor B)^o &:=\; !A^o \,\oplus\, !B^o \\
    r'_\Rightarrow & = r_\multimap \circ (r_! \times id)
      & (A \Rightarrow B)^o &:=\; !A^o \multimap B^o \\
    r'_\exists &  =  r_\exists\circ r_!
      & (\exists x.\, A)^o &:= \exists x.\, ! A^o \\
    r'_\forall &  =  r_\forall
      & (\forall x.\, A)^o &:= \forall x.\,  A^o
  \end{align*}
\end{example}

\subsubsection{Example: Computing the set of free variables}

The recursion principle can be used to build substitution-safe translations
of languages, but it can also be applied to models that are not arising from
languages.
For instance, we can compute the free variables of a lambda term by
building an appropriate model \cite[Example 9.2]{PresentableSignatures21}:

\begin{example}[The model of free variables]
  \label{ex:free-var}
  The covariant powerset functor  $\mc{P} : \Set \to \Set$ is a monoid for
  the natural transformations $\eta',\mu'$:
  \begin{align*}
    \begin{array}{lcccc}
      \eta'_\Gamma &:& X &\longrightarrow& \mc{P}(\Gamma) \\
      \eta'_\Gamma &:& x &\longmapsto& \{ x \}
    \end{array}
    &&
    \begin{array}{lcccc}
      \mu'_\Gamma &:& \mc{P}(\mc{P}(\Gamma)) &\longrightarrow& \mc{P}(\Gamma) \\
      \mu'_\Gamma &:& U &\longmapsto& \bigcup\limits_{u : U} u
    \end{array}
  \end{align*}
  It forms a model of the untyped lambda calculus, when completed with the
  module morphisms $\app'_X = (U,V) \longmapsto U \cup V$ and $\abs'_X = U
  \longmapsto U \cap X$.
\end{example}

\noindent Then, by initiality there is a unique morphism of models
$\mrm{free} : \ol{\Lambda} \to \mc{P}$ that computes the free variables in
lambda terms.
Compared to the usual definition, here, we additionally know how to compute
the free variables on substitution:
\[
  \trm{free}(u[\star := v ])
    \;=\; \bigcup_{y \in \mc{P}({v})(\trm{free}(u))} y
    \;=\; \left\{ \begin{array}{ll}
                \trm{free}(u) \setminus \{\star\}
                  &\trm{ if } \star \not\in u \\
                \trm{free}(u) \setminus \{\star\} \cup \trm{free}(v)
                  &\trm{ if } \star \in u
              \end{array}
      \right.
\]
For instance, as $\mrm{free}((\lambda \star.u)v) = \trm{free}(u) \setminus
\{\star\} \cup \mrm{free}(v)$, it directly enables us to see that free
variables are not preserved under beta reduction.

%% file: Sections/example_F_Set.tex
\section{Untyped Languages: The Instance $[\mb{F},\Set]$}
\label{sec:example_F_set}

We have explained in detail how our models work on the instance $[\Set,\Set]$.
However, that is not the only possibility to model untyped languages with variable
binding.
Another important example in the literature is the category $[\mb{F},\Set]$
\cite{FPT99}, where $\mb{F}$ is the category with the natural numbers as
objects and functions $\llbracket n \rrbracket \to \llbracket m
\rrbracket$ as morphisms $n \to m$.%
\footnote{In other words, $\mb{F}$ is the
skeleton of the category of finite sets.}
Using $\mb{F}$ to model contexts rather than $\Set$ has some pro and cons.
Compared to $\Set$, $\mb{F}$ has the advantage to contain the minimum data
required to model untyped contexts: $\mb{F}$ is generated by a terminal
object and binary coproducts, and as such does not represent infinite contexts.
A useful consequence is that $\mb{F}$ is a small category whose objects each have decidable
equality, which is important when working constructively.
However, compared to $[\Set,\Set]$, building a monoidal structure for
$[\mb{F},\Set]$ --- and the obtained one --- is more intricate as we can not
simply compose the functors, as we did for $[\Set,\Set]$.
The framework is entirely parametric in the monoidal category by which it is parameterized;
instantiating with a more complicated monoidal category thus impacts all of the structure.
As the framework has been explained in great details in \cref{sec:example_set_set},
we follow here the same outline, but we only explain what changes with the
monoidal category.

\subsection{The instance $[\mb{F},\Set]$}
\label{subsec:assumptions_F_set}

To instantiate the \hyperref[thm:initiality-theorem]{initiality theorem}
with the base category $[\Set,\Set]$, the category must must have a monoidal
structure $\Cmon$, have an initial object, binary coproducts, and
$\omega$-colimits and precomposition by $\_ \otimes Z$ for $Z : [\Set,\Set]$
must preserve them.

Building a monoidal structure on $[\mb{F},\Set]$ is not as trivial as for
$[\Set,\Set]$, as our functors are no longer endofunctors and as such can
not be composed.
This problem is studied in \cite{RelativeMonads15} for the functor category $[\mb{J},\mc{C}]$.
The authors prove that it suffices for a ``well-behaved'' functor $J : \mb{J}
\to \mc{C}$ admitting left Kan extensions to exist, in order for $[\mb{J},\mc{C}]$
to admit a monoidal structure.
In this case, $J$ is taken as the unit, the left Kan extensions are used to
build a monoidal product, and the well-behavedness conditions are used to
prove that this does form a monoidal structure.
In our case, there is an embedding $J : \mb{F} \to \Set$ by $J(m) :=
\br{m}$, which is well-behaved as proved in \cite[Corollary 4.3]{RelativeMonads15}.
Moreover, since $\mb{F}$ is small and $\Set$ is cocomplete, by \cref{prop:LKE_and_coends},
$J$ admits left Kan extensions $\Lan{J}{\_} : [\mb{F},\Set] \to [\Set,\Set]$.
Hence, $[\mb{F},\Set]$ admits a monoidal structure, where $J$ is the unit,
and where the monoidal product is computed as:
\[ F \otimes G := (\Lan{J}{F} \circ G)(k)
                = \int^{n : \mb{F}} F(n) \times G(k)^n \]
As a functor category, $[\mb{F},\Set]$ inherits limits and colimits from
$\Set$; thus, it is complete and complete, and has the appropriate colimits.
Furthermore, as precomposition, and left Kan extensions as left adjoint,
preserve colimits, $\_ \otimes Z$ preserves colimits.

\subsection{Signatures on $[\mb{F},\Set]$}
\label{subsec:sig_F_set}

As for $[\Set,\Set]$, we need to prove that $\omega$-cocontinuous signatures
with strength encompass a wide class of untyped higher-order languages for
the initiality theorem to be useful.
In practice, the same languages can be represented on $[\mb{F},\Set]$ as on
$[\Set,\Set]$.

By \cref{prop:sigstrength_omega_cocomplete}, the signatures under
consideration are closed under products, which enables us to build our
languages \emph{modularly}, constructor by constructor.
Furthermore, it can be shown that for all $Z : \mc{C}, Z \otimes \_$
preserves finite products\footnote{
  This is done using the universal property of coends, similarly to defining
  a strength for variable binding.
  },
hence by \cref{ex:presheaves-limits,prop:omega-limits,prop:sigstrength_omega_complete},
they are also closed under finite products, and there is a terminal signature $\Theta^0$.
As there is always a trivial signature $Theta$, and our signatures are also
closed under composition, it suffices to represent unary variable binding to
represent \hyperref[def:untyped-alg-sig]{algebraic signatures}.

The functor underlying variable binding is the same, since the category $\mb{F}$ models
contexts in the same way the category $\Set$ does.
However, the strength needs to be adapted to the new monoidal structure:

\begin{proposition}[Variable binding]
  In the category $[\mb{F},\Set]$, there is an $\omega$-cocontinuous
  signature with strength denoted $\Theta^{(1)}$ representing variable binding,
  for the underlying functor $\delta : [\mb{F},\Set] \to [\mb{F},\Set]$,
  defined as $\delta\; X\; n := X(n + 1)$.
\end{proposition}
\begin{proof}
  The functor $\delta$ is $\omega$-cocontinuous as colimits are computed
  pointwise in a functor category.
  Given $X,Y : [\mb{F},\Set]$, and $y : J \to Y$, defining a strength for
  $\delta$ amounts to defining, for any $m : \mb{F}$, a function
  $(\delta X \otimes Y)(m) \to (\delta (X \otimes Y))(m)$, which unfolds to:
  \[ \int^{n : \mb{F}} X(n+1) \times Y(m)^n \quad\longrightarrow\quad \int^{n : \mb{F}} X(n) \times Y(m+1)^n .\]
  By the universal property of coends, this is equivalent to defining, for
  any $n : \mb{F}$, a map
  \[ X(n+1) \times Y(m)^n \;\;\longrightarrow\;\; \int^{n : \mb{F}} X(n) \times Y(m+1)^n. \]
  Such a function can be defined as the following composition:
  \begin{align*}
      X(n+1) \times Y(m)^n \;\;
        & \longrightarrow \;\; X(n+1) \times Y(m+1)^n \times 1 \\
        & \longrightarrow \;\; X(n+1) \times Y(m+1)^n \times Y(1) \\
        & \longrightarrow \;\; X(n+1) \times Y(m+1)^{n+1} \\
        & \longrightarrow \;\; \int^{n : \mb{F}} X(n) \times Y(m+1)^n
    \end{align*}
  The remaining properties of a strength are easily provable.
\end{proof}

\noindent Consequently, as before, we can represent algebraic signatures.
For instance, the lambda calculus can simply be defined by $\Theta \times
\Theta + \Theta^{(1)}$.

More generally, $\omega$-cocontinuous signatures with strength over
$[\mb{F},\Set]$ are closed under colimits, so they still encompass
presentable signatures.
By switching to the subcategory of $\omega$-cocontinuous functors of
$[\mb{F},\Set]$, and defining $X \mapsto X \otimes X$, one could also
consider explicit flattening.

\subsection{Models on $[\mb{F},\Set]$}
\label{subsec:model_F_set}

We have seen that a model on $[\Set,\Set]$ amounts to a function associating
terms to contexts, with a variable constructor and languages specific
constructors, equipped with a well-defined renaming and flattening operation
that computes recursively.
Moreover, by combining renaming and flattening, it is possible to
derive a simultaneous substitution that satisfies the Kleisli laws of monads;
additionally, this substitution computes recursively in the case of algebraic signatures.
Models on $[\mb{F},\Set]$ essentially behave like the models in $[\Set,\Set]$, up
to some differences regarding substitution; this is due to the more complicated
monoidal structure.

\subsubsection{Language, Constructors, and Renaming}

While the monoidal structures of $[\mb{F},\Set]$ and
$[\Set,\Set]$ are quite different, both categories are functors categories.
Consequently, models of signatures over either category look quite similar.
A model of a signature $\Sigma$ has, as an underlying
structure, a functor $R : \mb{F} \to \Set$, with two natural transformations
$\eta_{\,\Gamma} : J(\Gamma) \to R(\Gamma)$, and $r_{\,\Gamma} : \Sigma(R)(\Gamma) \to R(\Gamma)$.
As before, $R$ is a functor associating to any --- now finite --- context $\Gamma$, the
set of well-scoped lambda terms in context $\Gamma$; its functorial
action is renaming.
Furthermore, the natural transformation $\eta_{\,\Gamma}$ corresponds to the
variable constructor, $r$ to the language-specific constructors, and the
naturality condition to commutation with renaming.
Note however a slight difference: in $\eta$, $J$ appears for ``homogeneity'',
as otherwise $\eta$ would not be well-typed, as $\Gamma : \mb{F}$ but
$R(\Gamma) : \Set$.

\subsubsection{Flattening}

Consider the lambda calculus.
The flattening operation $\mu : \Lambda \otimes \Lambda \to \Lambda$ is a
natural transformation --- hence compatible with renaming --- and is fully
axiomatised by the monoid laws, and the module structure, as described for
$[\Set,\Set]$.
Nevertheless, compared to the case $[\Set,\Set]$, the name flattening is
debatable as $\Lambda \otimes \Lambda (k) := \int^{l : \mb{F}} \Lambda(l)
\times \Lambda^{l}(k)$ does not obviously corresponds to lambda terms of
lambda terms with $k$ variables.
To give more intuition about it, we unfold and explain the definition of
$\Lambda \otimes \Lambda$.

It is not directly possible to consider lambda terms of lambda terms as
$\Lambda(\Lambda(\Gamma))$ is ill-typed, as $\Lambda(\Gamma) : \Set$ but
$\Lambda : \mb{F} \to \Set$.
Instead, we consider terms of the form $\Lambda(l) \times \Lambda^{l}(k)$
which we see as a ``rough'' explicit substitution.
Indeed, if one were to substitute the $l$ variables by the $l$ terms, one
would get a term in the context with $k$ variables.
Therefore, to define the lambda terms of lambda terms, we need to consider
the elements of $(\Lambda(l) \times \Lambda^{l}(k))_{l:\mb{F}}$ all
together.
The naïve candidate would be the coproduct $\bigsqcup_{l:\mb{F}}\,
\Lambda(l) \times \Lambda^{l}(k)$, but this is not quite right.
Indeed, each element of $\Lambda(l) \times \Lambda^{l}(k)$ yields a distinct
element in the coproduct, even though two elements of $\Lambda(l) \times
\Lambda^{l}(k)$ can actually have the same lambda terms of lambda terms.
For instance, permuting the $l$ variables and taking the associated term of
terms yield the same term as permuting the inputs and taking the
associated term of terms.
Thankfully, accounting for the coherence conditions in $l$ that the
coproduct fails to encompass is exactly what coends enable us to do.
Indeed, the coend is, by definition the smallest object, with injections
$i_{l} : \Lambda(l) \times \Lambda^{l}(k) \to \int^{l : \mb{F}} \Lambda(l) \times \Lambda^{l}(k)$
respecting the actions in $l$, as shown below:
\[
  \begin{tikzcd}
      & \Lambda(n) \times \Lambda^{n}(k)
          \ar[dr, start anchor=south east, end anchor=north west, "i_n"]
          \ar[drr, start anchor=5, end anchor=north west, bend left=20, "x_C"]
      &
      & \\
    \Lambda(n) \times \Lambda^{m}(k)
          \ar[ur, start anchor=north east, end anchor=south west]
          \ar[dr, start anchor=south east, end anchor=north west]
      &
      & \int^{l : \mb{F}} \Lambda(l) \times \Lambda^{l}(k) \ar[r, dashed, "\exists ! h"]
      & X \\
      & \Lambda(m) \times \Lambda^{m}(k)
          \ar[ur, start anchor=north east, end anchor=south west, swap, "i_{m}"]
          \ar[urr, start anchor=-5, end anchor=south west, swap, bend right=20, "x_{C'}"]
      &
      &
  \end{tikzcd}
\]
\noindent Thus, we can consider $\Lambda \otimes \Lambda := \int^{l : \mb{F}}
\Lambda(l) \times \Lambda^{l}(k)$ as lambda terms of lambda terms.
Moreover, by identifying  $\Lambda^l(k)$ with $J(l) \to \Lambda(k)$, we
can consider $i : (J(l) \to \Lambda(k)) \to (\Lambda(l) \to (\Lambda \otimes
\Lambda) (k))$ as renaming from variables to terms.
This point of view generalizes to a functor $F : \mb{F} \to \Set$:
we consider $F \otimes \Lambda$ as $F$-terms of lambda terms, and
$i_F : (J(l) \to \Lambda(k)) \to (F(l) \to (F \otimes
\Lambda) (k))$ as renaming from variables to terms.

Having identified $\Lambda \otimes \Lambda$ with lambda terms of lambda terms,
the rest of the framework is as for $[\Set,\Set]$: the monoid laws specify
how flattening behaves regarding variables, while the module morphisms
specify how flattening behaves on language-specific constructors.

\subsubsection{Simultaneous substitution}
To derive a well-defined simultaneous substitution structure for
$[\Set,\Set]$, we use an equivalence between monoids in endofunctor
categories and Kleisli triples.
As for the definition of $\eta$, this is not directly possible as Kleisli
triples are ill-typed on $[\mb{F},\Set]$, but they are generalisable by
inserting $J$ in the right places for homogeneity, and so is the
equivalence.
In \cite{RelativeMonads15}, Altenkirch, Chapman, and Uustalu introduce ``relative
monads'' --- generalised Kleisli Triples --- and establish that for a
well-behaved functor $J : \mb{J} \to \mc{C}$ admitting left Kan extensions,
there is an equivalence between monoids --- in the monoidal
category defined above --- and relative monads.

\begin{definition}[Relative Monads]
  A \emph{relative monad} over $J : \mb{J} \to \mc{C}$ is given by
  \begin{itemize}[label=$-$]
    \setlength\itemsep{-1pt}
    \item A function $T : \mrm{ob}(\mb{J}) \to \mrm{ob}(\mc{C})$ on objects.
    \item For all $\Gamma : \mb{J}$, a morphism $\eta_\Gamma : J(\Gamma) \to T(\Gamma)$
    \item For all $\Gamma,\Delta : \mc{C}$, a morphism
    $\sigma_{\, \Gamma,\Delta} : \mc{C}(J(\Gamma),\, T(\Delta)) \to \mc{C}(T(\Gamma),\, T(\Delta))$
  \end{itemize}
  satisfying the following equations:
  \begin{align*}
    \begin{tikzcd}[ampersand replacement=\&]
      J(\Gamma) \ar[r, "\eta_{\,\Gamma}"] \ar[dr, swap, "c"]
        \& T(\Gamma) \ar[d, "\sigma(c)"] \\
        \& T(\Delta)
    \end{tikzcd}
    &&
    \begin{tikzcd}[ampersand replacement=\&]
      T(\Gamma) \ar[dr, bend left, "\sigma_{\,\Gamma,\Gamma}(\eta_{\,\Gamma})"]
                \ar[dr, swap, bend right, "\Id"]
        \& \\
        \& T(\Gamma)
    \end{tikzcd}
    &&
    \begin{tikzcd}[ampersand replacement=\&]
      T(\Gamma) \ar[r, "\sigma_{\,\Gamma,\Delta}(c)"] \ar[dr, swap, "\sigma_{\,\Gamma,\Upsilon}(\sigma_{\Delta,\Upsilon}(c') \circ c)"]
        \& T(\Delta) \ar[d, "\sigma(c')"] \\
        \& T(\Upsilon)
    \end{tikzcd}
  \end{align*}
\end{definition}

\noindent One can easily check that relative monads provide us with exactly
the same axiomatisation of simultaneous substitution as Kleisli triples up
to homogeneity.
In the previous equivalence, the substitution
$\sigma : (\Gamma \to \Lambda(\Delta)) \to (\Lambda(\Gamma) \to \Lambda(\Delta))$
was defined using renaming as $\sigma(c) := \mu_\Delta \circ \Lambda(c)$.
Here, as $\Lambda(c)$ is ill-typed, it is replaced by its alternative
$i_\Lambda : (J(\Gamma) \to \Lambda(\Delta)) \to (\Lambda(\Gamma) \to
(\Lambda \otimes \Lambda) (\Delta))$ and substitution $\sigma : (J(\Gamma)
\to \Lambda(\Delta)) \to (\Lambda(\Gamma) \to \Lambda(\Delta))$ is defined
as:
\[
  \begin{tikzcd}
    \Lambda(\Gamma) \ar[r, "i_\Lambda(c)"]
      & (\Lambda \otimes \Lambda)(\Delta) \ar[r, "\mu_\Delta"]
      & \Lambda(\Delta)
  \end{tikzcd}
\]

It remains to specify how simultaneous substitution behaves on
constructors such as $\app$ and $\abs$.
In the general case, using that $i_{(\_)}(c)$ is natural, as simultaneous
substitution is defined in terms of flattening and $i$, it can be computed
through them:
\[
  \begin{tikzcd}[column sep=huge]
    H(T)(\Gamma)         \ar[r, "r"] \ar[d, swap, "i_{H(T)}(c)"]
                         \ar[ddd, swap, shift right=27pt, shorten=-5pt,
                             bend right=42, "\sigma(?)", red]
      & T(\Gamma)        \ar[d, "i_{T}(c)"]
                         \ar[ddd, shift left=15pt, shorten=-5pt, bend left=40, "\sigma(c)"] \\
    (H(T) \otimes T)\,(\Delta)    \ar[r, dashed, "r \otimes T"]
                                  \ar[d, swap, "\theta"]
      & (T \otimes T)\, (\Delta)  \ar[dd, "\mu"]\\
    H(T \otimes T)\, (\Delta)     \ar[d, swap, "H(\mu)"]
      & \\
    H(T)(\Delta)                  \ar[r, swap, "r"]
      & T(\Delta)
  \end{tikzcd}
\]
\noindent Yet, as for the case $[\Set,\Set]$, there is no particular reason
why the left-hand side would factor in a substitution.
However using that coends preserves product and the universal property of
$i$, it is possible for \emph{algebraic} signatures.
For instance, for the lambda calculus, we get the usual laws:
\begin{align*}
  \begin{tikzcd}[ampersand replacement=\&]
    \Lambda(\Gamma) \times \Lambda(\Gamma) \ar[r, "\app_\Gamma"]
                                           \ar[d, swap, "\sigma_{\,\Gamma,\Delta}(c) \times \sigma_{\,\Gamma,\Delta}(c)"]
      \& \Lambda(\Gamma)                    \ar[d, "\sigma_{\,\Gamma,\Delta}(c)"] \\
    \Lambda(\Delta) \times \Lambda(\Delta) \ar[r, swap, "\app_\Delta"]
      \& \Lambda(\Delta)
  \end{tikzcd}
  &&
  \begin{tikzcd}[ampersand replacement=\&]
    \Lambda(\Gamma + 1) \ar[r, "\abs_\Gamma"]
                        \ar[d, swap, "\sigma_{\,\Gamma + 1,\Delta + 1}(\ol{c})"]
      \& \Lambda(\Gamma) \ar[d, "\sigma_{\,\Gamma,\Delta}(c)"] \\
    \Lambda(\Delta+1)   \ar[r, swap, "\abs_\Delta"]
      \& \Lambda(\Delta)
  \end{tikzcd}
  \hfill\null
\end{align*}
where $\ol{c} := \lbrack \Lambda(\mrm{inl}_{\Delta,1}) \circ c,
\eta_{\Delta+1} \circ \mrm{inr}_{\Delta, 1}\rbrack : J(\Gamma +1) \to
\Lambda (\Delta+1)$ is the usual weakening. Here, we use that $J(X)+1$ is
strictly equal to $J(X+1)$.

\subsection{The Recursion Principle on $[\mb{F},\Set]$}
\label{subsec:rec_F_set}

The recursion principle for $[\mb{F},\Set]$ works exactly as for $[\Set,\Set]$.
Indeed, a morphism of models respects the constructors by design.
Furthermore, it can be shown that it also respects substitution.
For instance, for two models of the lambda calculus, one has:
\[
  \begin{tikzcd}
    \Lambda_1(\Gamma) \ar[r, "i_{\Lambda_1}(c)"] \ar[d, swap, "f"]
                    \ar[rr, shift left=8pt, shorten=-5pt, bend left=25, "\sigma_1(c)"]
      & \Lambda_1(\Lambda_1(\Delta)) \ar[r, "\mu_1"] \ar[d, "f \circ f"]
      & \Lambda_1(\Delta ) \ar[d, "f"]\\
    \Lambda_2(\Gamma) \ar[r, swap, "i_{\Lambda_2}(f \circ c)"]
              \ar[rr, swap, shift right=8pt, shorten=-5pt, bend right=25, "\sigma_2(f \circ c)"]
      & \Lambda_2(\Lambda_2(\Delta)) \ar[r, swap, "\mu_2"]
      & \Lambda_2(\Delta)
  \end{tikzcd}
\]
Then, by initiality, building models provide the initial object with
the usual recursion principle strengthened to respect substitution.
The difference in the monoidal structure makes little difference if one is
using the recursion principle to translate languages.
Indeed, in this case, we rely on a monoid structure that already exists as
it is provided by the initiality theorem, as in \cref{ex:first-order-to-linear}.
However, it makes a difference if one is building a monoid for a specific
purpose, such  as computing the set of free variables as is done in \cref{ex:free-var}.
Indeed, one has to account for the more complicated monoidal product, and the
fact that the natural isomorphisms $\alpha,\lambda,\rho$ are no longer trivial.
Thankfully, the adjunction $\Lan{J}{\_} \dashv J^*$ is monoidal, and implies
a monoidal adjunction between the categories $\Mon([\mb{F},\Set])$ and
$\Mon([\Set,\Set])$, which helps us build such monoids.

%% file: Sections/example_SetT_SetT.tex
\section{Simply-Typed Languages: The Instance $[\Set^T,\Set^T]$}
\label{sec:example_setT_setT}

We have fully justified in \cref{sec:example_set_set,sec:example_F_set} that
our models appropriately model the substitution structure of untyped
higher-order languages, and encompass different mathematical models of
contexts.
We justify here that that the framework also encompasses simply-typed
higher-order languages.
To do so, given a set of types $T : \Set$ with decidable equality, we focus
on the monoidal category $[\Set^T,\Set^T]$ --- where $T$ is seen as a
discrete category --- popular in the literature for its simple monoidal
structure \cite{HirschowitzMaggesi07,ZsidoPhd10,HssTypedUnimath22,HssNonWellfounded23}.
Other instances are possible, such as $[\mb{F} \downarrow T, \Set]^T$ as in
\cite{Cbn02}, \cite[Chapter 5]{ZsidoPhd10}.
For that case, we refer to \cref{sec:example_F_set} to deal with the
monoidal structure.
As the monoidal structures have been explained in depth in previous sections,
we follow here the same outline but focus on what changes when scaling up
to simply-typed languages.

\subsection{The Instance $[\Set^T,\Set^T]$}
\label{subsec:assumptions_setT_setT}

Given a set $T : \Set$ with decidable equality, we need to prove that
$[\Set^T,\Set^T]$ satisfies the hypotheses of the
\hyperref[thm:initiality-theorem]{initiality theorem}.
First, $[\Set^T,\Set^T]$ is well-defined as $T$ has decidable equality, and
therefore can be equipped with the structure of a discrete category without
the law of excluded middle.
Then, the arguments are exactly the same as in \cref{subsec:assumptions_set_set}.
As an endofunctors category $[\Set^T,\Set^T]$ has a monoidal structure for composition
and the identity.
It is complete and cocomplete as a functor category, and since (co)limits are
computed pointwise, they are preserved by precomposition $\_ \circ Z$ for $Z
: [\Set,\Set]$.

\subsection{Signatures on $[\Set^T, \Set^T]$}
\label{subsec:sig_SetT_SetT}

By construction, our models rely on the underlying monoidal category to
represent contexts.
This forces us to hardcode the type system $T$ in the monoidal category, as
in $[\Set^T, \Set^T]$.
Therefore, the type system $T$ is to be chosen depending on the language we
want to represent; and once $T$ is fixed, we can only consider simply-typed
languages on it.
For instance, if one wishes to represent the simply-typed lambda calculus
with base types $B : \Set$, then $T$ must at least have $T_{\Lambda_B} :=
b \in B \,|\, T_{\Lambda_B}  \to T_{\Lambda_B} $ as a subset.

Out of previously considered signatures, we mainly focus on simply-typed
algebraic signatures.
Like their untyped counterpart, simply-typed algebraic signatures are the
simplest signatures enabling us to specify simply-typed languages with
variable binding such as the untyped lambda calculus.
Moreover, like the untyped ones, they are quite expressive and enable us to
express many languages:

\begin{definition}[Algebraic Signatures]
  \label{def:simply-typed-alg-sig}
  A simply-typed algebraic signature is given by a set $I : \Set$, with an
  arity function $\mrm{ar} : I \to \List(\List(T)\times T) \times T$.
\end{definition}

\noindent For any $i : I$, $\mrm{ar}(i)$ is the arity of an independent
constructor.
The first component $\List(\List(T)\times T)$ specifies the inputs
of the constructor; each input being specified by its type $T$, and by the
list of variable types bound in it.
The second component specifies the output type of the constructor.

As usual by \cref{prop:sigstrength_omega_cocomplete}, $\omega$-cocontinuous
signatures with strength are closed under coproducts, so it suffices to
represent individual constructors to represent algebraic signatures.
Representing constructors is more delicate than in the untyped case as we
need to account for typing: constructors build terms of a specific type.
As an intermediate step, we first represent unary constructors before
considering $n$-ary ones.

Consider a unary constructor $((l,s),t)$, binding $l : \List(T)$ variables
in an input of type $s : T$, and returning a term of type $t : T$.
In our models, such a constructor is represented by a module morphism $M \to
\Theta$; which for all contexts $\Gamma : \Set^T$, and type $u : T$ yields a
map $M(\Gamma)(u) \to R(\Gamma)(u)$.
In the case $u := t$, the constructor is supposed to return a term of type
$t$ out of an input of type $s$ with $l$ free variables.
This leads us to define $M(\Gamma)(t)$ by $R(\Gamma + l)(s)$, to get a map
$R(\Gamma + l)(s) \to R(\Gamma)(t)$.
For all other types, the constructor should not exist; this leads us to define
$M(\Gamma)(u)$ by $\emptyset$ to get a map $\emptyset \to R(\Gamma)(u)$.
Hence, to represent unary constructors, we need to represent typed variable
binding, but also trivialised uninteresting fibers, and appropriately
re-index them.

\begin{definition}[Yoneda and Contexts]
  \label{def:yoneda-emb}
  The Yoneda embedding $y : T \to \Set^T$ defined as $y(t) := \mrm{Hom}(t,\_)$
  enables us to define the typed context with one variable as follows:
  \[ y(t)\; u \;
  = \left\{
      \begin{array}{cl}
        \{ \star \}  & \mrm{if}\; u := t \\
        \emptyset    & \mrm{otherwise}
      \end{array}
    \right.
\]
\end{definition}

\begin{proposition}[Variable binding]
  There is an $\omega$-cocontinuous signature with strength representing the
  binding of a fresh variable of type $t$ denoted $\Theta^{(t)}$.
  Its underlying endofunctor $\delta^{(u)} : [\Set^T,\Set^T] \to [\Set^T,\Set^T]$
  is defined as $\delta^{(u)}\; X\; \Gamma := X(\Gamma + y(t))$
\end{proposition}

\noindent As $\omega$-cocontinuous signatures with strength are closed under
composition, the binding of $l := [u_1, ...,u_n] : \List(T)$ variables can
be defined as $\Theta^{(l)} := \Theta^{(u_1)} \circ ... \circ \Theta^{(u_n)}
\circ \Theta$.
It would be possible to define a constructor $\Theta^{(l)} \to \Theta$, but
it would not be suitable since the input and output of the constructor would
have the same type.
To define unary constructors, we also need to define a ``Dirac'' function to
trivialise uninteresting fibers and a $\swap$ function to appropriately
re-index the fibers:

\begin{proposition}[Dirac]
  Given a type $s : T$, there is an $\omega$-cocontinuous signature with
  strength, that given $X : [\Set^T,\Set^T]$ trivialises all fibers in its
  image except the one in $t$.
  Its underlying functor $\delta_u^s : [\Set^T,\Set^T] \to  [\Set^T,\Set^T]$,
  is defined as:
  \[ \delta_u^s\; X\; \Gamma\; u \;
       = \left\{
    \begin{array}{ll}
      X(\Gamma)(t) & \mrm{if}\; u := s \\
      \emptyset    & \mrm{otherwise}
    \end{array} \right.
  \]
\end{proposition}

\begin{proposition}[Swap]
  Given two types $s,t : T$, there is an $\omega$-cocontinuous signature
  with strength, that given $X : [\Set^T,\Set^T]$ swaps the two fibers in
  $s$ and $t$, in its image.
  Its underlying functor $\swap_s^t : [\Set^T,\Set^T] \to  [\Set^T,\Set^T]$
  is defined as:
  \[ \swap_s^t\; X\; \Gamma\; u := \left\{
    \begin{array}{ll}
      X(\Gamma)(s) & \mrm{if}\; u := t \\
      X(\Gamma)(t) & \mrm{if}\; u := s \\
      X(\Gamma)(u) & \mrm{otherwise}
    \end{array} \right.
  \]
\end{proposition}

\begin{proposition}[Unary Constructors]
  A unary constructors $((l,s),t)$ can be represented by the
  $\omega$-cocontinuous signature with strength:
  \[ [\Theta^{l}_s]_t := \swap_s^t \circ \delta_u^s \circ \Theta^{(l)} \]
\end{proposition}

Similarly to the unary case, an $n$-ary constructor building terms of type
$t$ should have all its fiber trivial, except the one in $t$, which is
supposed to be the product of the fibers of all the inputs.
For instance, for the simply-typed lambda calculus $\Lambda_B$, for all $s,t
: T_{\Lambda_B} $, $\app_{s,t}$ should be a map $\Lambda_B(\Gamma)(s \to t)
\times \Lambda_B(\Gamma)(s) \to \Lambda_B(\Gamma)(t)$ in $t$, and $\emptyset
\to \Lambda_B(\Gamma)(u)$ otherwise.
In the zero-ary case, this translates to:

\begin{proposition}[Zero-ary constructors]
  A zero-ary constructor of output type $t$ can be represented by the
  $\omega$-cocontinuous signature with strength:
  \[ [\Theta_0]_t := \delta_u^t \circ \Theta_0 \]
\end{proposition}

\noindent For the $n$-ary case, with $n > 0$, as products in a functor
category are computed pointwise, it suffices to take the product of all the
unary inputs, to represent $n$-ary constructors.
This is possible by \cref{ex:presheaves-limits,prop:omega-limits,prop:sigstrength_omega_complete}:

\begin{proposition}[Constructors]
  An $n$-ary constructor $([(l_1,s_1),...,(l_n,s_n)], t)$ can be represented
  by an $\omega$-cocontinuous signature with strength:
  \[ [\Theta^{l_1}_{s_1}]_t \times ... \times [\Theta^{l_n}_{s_n}]_t \]
\end{proposition}

Having represented constructors, we can now represent simply-typed algebraic
signatures, which enables us to represent many languages, such as the
simply-typed lambda calculus or PCF:

\begin{example}[Algebraic Signatures]
  For a fixed type system $T$, algebraic signatures $(I,\mrm{ar})$ are
  represented by $\omega$-cocontinuous signature with strength of the form:
  \[ \bigplus_{i : I}\; [\Theta^{l_1}_{s_1}]_{t_i} \times ... \times [\Theta^{l_{n_i}}_{s_{n_i}}]_{t_i} \]
\end{example}

\begin{example}[Simply-typed Lambda calculus]
  \label{ex:STLCB}
  The simply-typed lambda calculus with base type $B$, hence with typing
  system $T_{\Lambda_B} := b \in B \,|\, T_{\Lambda_B}  \to T_{\Lambda_B}$,
  and constants $(C_b : \Set)_{b : B}$, can be represented by the
  $\omega$-cocontinuous signature with strength:
  \[ \left( \bigplus_{b : B} \bigplus_{k : C_c} [\Theta_0]_b \right)
     \;\; \boldsymbol{+} \;\;
     \left( \bigplus_{s,t:T_{\Lambda_B}}\;\; [\Theta_{s \to t}]_t \times [\Theta_{s}]_t
     \; \boldsymbol{+} \;
     [\Theta^{s}_{t}]_{s \to t} \right)
\]
\end{example}

\begin{example}[PCF]
  PCF is a simply-typed higher-order programming languages able to handle
  logic, arithmetic, and recursion.
  Its type system is defined as system $T_\PCF := \Nat \,|\, \Bool \,|\,
  T_\PCF  \to T_\PCF$.
  For logic, it has the constructors: two constants $\top,\bot : \Bool$, two
  "if" operators $\mrm{if_\Bool}: \Bool \to \Bool \to \Bool \to \Bool$ and
  $\mrm{if_\Nat}: \Bool \to \Nat \to \Nat \to \Nat$.
  For arithmetics, it has the constructors: for all $n : \N$ a constant $k_n
  : \Nat$, $\mrm{Succ},\mrm{Pred} : \Nat \to \Nat$, and a zero test,
  $\mrm{zero?} : \Nat \to \Bool$.
  It also has variables, $\app$, $\abs$ and a fixpoint combinator $Y$.
  Its can be represented by an $\omega$-cocontinuous signature defined as
  \begin{gather*}
  [\Theta_0]_\Bool
    \;\; \boldsymbol{+} \;\;
  [\Theta_0]_\Bool
    \;\; \boldsymbol{+} \;\;
  [\Theta_{\Bool \to \Bool \to \Bool}]_\Bool
    \;\; \boldsymbol{+} \;\;
  [\Theta_{\Bool \to \Nat \to \Nat}]_\Nat \\
  \;\; \boldsymbol{+} \;\;
  \left( \bigplus_{n:\N}\;\; [\Theta_0]_\Nat \right)
    \;\; \boldsymbol{+} \;\;
  [\Theta_{\Nat}]_\Nat
    \;\; \boldsymbol{+} \;\;
  [\Theta_{\Nat}]_\Nat \\
  \;\; \boldsymbol{+} \;\;
  \left( \bigplus_{s,t:T_\PCF}\;\; [\Theta_{s \to t}]_t \times [\Theta_{s}]_t
                                \; \boldsymbol{+} \;
                             [ \Theta^{[s]}_{t}]_{s \to t} \right)
  \;\; \boldsymbol{+} \;\;
  \left( \bigplus_{s:T_\PCF}\;\; [\Theta_{s \to s}]_s \right)
  \end{gather*}
\end{example}

\begin{remark}
  While it suffices for our underlying category to be closed only under finite products, it is
  required to be closed under general coproducts.
  Indeed, for simply-typed languages, we usually have an infinite numbers of
  constructors.
  For instance, for the simply-typed lambda calculus, we have a constructor
  $\app_{s,t} : (s \to t) \times s \to t$ for all $s,t \in T_{\Lambda_B}$.
\end{remark}

More generally, it would also be possible to consider presentable signatures
-- as our $\omega$-cocontinuous signatures with strength are closed under
colimits --- or to consider explicit flattening, as done in \cref{subsec:sig_set_set}.

\subsection{Models on $[\Set^T, \Set^T]$}
\label{subsec:models_SetT_SetT}

The models for $[\Set^T,\Set^T]$ and simply-typed algebraic signatures are
highly similar to the ones of $[\Set,\Set]$ in \cref{subsec:model_set_set}.
We briefly explain the differences using the simply-typed lambda calculus
with base type $B$ but without constants as an example.

\subsubsection{Language and Constructors}

A model of the simply-typed lambda calculus gives us a functor $\Lambda_B :
\Set^T \to \Set^T$ that associates, to a any typed context $\Gamma : \Set^T$,
the functor $\Lambda_B(\Gamma) : \Set^T$ of lambda terms over it.
Since $T$ is a discrete category, the functor $\Lambda_B(\Gamma) : \Set^T$, is
simply a familly of sets $(\Lambda_B(\Gamma)(u))_{u : T_{\Lambda_B}}$.
Hence, given a type $u$, $\Lambda_B(\Gamma)(u)$ is then the set of
well-scoped lambda-terms of type $u$ in context $\Gamma$.
As usual, the natural transformation $\eta$ of the monoid gives us a
variable constructor $\var_\Gamma : \Gamma \to \Lambda_B(\Gamma)$.
Moreover, for all $s,t : T_{\Lambda_B}$, there are module morphisms
$\app_{s,t} : [\Theta_{s \to t}]_t \times [\Theta_s]_t \to \Theta$ and
$\abs_{s,t} : [\Theta^{[s]}_{t}]_{s \to t} \to \Theta$.
Forgetting about the module structure, they are given by natural transformations.
In $t$, they provide us with the two usual constructors $\app_{s,t} :
\Lambda(\Gamma)(s \to t) \times \Lambda(\Gamma)(s) \to \Lambda(\Gamma)(t)$
and $\abs_{s,t} : \Lambda(\Gamma + y(s))(t) \to \Lambda(\Gamma)(s \to t)$,
and in all other types $u$, they provide us with the trivial map
$\emptyset \to \Lambda(\Gamma)(u)$.

\subsubsection{Renaming}

As before, the functorial action of $\Lambda_B : \Set^T \to \Set^T$ is
renaming.
Given a natural transformation $c : \Gamma \to \Delta$ renaming labels, it
associates the natural transformations $\Lambda_B(c) : \Lambda(\Gamma) \to
\Lambda(\Delta)$ renaming terms.
As $T$ is discrete, the naturality condition is vacuous, and a morphism $c :
\Gamma \to \Delta$ is simply a family of functions $(\Gamma(u) \to \Delta(u))_{(u :T_{\Lambda_B})}$.
Hence, renaming for typed languages is the usual independent renaming for
each type of labels.
Consequently, the naturality of the constructors $\var,\app_{s,t},\abs_{s,t}$
describes, type by type, how the constructors commute with renaming.
For $\var$, we get the usual diagram, whereas for $\app_{s,t}$ and $\abs_{s,t}$,
we get the usual diagram in $t$, and a trivial diagram otherwise, as in
this case, $\app_{s,t}$ and $\abs_{s,t}$ are trivial.
For instance, for $\app_{s,t}$ it yields:
\begin{align*}
  \begin{tikzcd}[ampersand replacement=\&, column sep=huge]
    \Lambda_B(\Gamma)(s \to t) \times \Lambda_B(\Gamma)(s) \ar[r, "\app_\Gamma(t)"]
      \ar[d, swap, "\Lambda_B(c)(s \to t) \times \Lambda_B(c)(s)"]
      \& \Lambda_B(\Gamma)(t) \ar[d, "\Lambda_B(c)(t)"] \\
    \Lambda_B(\Delta)(s \to t) \times \Lambda_B(\Delta)(t) \ar[r, swap, "\app_\Delta(t)"]
      \& \Lambda_B(\Delta)(t)
  \end{tikzcd}
  &&
  \begin{tikzcd}[ampersand replacement=\&, column sep=huge]
    \emptyset \ar[r] \ar[d]
      \& \Lambda_B(\Gamma)(u) \ar[d] \\
    \emptyset \ar[r]
      \& \Lambda_B(\Delta)(u)
  \end{tikzcd}
\end{align*}

\subsubsection{Flattening}

The monoid structure provides us with a natural transformation representing
flattening $\mu : \Lambda_B \circ \Lambda_B \to \Lambda_B$.
The monoid laws specify its behaviour regarding itself and variables.
The behaviour of flattening on constructors is specified by the module structure.
As for renaming, this provides us with the usual flattening diagram for the
type $t$, and a trivial diagram for any other type.
For instance, for $\app_{s,t}$, it gives us:
\begin{align*}
  \begin{tikzcd}[ampersand replacement=\&, column sep=large]
    \Lambda_B(\Lambda_B(\Gamma))(s \to t) \times \Lambda_B(\Lambda_B(\Gamma))(s)
          \ar[r, "\app_{\Lambda_B(\Gamma)}"]
          \ar[d, swap, "\theta^\app := \Id"]
      \& \Lambda_B(\Lambda_B(\Gamma))(t) \ar[dd, "\mu_\Gamma(t)"] \\
    \Lambda_B(\Lambda_B(\Gamma))(s \to t) \times \Lambda_B(\Lambda_B(\Gamma))(s)
        \ar[d, swap, "\mu_\Gamma(s \to t) \times \mu_\Gamma(s)"]
      \& \\
    \Lambda_B(\Gamma)(s \to t) \times \Lambda_B(\Gamma)(s) \ar[r, swap, "\app_\Gamma"]
      \& \Lambda_B(\Gamma)(t)
  \end{tikzcd}
  &&
  \begin{tikzcd}[ampersand replacement=\&]
    \emptyset \ar[r] \ar[d]
      \& \Lambda_B(\Lambda_B(\Gamma))(u) \ar[dd] \\
    \emptyset \ar[d]
      \& \\
    \emptyset \ar[r]
      \& \Lambda(\Gamma)(u)
  \end{tikzcd}
\end{align*}

\subsubsection{Substitution}

The substitution works exactly as for $[\Set,\Set]$, since $[\Set^T,\Set^T]$
is also an endofunctor category.
Indeed, a monoid in any endofunctor category is equivalent to
\hyperref[def:kleisli_triples]{Kleisli triples}, which provide us with a
proper axiomatised substitution
$\sigma_{\,\Gamma,\Delta}: (\Gamma \to \Lambda_B(\Delta)) \to (\Lambda_B(\Gamma) \to \Lambda_B(\Delta))$
defined as $\sigma_{\,\Gamma,\Delta}(c) := \mu_{\Lambda_B(\Delta)} \circ \Lambda_B(c)$.
As before, since it is defined by flattening and renaming, it can be computed on
constructors through them.
However, there is no reason why it would factor into a substitution in the
general case.
Yet, in type $t$, for simply-typed algebraic signatures, it can be expressed
as a substitution.
Indeed, in type $t$, the substitution computes on constructors as in the
$[\Set,\Set]$ case, by a succession of renaming, strength and flattening, and
thus can be factored.
The other cases do not matter, as the constructors are trivial for any other
type $u$.
For instance, for $\app_{s,t}$ we get:
\begin{align*}
  \begin{tikzcd}[ampersand replacement=\&, column sep=large]
    \Lambda_B(\Gamma)(s \to t) \times \Lambda_B(\Gamma)(s)
            \ar[r, "\app_\Gamma(t)"]
            \ar[d, swap, "\sigma(c)(s \to t) \times \sigma(c)(s)"]
      \& \Lambda_B(\Gamma)(t)
            \ar[d, "\sigma(c)(t)"] \\
    \Lambda_B(\Delta)(s \to t) \times \Lambda_B(\Delta)(s)
            \ar[r, swap, "\app_\Delta(t)"]
      \& \Lambda_B(\Delta)(t)
  \end{tikzcd}
  &&
  \begin{tikzcd}[ampersand replacement=\&]
    \emptyset
            \ar[r]
            \ar[d]
      \& \Lambda_B(\Gamma)(t)
            \ar[d] \\
    \emptyset
            \ar[r]
      \& \Lambda_B(\Delta)(t)
  \end{tikzcd}
\end{align*}

\subsection{The Recursion Principle on $[\Set^T,\Set^T]$}
\label{subsec:rec_SetT_SetT}

The recursion principle for $[\Set^T,\Set^T]$ works as for $[\Set,\Set]$:
morphisms of models respect the constructors and substitution, initiality
enables us to build morphisms of models by building models, and doing so
amounts to the usual recursion principle.

\subsubsection{A limited recursion principle}

However, applying the recursion principle to translate simply-typed
languages as in the untyped case is limited.
In the untyped case, to build a translation between two languages, we used
that both languages have a model in the \emph{same} monoidal category
$[\Set,\Set]$.
Indeed, by initiality, it suffices to build a model of the first language on
top of the monoid of the second language, that is completing the monoid with
module morphism corresponding to the first language, to build such a
translation.
For instance, in \cref{ex:first-order-to-linear}, to translate first-order
logic to linear logic, we built a model of first-order logic on top of the
monoid of linear logic.

In the simply-typed case, this is very limited as the type system $T$ is
hardcoded in the monoidal category, as in $[\Set^T,\Set^T]$.
Indeed, by the initiality theorem, two different languages defined in two
different type systems $T$ and $T'$ yield models, and in particular
monoids, in different monoidal categories $[\Set^T,\Set^T]$ and
$[\Set^{T'},\Set^{T'}]$.
This prevents  us from applying the same method as in the untyped case,
as soon as $T \neq T'$.
Moreover, note that it not possible to simply consider our languages as
languages of the typing system $T + T'$.
Indeed, even though we would get for both languages monoids in the same
category, the constructors would still belong to different type systems, and
it would not be possible to represent the constructors of one using the
constructors of the other.

\subsubsection{An alternative recursion principle}\label{sec:alt-rec-principle}

This is quite limiting as most interesting translation of simply-typed
languages are between languages with different type systems.
For instance, consider the translation from PCF to the untyped lambda
calculus, of the translation of proofs in classical logic to intuitonistic
logic, following double negation, as described in \cite{ExtendedInitiality12}.
Yet, it is not very surprising as a translation of simply-typed languages
should first rely on a translation of typing systems $g : T \to T'$.
A possible solution to this issue is developed in \cite{ExtendedInitiality12},
but it is instance specific and requires changes to the whole framework.
We develop here instead an informal solution, based on the parametricity
in the monoidal category, that does not requires any modifcation to the
framework.

Asume given a translation of typing systems $T \to T'$.
To translate monoids from $[\Set^{T'},\Set^{T'}]$ to $[\Set^T,\Set^T]$,
we start by translating contexts and endofunctors.
By \cref{prop:LKE_and_coends}, $g$ admits a global left Kan extension, denoted
$\vec{g} := \Lan{g}{\_}$, that computes as below.
It entails an adjunction on contexts, i.e. between $\Set^T$ and $\Set^{T'}$:
\begin{align*}
  \begin{tikzcd}[ampersand replacement = \&, column sep=large]
      \Set^T
          \ar[r, bend left, "\vec{g}"]
          \ar[r, phantom, "\perp"]
        \& \Set^{T'} \ar[l, bend left, "g^*"]
  \end{tikzcd}
  &&
  \vec{g}(\Gamma)(t') \;:= \bigsqcup_{\substack{t : T \\ g(t) = t'}} \Gamma(t)
\end{align*}
This adjunction can be lifted to the endofunctors categories:

\begin{proposition}
  Given an adjunction $A \to B : k \vdash l : B \to A$ with unit $\eta : \Id
  \to l \circ k$ and counit $\epsilon : k \circ l \to \Id$, there is an
  adjunction on endofunctor categories:
  \begin{align*}
    \begin{tikzcd}[ampersand replacement = \&, column sep=large]
      A \ar[r, bend left, "k"]
                \ar[r, phantom, "\perp"]
        \& B \ar[l, bend left, "l"]
    \end{tikzcd}
    &&
    \begin{tikzcd}[ampersand replacement = \&, column sep=large]
        {[}A{,}A{]} \ar[r, bend left, "X_A \;\mapsto\; k \circ X_A \circ l"]
                  \ar[r, phantom, "\perp"]
          \& {[}B{,}B{]} \ar[l, bend left, "X_B \;\mapsto\; l \circ X_B \circ k"]
    \end{tikzcd}
  \end{align*}
  Moreover, the functor $G := X_B \;\mapsto\; l \circ X_B \circ k$ is
  \hyperref[def:monoidal-functor]{monoidal}, with $G_0 : \Id \to l \circ k
  := \eta$ and $G_2 : l \circ X_B \circ k \circ l \circ X_B' \circ k \to l
  \circ X_B \circ X_B' \circ k := l \circ X_B \circ \epsilon \circ X_B'\circ
  k$.
\end{proposition}

\noindent Importantly, the adjunction on endofunctors categories provides
us with a monoidal functor $G : [\Set^{T'},\Set^{T'}] \to [\Set^T,\Set^T]$
that enables us to translate monoids:

\begin{proposition}
  \label{prop:functorial-monoids}
  Every monoidal functor $(F,F_0,F_2) : (\mc{C},I,\otimes) \to
  (\mc{D},J,\bullet)$ entails a functor on monoids $\Mon(\mc{C}) \to
  \Mon(\mc{D})$.
  It associates to any monoid $(R,\eta,\mu) : \Mon(\mc{C})$ the monoid
  $(F(R),F(\eta) \circ F_0,F(\mu) \circ F_2) : \Mon(\mc{D})$, and to any
  morphism of monoid $f : R \to R'$, the morphism $F(f) : F(R) \to F(R')$.
\end{proposition}

\noindent Using this functor, it is possible to apply the recursion
principle to translate languages using different type systems, by first
translating the monoid obtained on $[\Set^{T'},\Set^{T'}]$ to
$[\Set^T,\Set^T]$, and then building a model on it as before.

We illustrate this by translating the \hyperref[ex:STLCB]{simply-typed lambda calculus}
with base type $B$ but without constants to the \hyperref[ex:ULC]{untyped lambda calculus},
by erasing types.
To do so, we see the untyped lambda calculus as a simply-typed language on
the trivial type system $T' := \{ \star \}$, which forces $g : T_{\Lambda_B}
\to \{ \star \}$ to be constant.
By the previous discussion, $g^*$ has a left adjoint $\vec{g} : \Set^{T_{\Lambda_B}} \to \Set^{*}$
defined as $\vec{g}(\Gamma) := \star \mapsto \sqcup_{t : T_{\Lambda_B}}\, \Gamma(t)$,
and there is an associated monoidal functor $G$ that translates monoids.
Hence, to translate the simply-typed lambda calculus to the untyped one, we
need to build a model of the simply-typed lambda calculus on top of the
monoid $G(\Lambda)$; that is, we need to build module morphisms $[\Theta_{s \to t}]_t
\times [\Theta_{s}]_t$ and $[\Theta^{s}_{t}]_{s \to t}$ for all $s,t :
T_{\Lambda_B}$.
By definition, building the natural transformations underlying the module
morphism amounts to building for all $\Gamma : T_{\Lambda_B}$, natural morphisms
$G(\Lambda)(\Gamma)(s \to t) \times G(\Lambda)(\Gamma)(s) \to G(\Lambda)(\Gamma)(t)$
and $G(\Lambda)(\Gamma + y(s))(t) \to G(\Lambda)(\Gamma)(s \to t)$.
It unfolds to building:
\begin{align*}
  \Lambda(\vec{g}\;\Gamma)(*) \times \Lambda(\vec{g}\;\Gamma)(*) \to \Lambda(\vec{g}\;\Gamma)(*)
  &&
  \Lambda(\vec{g}\;(\Gamma + y(s)))(*) \to \Lambda(\vec{g}\;\Gamma)(*)
\end{align*}
The first morphism obviously corresponds to $\app_{\vec{g}\;\Gamma}$.
Furthermore, as $\vec{g}$ is a left adjoint, $\vec{g}(\Gamma + y(s)) =
\vec{g}\; \Gamma + \vec{g}(y(s)) = \vec{g}\;\Gamma +1$, and the second one
corresponds to $\abs_{\vec{g}\;\Gamma}$.
Constants could be added to the source language, providing we provide lambda terms for them in the target language.
It remains to prove that they respect module substitutions --- this is quite tedious.
According to their definitions, it amounts to proving it in type $t$, and
can be done using the definition of the strengths and the naturality of the
unit and counit.
Thus, we get translation that erases types: it forgets the types of the
variables, and replaces all $\app_{s,t},\abs_{s,t}$ by $\app,\abs$.
Moreover, it respects substitution by design.

%% file: Sections/related_work.tex
\section{Related Work}
\label{sec:related-work}

The literature on initial semantics is quite prolific, and contains many
different approaches to initial semantics.
We focus on three traditions using different mathematical structures, and
named after them: $\Sigma$-monoids, modules over monads, and heterogeneous
substitution systems.
Unfortunately, the links between the different approaches is not laid out clearly in the literature,
and all the traditions contain different variations, which can be
confusing for newcomers to the field.
To bridge this gap in the literature, we have suitably combined and
abstracted the different approaches in
\cref{sec:models,sec:initiality_theorem,sec:building_initial_model}.

In the following, we first give a brief, and necessarily incomplete, chronological overview of the
different work on initial semantics in \cref{subsec:rw-overview}.
We then provide an extensive and detailed description of related work,
tradition by tradition; where we explain how the different approaches relate
to our framework, and therefore to each other.
In particular, we discuss the different existing variations in the
frameworks, and we justify the design choices of our framework.
We discuss work on $\Sigma$-monoids in \cref{subsec:rw-sigma-mon}, on
modules over monoids in \cref{subsec:rw-modules-over-monoids}, and on
\cref{subsec:rw-hss}.

\subsection{Overview of the related work}
\label{subsec:rw-overview}

The idea of using monads as an abstraction to reason about substitution was
first introduced by Bellegarde and Hook \cite{BellegardeHook94}.
To ease the construction of a monad, building on nested data types
\cite{NestedDataTypes98}, Bird and Paterson \cite{DeBruijnasNestedDatatype99}
defined the untyped lambda calculus intrinsically, and strengthened the usual
fold operation to ``generalised fold''.
To be able to handle more languages, they later offered a general treatment
of generalised fold for nested data types in \cite{GeneralisedFold99}.
As an alternative to generalised fold, Altenkirch and Reus considered instead
structural induction for nested data types in \cite{AltenkirchReus99}.

\subsubsection{Origins: 1999 - 2007}

Capturing the substitution structure of untyped higher-order languages as
initial objects was first achieved in \cite{FPT99} by Fiore, Plotkin, and
Turi, using the category $[\mb{F},\Set]$.
To do so, they introduced $\Sigma$-monoids, and suggested using strength and
monoidal categories to prove initiality results.
Fiore then investigated simply-typed higher-order languages using those
methods in \cite{Cbn02,MMCCS05}.

Matthes and Uustalu introduced heterogeneous substitution systems on
endofunctor categories in \cite{Hss04}, in order to prove that both
wellfounded and non-wellfounded syntax have a monadic substitution
structure.
In particular, they introduced signatures with strength as a formal notion
of signature, and proved generic theorems for them.

The framework based on monoidal categories and $\Sigma$-monoids was later detailed
by Fiore and Hamana, in \cite{SecondOrderDep08}, were they furthermore
considered the addition of meta-variables, and suggested a method to handle
dependently typed languages.
We do not deal with meta-variables in this work.

Modules over monads and associated models were introduced for endofunctor
categories by André Hirschowitz and Maggesi in \cite{HirschowitzMaggesi07}, in
order to capture the substitution properties of constructors.
This work was extended in \cite{HirschowitzMaggesi10}, were they proved,
using a proof specific to the base category of sets, that untyped higher-order languages
have an initial model on $[\Set,\Set]$.

\subsubsection{Consolidation: 2007 - 2015}

Equations for $\Sigma$-monoids were considered in Hur's dissertation \cite{HurPhd},
and published with Fiore in \cite{FioreHur07,FioreHur09,FioreHur08}.
Equations and meta-variables have also been considered in \cite{FioreMahmoud10,FioreHur10}.
We do not deal with semantics in this work, neither with equations, nor with operational semantics.

The links between the $\Sigma$-monoids approach and module over monads one
have been investigated during Zsidó dissertation \cite{ZsidoPhd10}.
In particular, Zsidó fully worked out the simply-typed instance for both traditions.
A variant of Zsidó's construction for modules over monads was formalized by Ahrens \cite{ISCoq10} in the proof assistant Coq.

A proposition to handle polymorphic languages was investigated by Hamana in
\cite{Polymorphism11}, and considered with meta-variables and equations
by Fiore and Hamana in \cite{PolymorphismEq13}.

The current concept of signatures for modules over monads was introduced by
Hirschowitz and Maggesi in \cite{HirschowitzMaggesi12}.
They additionally showed that these signatures generalise signatures with strength.

Modules over monads for syntax and semantics were investigated further in
Ahrens' dissertation \cite{AhrensPhd}.
He developed a framework to extend the initiality principle of
simply-typed languages to allow for translations across typing systems
\cite{ExtendedInitiality12}.
Ahrens also considered reductions rules using modules over monads and relative monads.
A framework for untyped languages was published in \cite{UntypedRelativeMonads16},
and for simply-typed languages in \cite{TypedRelativeMonads19}.

Heterogeneous Substitution Systems have been revisited in \cite{HssRevisited15},
by Ahrens and Matthes, where they extend the hss framework with an initiality
result and a formalisation.

$\Sigma$-monoids have also been revisited by Fiore and Saville in \cite{ListObjects17}.
They extended $\Sigma$-monoids to $T$-monoids, weakened the assumptions of
previous theorems, and provided more detailed proofs of the theorems.

\subsubsection{Recent Work: 2018 - Present}

Modules over monads and semantics were particularly investigated in
Ambroise Lafont's dissertation \cite{LafontPhd}, in collaboration
with other people, in particular, Ahrens, André Hirschowitz, Tom Hirschowitz, and Maggesi.
Equations were investigated through signatures in \cite{PresentableSignatures21},
and on their own in \cite{2Signatures19}.
Reductions rules and strategies were investigated in
\cite{ReductionMonads20} and \cite{TransitionMonads20,TransitionMonads22}, respectively.

Heterogeneous substitution systems were applied by Ahrens, Matthes, and
Mörtberg to handle untyped and simply-typed higher-order languages in Coq's
UniMath library in \cite{HssUntypedUniMath19,HssTypedUnimath22}

The framework using $\Sigma$-monoids was extended by Borthelle, Lafont, and
Tom Hirschowitz to skew-monoidal categories in \cite{CellularHoweTheorem20},
in order to study bisimilarity.

Another skew-monoidal category was investigated by André and Tom
Hirschowitz, Lafont and Magessi in \cite{NamelessDummies22}, to study
De Bruijn monads.

Fiore and Szamozvancev recently used the work on $\Sigma$-monoids,
meta-variables, and equation to design a framework to handle higher-order
languages in Agda \cite{FioreSzamozvancevPopl22}.

Very recently, independently from us and for different reasons,
heterogeneous substitution systems have been generalised to monoidal
categories in the study of non-wellfounded syntax in \cite{HssNonWellfounded23}.

\subsection{$\Sigma$-monoids}
\label{subsec:rw-sigma-mon}

\subsubsection{Origins}

Capturing higher-order languages with their substitution structure as
initial models was first achieved for untyped algebraic signatures, in
a seminal work by Fiore, Plotkin and Turi in \cite{FPT99}, using the category $[\mb{F},\Set]$.
They first showed that the pure syntax of untyped higher-order calculi,
specified by algebraic signatures, is modeled by particular algebras on
$[\mb{F},\Set]$.
They then introduced $\Sigma$-monoids,\footnote{
  $\Sigma$-monoids are a particular case of models for signatures with
  strength \cref{remark:model-sigma-monoids}.}
and stated that every binding signature yields an initial $\Sigma$-monoids on $[\mb{F},\Set]$ ---
thus, providing a framework for initial semantics.
No proofs were given in the extended abstract \cite{FPT99};
however, the authors suggested
it can be proven using that, since $[\mb{F},\Set]$ is a \emph{closed}\footnote{
  A monoidal category is (bi)closed, when for all $Z : \mc{C}$, the functors
  $\_ \otimes Z$ and $Z \otimes \_$ have right adjoints.
  }
monoidal category, every free $\Sigma$-algebra such that $\Sigma$ has a
pointed strength is parametrically free.

This claim was made more precise by Fiore in another extended abstract
\cite[Sections I.1.1 - I.2.2]{SecondOrderDep08}.
After giving an analysis of (pointed) strength in terms of actions over
monoidal categories, Fiore stated that if a monoidal category is closed and
has binary products, then every $(I + \Sigma + X \otimes \_)$-initial
algebra, such that $\Sigma$ has a pointed strength, yields by parametrised
initiality the free $\Sigma$-monoids over $X$.
Fiore furthermore suggested that if all of the $(I + \Sigma + X \otimes \_)$-initial algebras
exist,\footnote{
  This is the case as soon as the monoidal category additionally has initial
  object and $\omega$-colimits, and $\Sigma$ and $X \otimes \_$ are $\omega$-cocontinuous.
} then they assemble into a left adjoint to the forgetful functor
$U : \Sigma\trm{-monoid} \to C$.
Such a theorem yields an initiality theorem since left adjoints
preserve initial objects: the free $\Sigma$-monoid over the initial object
is an initial $\Sigma$-monoid.
Every binding signature having an initial model for $[\mb{F},\Set]$ is then
discussed as an instance of this result.
We explained this instance from a slightly different perspective in
\cref{sec:example_F_set}, using relative monads \cite{RelativeMonads15},
which were invented later.

$\Sigma$-monoids and the theorems above have been extended by Fiore and
Saville in \cite{ListObjects17}, to the larger class of $T$-monoids.
$T$-monoids are very similar to $\Sigma$-monoids, except that $T$ is assumed
to be a strong monad, which enables one to account for several universal algebra notions.
While we have no particular use for $T$-monoids in its generality, this work
is interesting in two regards.
First, they weakened the closedness conditions in the adjoint theorem to an
$\omega$-cocontinuity condition.
Second, they provided more detail on how to prove the adjoint theorem
from which stems the initial one.

We discuss the initiality theorem and the adjoint theorem further in
\cref{subsubsec:rw-co-vs-adj}, and the difference between the proof based on
parametrised initiality and the proof presented in \cref{subsec:building_initial_model}
based on hss, in \cref{subsubsec:rw-param-vs-hss} and \cref{subsubsec:rw-building-param-hss}.

\subsubsection{Cocontinuity vs.\ existence of adjoints}
\label{subsubsec:rw-co-vs-adj}

The above results are similar to the one presented in
\cref{sec:initiality_theorem}, up to two differences.

Firstly, as in \cite{ListObjects17}, but in contrast to prior work,
we do not require the monoidal product to be closed, but require
instead precomposition $\_ \otimes Z$ to preserve initiality, binary
products and $\omega$-colimits, and postcomposition $Z \otimes \_$ to
preserve $\omega$-colimits, and this for all $Z$.
This is a weaker assumption, as being $\omega$-cocontinuous is implied by
having a right adjoint.
This is an important step towards relating the different approaches, since, in contrast to
$[F,\Set]$, precomposition does not have a right adjoint for $[\Set,\Set]$.

Indeed, if precomposition by $Z$ had a right adjoint $R_Z : [\Set,\Set] \to [\Set,\Set]$,
then $\Set \to \Set$ would be locally small, as using the Yoneda lemma and that $\Id = y_1$,
we would have: $\mrm{Hom}(Z,G) \cong \mrm{Hom}(\Id \circ Z,G) \cong \mrm{Hom}(\Id,R_Z(G))
\cong \mrm{Hom}(y_1,R_Z  G) \cong R_Z(G)(1)$.
In this case, by a theorem by Freyd and Street \cite{FreydStreet95}, we get
that $\Set$ is essentially small\footnote{ A category is essentially small
if it is equivalent to a small category}, which is impossible as the cardinals
do not form a set, and are equal if equivalent.

Secondly, they proved the adjoint theorem (c.f.\ \cref{thm:adjoint-theorem})
and deduced the initiality theorem (c.f.\ \cref{thm:initiality-theorem})
from it, using that left adjoints preserve initial objects.
We have done the opposite; we proved the initiality theorem, and deduced the
adjoint theorem from it (c.f.\ \cref{subsec:building-adjoint}).
This enables us to prove the initiality theorem using the underlying functor
$I + H\_$ rather than the functor $I + H\_ + X \otimes \_$, which enables us to
remove the hypothesis that $X \otimes \_$ is $\omega$-cocontinuous in the
hypothesis of the initiality theorem.
This is important for dealing with some skew-monoidal categories as
discussed in \cref{subsubsec:rw-skew}.

\subsubsection{Parametrised initiality vs.\ heterogeneous substitution systems}
\label{subsubsec:rw-param-vs-hss}

Since the use of parametrised initiality for $\Sigma$-monoids is explained in little
detail in \cite{FPT99,SecondOrderDep08,ListObjects17}, the differences of
vernacular and presentation can lead one to believe that the proof based on
parametrised initiality is fundamentally different from the one based on hss
described in \cref{sec:building_initial_model}.

However, hss and parametrised initiality are actually very similar, and
modulo the difference between initiality theorem and adjoint theorem, the
proofs are strongly related; even though to the best of our knowledge this has
never been reported before.
To explain the differences, we focus on proving the initiality theorem:

\begin{definition}[Parametrised Initiality]
  Let $F : \mc{C} \times \mc{C} \to \mc{C}$ be a bifunctor with pointed
  strength $\mrm{st}$, and $U : \mc{C}$.
  An $F(U,\_)$-algebra $(R,r)$ is \emph{parametrised initial} if for
  any pointed object $Z : \mc{C}$ and $F(U \otimes Z,\_)$-algebra $(C,c)$,
  there is a unique morphism $h : R \otimes Z \to C$ such that the following diagram commutes.
  \[
    \begin{tikzcd}[column sep=large]
      F(U, R) \otimes Z \ar[r, "\mrm{st}"] \ar[d, swap, "r \otimes Z"]
        & F(U \otimes Z, R \otimes Z) \ar[r, dashed, "F(U \otimes Z{{,}} h)"]
        & F(U \otimes Z, C) \ar[d, "c"] \\
      R \otimes Z \ar[rr, swap, dashed, "h"]
        &
        & C
    \end{tikzcd}
  \]
\end{definition}

\noindent To prove an initiality theorem, we are interested in parametrised
initiality for $U := I$, and for the bifunctors of the form $F : (U,A)
\longmapsto U + H(A)$ such that $H$ is a functor with pointed strength $\theta$.
In this case, assuming that $\_ \otimes Z$ distributes over binary
products, parametrised initiality unfolds to \hyperref[def:hss]{hss}, except
that for parametrised initiality, the output of $h$ can be any $H$-algebra
$C$ with a map $f : Z \to C$, whereas it is fixed to be $T$ for hss.
\[
  \begin{tikzcd}[column sep=large]
    I \otimes Z \ar[r, "\eta \otimes Z"] \ar[dd, swap, "\lambda_Z"]
      & R \otimes Z \ar[dd, dashed, "h"]
      & H(R) \otimes Z \ar[l, swap, "r \otimes Z"]
                       \ar[d, "\theta_{R,e}"] \\
      &
      & H(R \otimes Z) \ar[d, dashed, "H(h)"] \\
    Z \ar[r , swap, "f"]
      & C
      & H(C) \ar[l, "r"]
  \end{tikzcd}
\]
Therefore, following the proof of \cref{subsec:hss_models}, both
parametrised initiality and hss yield models.

The main difference lies in the initiality part of the proof, more specifically
to prove that the initial algebra morphism respects the monoid multiplication.
For parametrised initiality, this can be directly proven by appropriately
instantiating $Z$ and $C$.
However, as $C$ is fixed to be $T$ in the definition of hss, it forces us
to use a fusion law as done in \cref{subsec:building_initial_model}.

This makes no difference for wellfounded syntax, but it is fundamental when
it comes to non-wellfounded syntax.
Indeed, as parametrised initiality automatically yields an initial model, it
can not be used to prove that non-wellfounded languages have a monadic
substitution structure; this is in contrast to hss which were designed to handle both
wellfounded and non-wellfounded syntax
\cite{Hss04,HssNonWellfounded23}.
To better relate the different approaches, we have decided to base our work
on hss in \cref{sec:building_initial_model}.

\subsubsection{Building parametrised initial algebras and heterogenous substitution systems}
\label{subsubsec:rw-building-param-hss}

It remains to understand how hss and parametrised initiality are built out
of initial algebras.
Let's first consider our case, where precomposition $\_ \circ Z$ is required
to be $\omega$-cocontinuous.

As detailed in \cref{subsec:building_hss} following \cite{HssUntypedUniMath19},
hss can be built out of initial algebras using generalised Mendler's style
iteration, by instantiating them with $F := I + H\_$, $L := \_ \circ Z$, $X
:= R$ and an appropriate $\Psi$.

\Mendler*

\noindent Similarly, generalised Mendler's style iteration could also be used
to derive parametrised initiality.

Rather than using Mendler's style iterations, Fiore and Saville
introduced instead in \cite[Theorem 4.7]{ListObjects17}, though without
proving it, a ``lax-uniformity property of initial algebra''.
That property is used to derive parametrised initiality, by
instantiating it with $F := F$, $G := F(\sim \circ Z, \_)$, $J := \_ \circ
Z$, $K := \Id$ and $t := \mrm{st}$.

\begin{theorem}
  \label{thm:FioreSaville}
  Let $\mc{A,B,C,D}$ be categories such that $\mc{C}$ has initial objects
  and $\omega$-colimits.
  Let $F,G,K,J$ be functors and $t : J \circ F \to G \circ (K \times J)$ a
  natural transformation as below left, such that for all $D : \mc{D}$,
  $F(D,\_)$ is $\omega$-cocontinuous with initial algebra $(R,r)$.
  Then, if $J$ preserves initiality and $\omega$-colimits, for any
  $G(KD,\_)$-algebra $(C,c)$ there exists a unique morphism $h : J(R) \to R$
  such that the following diagram (on the right) commutes.
  \begin{align*}
    \begin{tikzcd}[ampersand replacement=\&]
      \mc{D} \times \mc{C} \ar[r, "F"] \ar[d, swap, "K \times J"]
        \& \mc{C} \ar[d, "J"] \ar[dl, Rightarrow, shorten=13pt, swap, "t"] \\
      \mc{B} \times \mc{A} \ar[r, swap, "G"]
        \& \mc{A}
    \end{tikzcd}
    &&
    \begin{tikzcd}[ampersand replacement=\&]
      J(F(D,R)) \ar[r, "t"] \ar[d, swap, "r"]
        \& G(K(D),J(R)) \ar[r, dashed, "G(K(D){,} h)"]
        \& G(K(D),C) \ar[d, "c"]\\
      J(R) \ar[rr, swap, dashed, "h"]
        \&
        \& C
    \end{tikzcd}
  \end{align*}
\end{theorem}

\noindent This uniformity property can also be used to derive hss as they
are a particular case of parametrised initiality.

\Cref{thm:FioreSaville} presented above is actually a partial version
of \cite[Theorem 4.7]{ListObjects17}.
Indeed, \cite[Theorem 4.7]{ListObjects17} features additional assumptions and conclusions:
it additionally assumes that $\mc{A}$
has initial objects and $\omega$-colimits, and, given $B : \mc{B}$, for
$G(B,\_)$ to be $\omega$-cocontinuous;
and additionally concludes that there
is a natural transformation $t : J \circ \mu_F \to \mu_G \circ K$ associated
to the initial $G(KD,\_)$-algebra.
The additional content is actually a corollary of \cref{thm:FioreSaville}.
Hence, it can be separated from \cite[Theorem 4.7]{ListObjects17} to get the
stronger \cref{thm:FioreSaville}.

Doing so is important in two regards.
Firstly, as noticed in \cite{CoqPl2023MonCatHss}, the extra assumptions
on $\mc{A}$ and $G$ are not needed to prove the initiality theorem.
Secondly, though it might be surprising at first sight due to differences of
presentation, Mendler's style iterations and \cref{thm:FioreSaville} are
actually equivalent, hence the differences between the two proofs
boil down to the differences between hss and parametrised initiality.

In one direction,
as remarked in \cite{CoqPl2023MonCatHss}, \cref{thm:FioreSaville} can be
deduced from Mendler's style iteration by setting $F := F(D,\_)$, $L := J$,
$X := C$ and $\Psi\;h \mapsto c \circ G(KD,C) \circ t$.

Conversely, as noticed by Lafont, Mendler's style iteration can be deduced
from the \cref{thm:FioreSaville} by setting $\mc{B,D} := 1$, $\mc{C} :=
\mc{C}$, $\mc{A} := \Set^\op$, and $F := F$, $L := \mc{D}(L(\_),X)$; this
enables us to set $t := \Psi$ by seeing $\Psi$ as a natural transformation of type
$C \to \Set^\op$ rather than of type $C^\op \to \Set$.
Applying \cref{thm:FioreSaville} to the trivial algebra $1 : \Set^\op$ then
yields a diagram in $\Set^\op$, or in $\Set$ as right-below, which provides
us with Mendler's style iteration when evaluated in $\star : 1$.
\begin{align*}
  \begin{tikzcd}[ampersand replacement=\&]
    \mc{C} \ar[r, "F"] \ar[d, swap, "\mc{D}(L(\_){,}X)"]
      \& \mc{C} \ar[d, "\mc{D}(L(\_){,}X)"] \ar[dl, Rightarrow, shorten=13pt, swap, "\Psi"] \\
    \Set^\op \ar[r, swap, "\Id"]
      \& \Set^\op
  \end{tikzcd}
  &&
  \begin{tikzcd}[ampersand replacement=\&]
    \mc{D}(L(F(R)),X)
      \& \mc{D}(L(R),X)  \ar[l, swap, "\Psi"]
      \& 1 \ar[l, swap, dashed, "h"] \\
    \mc{D}(L(R),X) \ar[u, "(L(r)^*"]
      \&
      \& 1 \ar[u] \ar[ll, dashed, "h"]
  \end{tikzcd}
\end{align*}

In the less general case where precomposition $\_ \circ Z$ has a right adjoint
$R_Z$, no details are given in \cite{FPT99,SecondOrderDep08} on how to derive
parametrised initiality; however, a remark in \cite{ListObjects17} states that it ``is well-known, for a left-closed monoidal category [\ldots], the notions of initiality and of parametrised initiality
coincide''.
Indeed, in monoidal categories, parametrised initiality implies initiality by
setting $Z := I$.
Conversely, by the adjunction, building a morphism $R \circ Z \to C$ is
equivalent to building a morphism $R \to R_Z(C)$, which by initiality can be
done by building an algebra structure for $R_Z(C)$.
Building the appropriate algebra then yields parametrised initiality
when transported back along the adjunction.

The approach using hss \cite{Hss04,HssRevisited15} relies instead on a
variant of generalised Mendler's style iteration \cite[Theorem 2]{GeneralisedFold99},
that requires $L$ (instantiated to be precomposition) to have a right adjoint.
Investigating the proof \cite[Theorem 2]{GeneralisedFold99} yields the same
construction, therefore once again both proofs boil down to the difference
between hss and parametrised initiality.

\subsubsection{Applications}
As the framework above was defined from scratch for monoidal categories, it
can and has been applied to more involved instances than $[\mb{F},\Set]$ and
untyped languages.
However, note that most of the applications rely on closedness, which
restrict which monoidal categories can be used, as they were published
before \cite{ListObjects17}.

Simply-typed languages where first investigated in \cite[Section
II.1.1]{Cbn02}, where Fiore explains that the simply-typed lambda calculus
is an initial algebra on $[\mb{F} \downarrow T,\Set]^T$, and suggests that
substitution could also be accounted for using the framework of \cite{FPT99}.
This was briefly detailed in \cite[Section 1.3]{MMCCS05}, where Fiore
additionally discusses its monoidal structure.
This was fully worked out, and in great detail, by Zsidó.
In her dissertation \cite[Chapter 5]{ZsidoPhd10}, Zsidó fully proves,
without using any high-level theorem, that simply-typed algebraic
signatures yield strong endofunctors on $[\mb{F} \downarrow T,\Set]^T$, that
the category $[\mb{F} \downarrow T,\Set]^T$ is a left-closed monoidal category, and that these signatures yield initial
$\Sigma$-monoids on it.
In \cref{sec:example_setT_setT}, we have chosen to develop the case of
$[\Set^T,\Set^T]$ instead, for its simple monoidal structure, yet $[\mb{F} \downarrow
T,\Set]^T$ is also encompassed, and should have an accessible presentation
in terms of relative monads, as done in \cref{sec:example_F_set} for
$[\mb{F},\Set]$.

Hamana investigated polymorphic languages in \cite{Polymorphism11}.
For system $F$ and $F_\omega$, Hamana builds categories $\int G$ and $\int
H$ such that the system $F$ and $F_\omega$ are initial algebra on
$\Set^{\int G}$ and $\Set^{\int H}$.
He then introduces polymorphic and higher-order polymorphic counterparts to
algebraic signatures, and generalises the constructions to them.
This instance is also encompassed by our framework, but we refer to
\cite{Polymorphism11} for the detailed construction of the categories $\int
G$ and $\int H$.

Fiore also suggested an initial framework for dependently typed languages in
\cite[Section II]{SecondOrderDep08}.

\subsubsection{Extension to skew-monoidal categories}
\label{subsubsec:rw-skew}

Monoidal categories are practical but do not encompass all interesting
categories.
Therefore, to apply Fiore's framework of $\Sigma$-monoids to untyped languages on $[\N,\Set]$,
Borthelle, Tom Hirschowitz\footnote{ Be aware that both André Hirschowitz
and Tom Hirschowitz worked on initial semantics, sometimes in joint work. To
distinguish them, we always refer to them by their full name.} and Lafont
generalised it to skew-monoidal categories \cite{CellularHoweTheorem20}, and
formalised it in Coq.

To our knowledge, modules over monoids also generalise to skew-monoidal
categories, however, as of yet, the proofs based on hss and parametrised
initiality do not seem to generalise.
Indeed, we can no longer prove the monoid's associativity law that is an
equality of morphisms of type $(R \otimes R) \otimes R \to R$, by instantiating hss
and parametrised initiality for $Z := R \otimes R$.
The reason is that this instantiation provides uniqueness of a morphism of type $R \otimes (R \otimes R) \to
R$, but as the associator $\alpha$ is not invertible in skew-monoidal categories, this is
no longer equivalent to uniqueness of a morphism $(R \otimes R) \otimes R \to R$.
Thus, we could have generalised the framework presented here, by adapting
the proof of \cite{CellularHoweTheorem20} that repeatedly applies
\cref{thm:FioreSaville}, but it would be at the cost of not using hss in our
proofs.
We have chosen not to as monoidal categories still cover most examples, and
as our objective is to relate the different approaches.

Skew-monoidal categories were also considered in \cite{NamelessDummies22} by
André and Tom Hirschowitz, Lafont and Maggesi.
Therein, they study De Bruijn monads and De Bruijn S-algebras, for untyped
languages, and identify them as monoids and $\Sigma$-monoids in the skew-monoidal
category $[1,\Set]$, for the relative functor $J : * \mapsto \N$.

Nevertheless, the initiality theorem proven in \cite{CellularHoweTheorem20}
for skew-monoidal categories fails to apply to $[1,\Set]$ as it ``assumes
that the tensor product is finitary in the second argument'', and thus the authors need
to prove an initiality theorem from scratch.
This is because they directly proved an adjoint theorem, and deduced the
initiality theorem from it.
Indeed, as discussed before, proving the theorems in the opposite order enables us to remove the
assumption on $X \otimes \_$ in the initiality theorem, and therefore would
have enabled them to directly apply it to $[1,\Set]$.
The authors have also considered simply-typed languages, using the
skew-monoidal category $[1,\Set^T]$, for the relative functor $J : *,t
\mapsto \N$.
Their main theorem was formalised in Coq and also in HOL Light.

\subsubsection{Further work on $\Sigma$-monoids}

Though this is not the subject of this article, work on $\Sigma$-monoids
has been develop further to account for meta-variables and equations.
Meta-variables for $\Sigma$-monoids were first developed by Hamana in
\cite{HamanaMetavar04}, and by Fiore in \cite[Section I.2]{SecondOrderDep08}.
Equational systems were considered in Hur's dissertation \cite{HurPhd}
supervised by Fiore.
This work was published with Fiore in \cite{FioreHur07,FioreHur09,FioreHur08},
but is partially extended in Hur's dissertation.
Equations and second-order languages were also considered in
\cite{FioreHur10,FioreMahmoud10}, for polymorphic languages in
\cite{PolymorphismEq13}
This work was recently used to design an Agda framework for reasonig with
higher-order languages and substitution in \cite{FioreSzamozvancevPopl22}.

\subsection{Modules over monads}
\label{subsec:rw-modules-over-monoids}

\subsubsection{Origins}

Modules over monads were studied, in \cite{HirschowitzMaggesi07}, by
André Hirschowitz and Maggesi, in order to capture the substitution
structure of higher-order languages' constructors.
On $\Set$, the authors used them to define models --- defined to be monoids with
suitable module morphisms on them ---  and stated that every untyped
algebraic signature has an initial model.
They also considered how modules over monads could encompass more notions, such as
simply-typed syntax, or some form of semantic properties.
This work was refined in \cite{HirschowitzMaggesi10}, where it has been
enriched with a proof based on syntax trees, that is specific to
$[\Set,\Set]$ and to untyped algebraic signatures.

This approach was extended, by Zsidó in her Ph.D.~\cite[Chapter 6]{ZsidoPhd10},
to simply-typed languages on $[\Set^T,\Set^T]$.
Firstly, she has shown that modules over monads encompass simply-typed
algebraic signatures, and extended models to the simply-typed case.
Secondly, she extended the proof based on syntax trees to prove that every
simply-typed algebraic signature has an initial model.
It was then formalised in Coq by Ahrens and Zsidó in \cite{ISCoq10}
using Coq's built-in inductive types.

\subsubsection{Modules over monads vs Modules over monoids}
\label{subsubsec:module_monads_vs_monoids}

Compared to the original work on the subject, our framework does not rely on
modules over monads but on \hyperref[def:modules]{modules over monoids} in a monoidal category.
This is because we defined a general framework for monoidal categories,
that can handle $[\Set,\Set]$ just like $[\mb{F},\Set]$; but unlike modules
over monoids, modules over monads can be defined in functor
categories that are not necessarily endo:
\begin{definition}[Modules over monads]
  Given a monad on $\mc{C}$, $(R,\eta,\mu) : \Mon([\mc{C},\mc{C}])$,
  a module over $R$ with codomain $\mc{D}$ is a tuple $(M,p^M)$ such that
  $M : \mc{C} \to \mc{D}$ is a functor and $p : M \circ R \to M$ is natural
  transformation compatible with $\eta$ and $\mu$ as in \cref{def:modules}.
\end{definition}
\noindent Indeed, for a monoidal category as $\_ \otimes \_ : \mc{C} \times \mc{C} \to
\mc{C}$, $M \otimes R$ is only well-defined for $M : \mc{C}$, and thus $M$
and $R$ can not be of different types.
Moreover, even for functor categories $[\mc{B},\mc{C}]$, it is not
definable as $M \circ R : \mc{B} \to \mc{D}$ and $M : \mc{C} \to \mc{D}$
are of different types.

In the case of endofunctor categories $[\mc{C},\mc{C}]$, modules over monads
are more general than modules over monoids, as modules over monoids are
exactly modules over monads with codomain $\mc{C}$.
The added power of choosing $\mc{D}$ was thought to handle the simply-typed
case, and indeed, in \cite[Chapter 6]{ZsidoPhd10} modules over monads on
$[\Set^T,\Set^T]$ with codomain $\Set$ are used to model simply-typed algebraic
signatures.
The idea being that the output set of modules can be used to represent the
typed terms of the inputs and output of the constructors.

Nevertheless, in practice modules over monads do not seem to add any
expressive power.
In the untyped case, for the category $[\Set,\Set]$
\cite{HirschowitzMaggesi07,HirschowitzMaggesi12,PresentableSignatures21,ReductionMonads20},
$\mc{D}$ is always chosen to be $\Set$, in which case both notions
coincides.
In the simply-typed case, as seen in \cref{sec:example_setT_setT}, the added
flexibility of modules over monads is not necessary to provide a language
with an appropriate initial model.
Moreover, as explained in \cite[Section 2.4]{TransitionMonads22}, for any
type $t : T$ there is an adjunction on the functor categories, that yields
an adjunction between modules over monads and modules over monoids:
\begin{align*}
  \begin{tikzcd}[ampersand replacement = \&, column sep=large]
    [\Set^T{,}\Set]            \ar[r, bend left, "y(t) \circ \_"]
                               \ar[r, phantom, "\perp"]
      \& {[}\Set^T{,}\Set^T{]} \ar[l, bend left, "(\_)_t \circ \_"]
  \end{tikzcd}
  &&
  \begin{tikzcd}[ampersand replacement = \&, column sep=large]
    \Mod(\Set^T{,}\Set)
        \ar[r, bend left]
        \ar[r, phantom, "\perp"]
      \& \Mod(\Set^T{,}\Set^T) \ar[l, bend left]
  \end{tikzcd}
\end{align*}
Here $y(t) : T \to \Set^T$ is the \hyperref[def:yoneda-emb]{Yoneda
embedding}, and $(\_)_t : \Set^T \to T$ is the projection.
This adjunction implies that modelling an algebraic constructor using
modules over monoids as $[\Theta_{s_1}^{l_1}]_{t_1} \times ... \times
[\Theta_{s_n}^{l_n}]_{t_n} \to \Theta$, or defining them using modules over
monads with codomain $T$ as $[\Theta^{l_1}]_{s_1} \times ... \times [\Theta^{l_n}]_{s_n} \to
[\Theta]_t$, as done in \cite[Chapter 6]{ZsidoPhd10} and \cite{ISCoq10}, is
equivalent.
Therefore, in practice, there is no difference in terms of models between
using modules over monads or modules over monoids.

However, using modules over monoids makes a real difference when it comes to
building an initial model.
In the case of modules over monoids, using that all the outputs of the
constructors can be set to the trivial signature $\Theta$, we summed the
inputs into one single signature $\Sigma$, such that we can represent our
constructors as a single morphism of modules $\Sigma \to \Theta$.
This is required to apply the proof of \cref{sec:building_initial_model} based on hss,
as it requires an initial algebra for the functor $I + \Sigma$.
It is not possible for modules over monads, as already for algebraic signatures
the modules describing the outputs $[\Theta]_t$ depend on the output types
$t$, and hence are all different.
Therefore, it is not possible to directly apply the proof of
\cref{sec:building_initial_model}.
This is of particular importance as the proof using syntax trees is less
modular and significantly longer than the one using hss.
Moreover, the proof using hss applies to any appropriate monoidal category
whereas the one using syntax tree is specific to $[\Set^T,\Set^T]$.

Lastly, modules over monads have the advantage of directly connecting to the
work of Fiore and hss as described in \cref{remark:model-sigma-monoids}.
For all of those reasons, we prefer modules over monoids to modules
over monads.

\subsubsection{Earlier attempt at comparing traditions: Zsidó's Ph.D.\ thesis}

Though little-known, one of the most important works, for our purpose of
relating the different traditions, is Zsidó's dissertation \cite{ZsidoPhd10}.
There, Zsidó relates Fiore's work
\cite{FPT99,Cbn02} to Hirschowitz and Maggesi's work
\cite{HirschowitzMaggesi07,HirschowitzMaggesi10} on initial semantics; those use,
$[\mb{F},\Set]$ and $[\Set,\Set]$, respectively, in the untyped case, and
$[\mb{F}\downarrow T]^T$ and $[\Set^T,\Set^T]$ in the simply-typed case.

In the untyped case, where modules over monads coincide with modules over
monoids, Zsidó generalises modules over monoids and models from $[\Set,\Set]$
to monoidal categories, much like in \cref{subsec:modules,subsec:models}, in
order to express the work on $[\mb{F},\Set]$ and $[\Set,\Set]$ in the same
language.
She then proves, by appropriately propagating the adjunction between
$[\mb{F},\Set]$ and $[\Set,\Set]$ throughout the framework, that for any
algebraic signature, there is an initial model on $[\mb{F},\Set]$ if and
only if there is one for $[\Set,\Set]$, and that they can be constructed
from one another.

In the simply-type case, she starts by reviewing the work of Fiore
\cite{Cbn02} in \cite[Chapter 5]{ZsidoPhd10}, and by completing
the work of André Hirschowitz and Maggesi sketched in
\cite[Section 6.3]{HirschowitzMaggesi07} in \cite[Chapter 6]{ZsidoPhd10}.
However, in her work on simply-typed syntax, Zsidó uses modules over monads
with codomain $T$, preventing her from defining both frameworks in the same
language, and fully relating the two approaches.
Consequently, in \cite[Chapter 7]{ZsidoPhd10}, she only provides
an adjunction between $\Sigma$-monoids and models.
Furthermore, the adjunction on models no longer comes from parametricity in
the input monoidal category but is now lifted ad hoc.

\subsubsection{Signatures}
\label{subsubsec:rw-modules-sig}

In all previous discussed work on modules over monads, all the signatures
considered were purely syntactic, either \hyperref[def:untyped-alg-sig]{untyped algebraic signatures}
or \hyperref[def:simply-typed-alg-sig]{simply typed algebraic signatures}.
The general notion of signatures, as in \cref{subsec:signatures}, was
introduced by André Hirschowitz and Maggesi in \cite{HirschowitzMaggesi12}
on $[\Set,\Set]$.
Still on $[\Set,\Set]$, the authors provided their signatures with the
modularity property described in \cref{subsec:models}, and established a
connection with signatures with strength as in
\cref{subsec:sigstrength-to-sig}.

This work was expanded on and formalised in Coq in
\cite{PresentableSignatures21} , still on $[\Set,\Set]$, by Ahrens, André
Hirschowitz, Lafont and Maggesi, where they introduced
\hyperref[def:pres-sig]{presentable signatures}.
Presentable signatures are interesting as they are representable but form
a larger class than purely syntactic signatures.
In practice examples of presentable signatures are covered by signatures
with strength but a more precise link is still to be established.

In two previously cited papers, the authors worked on $[\Set,\Set]$ where
modules over monoids and modules over monads coincide, and where
constructors can be represented by module morphisms $\Sigma \to \Theta$.
Since this is not possible for modules over monads, many authors
\cite{ZsidoPhd10,ISCoq10,ExtendedInitiality12,UntypedRelativeMonads16,TypedRelativeMonads19}
only considered syntactic signatures, and defined a model to be not a monoid
with a single module morphisms, but a monoid with a family of module morphisms.
As we have shown, it is not necessary for our purpose, and not doing so
enables us to consider more general notions of signature.

More generally, it would also be possible to require a second signature to
specify the output rather than using $\Theta$.
Yet, as we have no use for non trivial outputs, and as it simplifies the
framework, we limit ourself to $\Theta$ for the output signature.

\subsubsection{A generalized recursion principle for simply-typed languages}

As explained in \cref{subsec:rec_SetT_SetT}, the recursion principle for the
simply-typed case is less directly applicable than in the untyped case.
To provide it with a better recursion principle, in a technical paper
\cite{ExtendedInitiality12}, Ahrens reworked the entire framework.
He internalised the typing system $T$ by replacing monads by ``$T$-monads'',
adapted modules, signatures, and models, before proving an initiality theorem.
While this framework is interesting, it is very specific to endofunctor
categories $[\Set^T,\Set^T]$ and it is unclear how it relates to the
framework presented here.
He also formalised this framework in Coq and used it to provide a verified
translation of PCF into the untyped lambda calculus.

\subsubsection{Further work on initial semantics}

Though this is not the subject of this article, modules over monads have
been used further to add semantics on top of the raw-syntax framework
that is described in this work.
Modules over relative monads and a notion of 2-signature were used in
\cite{UntypedRelativeMonads16,TypedRelativeMonads19}, to model context-passing reduction
rules for untyped and simply-typed algebraic signatures.
Similar notions were used in \cite{2Signatures19} to add equations in the
untyped case.
The original work on reduction rules by Ahrens, was extended in the untyped
case to handle conditional rules as top-level $\beta$-reduction
\cite{ReductionMonads20},
and later extended to account simply-typed languages and strategies e.g.
call-by-value reduction in \cite{TransitionMonads20,TransitionMonads22}.

\subsection{Heterogeneous Substitution Systems}
\label{subsec:rw-hss}

\subsubsection{Origins}

\hyperref[def:hss]{Heterogeneous substitution systems} (hss) were introduced on
endofunctor categories $[\mc{C},\mc{C}]$ by Matthes and Uustalu in \cite{Hss04}.
Studying pointed strong functors as signatures, they design hss as intermediate
abstraction to prove that both wellfounded or non-wellfounded higher-order
languages have a well-behaved substitution structure, that is form monads.
In the well-founded case, they have proved that assuming that precomposition $\_ \circ Z$ has a
right-adjoint for all $Z$, then any signature with strength that has an initial
algebra has an associated hss, and as such a monad structure.
In the non-wellfounded case, they have proved then any signature with
strength that has a cofinal algebra has an associated hss, and as such a
monad structure.

This work was later strengthened into a framework for initial semantics by
Ahrens and Matthes in \cite{HssRevisited15}.
They assemble signatures with strength and hss in categories, prove a
\hyperref[thm:fusion-law]{fusion law} for generalised Mendler's style iteration,
and use it to prove that the hss built in \cite{Hss04} is actually initial as an
hss.
Thus, defining a framework for initial semantics based on hss, which was
formalised in Coq's UniMath library.
Compared to them, we have chosen to keep hss as an intermediate abstraction
in the proof rather than making it our notion of model.
Firstly, it enables us to have a common notion of model connecting the work
of Fiore, modules over monads, and hss.
Secondly, hss provide less explicit information on substitution than
monoids, and importantly have a stricter to apply / stronger recursion
principle, that we have found no specific use for.

Hss were applied, by Ahrens, Matthes, and Mörtberg in \cite{HssUntypedUniMath19},
to untyped algebraic signatures to construct inductive types and untyped
higher-order languages from elementary type constructors in the UniMath library.
The proof in \cite{Hss04} is based on left-closedness and the associated version
of Mendler's style iterations \cite[Theorem 2]{GeneralisedFold99}.
However, as discussed in detail in \cref{subsec:rw-sigma-mon}, $[\Set,\Set]$ is
not left-closed, therefore they turn to $\omega$-cocontinuity and switch to
the associated version of Mendler's style iterations \cite[Theorem 1]{GeneralisedFold99}.
The proof presented in \cref{sec:building_initial_model} is based on it.
We refer to \cref{subsec:rw-sigma-mon} for more detail on left-closedness
compared to $\omega$-cocontinuity.

\subsubsection{Generalised strength}

This work was applied to simply-typed algebraic signatures on $[\Set^T,\Set^T]$
in \cite{HssTypedUnimath22}.
To be able to do so modularly, they introduced generalised strength:
\begin{definition}[Generalised Strength]
  A \emph{generalised strength} for a functor $H : [\mc{C},\mc{D}'] \to [\mc{C},\mc{D}]$
  is a natural transformation $\theta$ such that for all $A : [\mc{C},\mc{D}']$,
  and pointed object $b : \Id \to B$ with $B : [\mc{C},\mc{C}]$:
  \[ \theta_{A,b} : H(A) \circ B \longrightarrow H(A \circ B) \]
  that verifies associativity and the unit as in \cref{def:sig-strength}.
\end{definition}
\noindent Just like strengths correspond to modules over monoids, by setting
$\mc{D}' := \mc{C}$, generalised strengths correspond to modules over
monads.
Indeed, given an endofunctor $H : [\mc{C},\mc{C}] \to [\mc{C},\mc{D}]$ with
generalised strength $\theta$, then for any monad $(R,\eta,\mu) : \Mon([\mc{C},\mc{C}])$,
$H(R)$ with the action $H(R) \circ R \xrightarrow{\theta_{R,\eta}} H(R \circ R)
\xrightarrow{H(\mu)} H(R)$ is a module over the monad $R$ with codomain $\mc{D}$.
Nonetheless, their proofs do no amount to working with modules over monads as
they only use generalised strengthened to build regular strength more
modularly, and only use regular strength in the end.
As shown in \cref{sec:example_setT_setT}, the extra structure of
generalised strengths is not necessary for modularity.

\subsubsection{Hss and monoidal categories}

Hss have been applied to build a substitution structure for simply-typed
\emph{coinductive} higher-order languages by Matthes, Wullaert, and Ahrens in
\cite{HssNonWellfounded23}.
To do so they have provided an extensive analysis of signatures with strength in
terms of actions and actegories, and used actegories as an abstract framework to
equip signatures with appropriate strengths.
While this can ease formalisation, it is quite technical, and we decided to
use a more direct approach to ease understanding.

They have also independently generalised hss to monoidal categories, and though
it is not the subject of their paper, provided a proof of the initiality theorem
basically as in \cref{sec:building_initial_model}.
Compared to us, the main difference is that their definitions are stated using
the \emph{reversed} monoidal category of ours \cite[Example 1.2.9]{2DimensionalCategories20}.
That is the monoidal category with the reversed monoidal product, i.e. such that
$X \otimes' Y := Y \otimes X$.
For instance, for them, strength are natural transformations $\theta : A
\otimes' H(B) \to H(A \otimes' B)$.
In practice it makes no difference as all monoidal categories are reversible,
and as they instantiate their framework with reversed categories too.

%% file: Sections/conclusion.tex
\section{Conclusion}
\label{sec:conclusion}

We have provided an introduction to different approaches
\cref{sec:models,sec:initiality_theorem,sec:building_initial_model},
and related it to the existing literature on this topic in \cref{sec:related-work}.
Doing so we have shown that modules over monoids and associated concepts
provide an abstraction to talk about constructors and substitution, that
restricting signatures to signatures with strength enables us to state an
initiality theorem (\cref{thm:initiality-theorem}) and an adjoint theorem
(\cref{thm:adjoint-theorem}), and that hss are a tool to prove
the initiality theorem.
Moreover, we have established precise links with existing frameworks and
their different variations, including a precise link between proofs based on
hss and the ones based on parametrized initiality.
To illustrate that our framework is appropriate for initial semantics, we
have also provided fully worked out examples for basic instances in the
literature: untyped higher-order languages on $[\Set,\Set]$ and
$[\mb{F},\Set]$ in \cref{sec:example_set_set,sec:example_F_set}, and
simply-typed ones on $[\Set^T,\Set^T]$ in \cref{sec:example_setT_setT}.
With this work, we aim to solidify and conceptualize the foundations of
initial semantics, and to provide an accessible presentation to concepts and
proof scattered throughout many papers, in order to provide a basis for
generalizing these results to more advanced notions and languages.

Nevertheless, there are still many open research questions in the area of
initial semantics; to consolidate the current knowledge on the subject, but
also to keep developing the framework.

Though we have set a basis for a better understanding, there is still work
to do to fully unify the existing frameworks:
\begin{itemize}
  \setlength\itemsep{-1pt}
  \item We have discussed and related signatures and signature with
        strength, as well as signatures with strength and presentable
        signatures.
        Yet, more precise links between those notions are still to be
        established.
  \item In this paper, we have only related different approaches on syntax.
        It remains to understand how the different approaches to semantics relate.
        For instance, how does the work on equations of \cite{FioreHur10} or
        \cite{FioreSzamozvancevPopl22} relate to the one in
        \cite{2Signatures19}.
  \item Similarly, both \cite{UntypedRelativeMonads16} and
        \cite{ReductionMonads20,TransitionMonads22} are concerned with
        reductions rules and have an interpretation in terms of relative
        monads.
        Could both works be encompassed by a more general framework for
        reduction rules?
\end{itemize}

A main challenge towards generalizing the framework is to be able to
appropriately handle advanced languages, with advanced type systems:
\begin{itemize}
  \setlength\itemsep{-1pt}
  \item There is little work on type systems with polymorphism, such as
        System F.
        Exceptions are \cite{Polymorphism11,PolymorphismEq13}, but those
        short conference papers do not contain any proofs, and involve
        rather complicated monoidal structure.
        Could a more rigorous account of polymorphic languages be given, and
        in particular of some modern languages?
        For instance using relative monads or endofunctor categories?
  \item As explained in \cref{subsec:rec_SetT_SetT}, the recursion principle
        for simply-typed languages, and more generally for typed languages,
        is not currently fully satisfying as its naive application is
        confined to languages with the same type systems.
        Could this principle be extended either by a more suitable
        application as suggested in \cref{subsec:rec_SetT_SetT}, or would
        this require a complete refinement of the framework as done in
        \cite{ExtendedInitiality12} ?
  \item As discussed in \cref{subsec:rw-sigma-mon}, the framework presented
        here could also by generalized from monoidal categories
        to skew-monoidal categories by adopting the proof of \cite{NamelessDummies22}.
        This would offer a new interpretation of models, for instance in
        terms of De Bruijn monads as in \cite{NamelessDummies22}.
        However, as of yet, hss and parametrized initiality do not seem to
        easily generalise to skew-monoidal categories.
        Generalizing them would be interesting both to modularize proofs,
        and for non-wellfounded syntax.
\end{itemize}